\documentclass[11pt, a4paper]{article}

\usepackage{vicari-mathematics}

\title{Simplex based Steiner tree instances yield large\texorpdfstring{\\}{ }integrality gaps for the bidirected cut relaxation}
\author{
	Robert Vicari%\footnote{Email: \texttt{vicari@uni-bonn.de}}%\\[6pt]
}
\date{}

\usepackage[left=2.5cm, right=2.5cm, top=3cm, bottom=3cm]{geometry}
\newcommand{\BCRimproved}{\hyperref[lp:BCRimproved]{\(\operatorname{BCR^{+}}\)}}
\newcommand{\MCFRimproved}{\hyperref[lp:MCFRimproved]{\(\operatorname{MCFR^{+}}\)}}

\newcommand{\BCRclass}{\hyperref[lp-class:BCR]{\(\operatorname{\mathcal{BCR}}\)}}
\newcommand{\BCRimprovedclass}{\hyperref[lp-class:BCRimproved]{\(\operatorname{\mathcal{BCR}^{+}}\)}}
\newcommand{\HYP}{\hyperref[lp-class:HYP]{\(\operatorname{\mathcal{HYP}}\)}}

\newcommand{\SGds}{\hyperref[graph:SGds]{\(\operatorname{SG_{d, s}}\)}}

\newcommand{\SIds}{\hyperref[instance:SIds]{\(\operatorname{SI_{d, s}}\)}}
\newcommand{\SIdsdelta}{\hyperref[instance:SIdsdelta]{\(\operatorname{SI_{d, s, \delta}}\)}}

\begin{document}

\maketitle

\begin{abstract}
	The \textit{bidirected cut relaxation} is the characteristic representative of the \textit{bidirected relaxations} (\(\operatorname{\mathcal{BCR}}\)) which are a well-known class of equivalent \(\operatorname{LP}\)-relaxations for the \(\operatorname{NP}\)-hard \textit{Steiner Tree Problem in Graphs} (\(\operatorname{STP}\)).
	Although no general approximation algorithm based on \(\operatorname{\mathcal{BCR}}\) with an approximation ratio better than \(2\) for \(\operatorname{STP}\) is known, it is mostly preferred in integer programming as an implementation of \(\operatorname{STP}\), since there exists a formulation of compact size, which turns out to be very effective in practice.
	
	It is known that the integrality gap of \(\operatorname{\mathcal{BCR}}\) is at most \(2\), and a long standing open question is whether the integrality gap is less than \(2\) or not.
	The best lower bound so far is \(\frac{36}{31} \approx 1.161\) proven by Byrka et al.\ \cite{BGRS13}.
	Based on the work of Chakrabarty et al.\ \cite{CDV11} about embedding \(\operatorname{STP}\) instances into simplices by considering appropriate dual formulations, we improve on this result by constructing a new class of instances and showing that their integrality gaps tend at least to \(\frac{6}{5} = 1.2\).
	
	More precisely, we consider the class of equivalent LP-relaxations \(\operatorname{\mathcal{BCR}}^{+}\), that can be obtained by strengthening \(\operatorname{\mathcal{BCR}}\) by already known straightforward Steiner vertex degree constraints, and show that the worst case ratio regarding the optimum value between \(\operatorname{\mathcal{BCR}}\) and \(\operatorname{\mathcal{BCR}}^{+}\) is at least \(\frac{6}{5}\).
	Since \(\operatorname{\mathcal{BCR}}^{+}\) is a lower bound for the \textit{hypergraphic relaxations} (\(\operatorname{\mathcal{HYP}}\)), another well-known class of equivalent LP-relaxations on which the current best \((\ln(4) + \varepsilon)\)-approximation algorithm for \(\operatorname{STP}\) by Byrka et al.\ \cite{BGRS13} is based, this worst case ratio also holds for \(\operatorname{\mathcal{BCR}}\) and \(\operatorname{\mathcal{HYP}}\).
	
	\remarks{Keywords}{Steiner Tree Problem \and Bidirected Cut Relaxation \and Integrality Gap}
	\remarks{Remarks}{This work is a revised version of the author's master's thesis \cite{Vicari2018}.}
\end{abstract}

	%\pagebreak

\section{Introduction}\label{sec:introduction}

In the \shortmathproblem[STP]{Steiner Tree Problem in Graphs} one is given an undirected connected graph \(G = (V, E)\), a set of \textit{required} vertices \(R \subseteq V\) and non-negative edge-costs \(c : E \to \mathbb{R}_{\geq 0}\).
We refer to the triple \((G, R, c)\) as \textit{Steiner tree instance} or sometimes just \textit{instance}.
The task is to compute a tree \(T\) in \(G\) spanning \(R\), i.e.\ \(R \subseteq V(T) \subseteq V\) and \(E(T) \subseteq E\), which has minimum edge-cost, i.e.\ \(c(E(T)) = \sum_{e \in E(T)} c(e)\) is minimum over all possible solutions.
We refer to a solution \(T\) as \textit{Steiner tree} and call the non-required vertices \(V(T) \setminus R\) \textit{Steiner vertices}.

\nameref{problem:STP} is NP-hard \cite{Karp1972} and actually cannot even be approximated within a factor smaller than \(\frac{96}{95} \approx 1.010\) unless \(\operatorname{P} = \operatorname{NP}\) \cite{CC08}.
A simple \(2\)-approximation can be obtained by computing a minimum spanning tree on \(R\) in the metric closure of \(G\) and \(c\) \cite[Sec.\ 3.1.1]{Vaz01}.
The currently best known approximation algorithm by Byrka et al.\ \cite{BGRS13} has a guarantee of \(\ln(4) + \varepsilon\) for arbitrarily small fixed \(\varepsilon > 0\) and is based on iteratively solving an LP-relaxation for \nameref{problem:STP}, called \textit{directed component-based relaxation} which is introduced in \cite{PD03}.
The LP is part of a class of equivalent LP-relaxations, commonly called \textit{hypergraphic relaxations} \label{lp-class:HYP}(\(\operatorname{\mathcal{HYP}}\)) \cite{CKP13}.
The integrality gap of these LPs is at most \(\ln(4)\) and they are strongly NP-hard to solve exactly, as shown in \cite{GORZ12}.
Although Goemans et al.\ \cite{GORZ12} introduce a method, using the notion of matroids, such that one only has to solve a hypergraphic relaxation once and still obtains a \((\ln(4) + \varepsilon)\)-approximation, solving the LP is still the bottleneck, as reported in \cite{FKOS16}, rendering the algorithm intractable for most applications.

Therefore Feldmann et al.\ \cite{FKOS16} study the relation to the efficiently solvable class of \textit{bidirected relaxations} (\(\operatorname{\mathcal{BCR}}\)) \cite{GM93}, for which the \textit{bidirected cut relaxation}, introduced in \cite{Wong1984}, is the characteristic representative.
In \cite{FKOS16} it is shown that the LP-relaxations in \(\operatorname{\mathcal{HYP}}\) are equivalent to the ones in \(\operatorname{\mathcal{BCR}}\) for so called \textit{Steiner claw-free} instances.
These are instances where there exists no non-required vertex with \(3\) or more neighbouring non-required vertices.

The main advantage of the bidirected cut relaxation is that there exists an extended formulation of compact size which has been established in practice as a useful base in integer programming for \nameref{problem:STP} \cite{KM98}.
Therefore, the LP-relaxations in \(\operatorname{\mathcal{BCR}}\) gained a lot of attention, but so far no approximation algorithm based on them with guarantee less than \(2\) is known.
One symptom of that is the lack of knowledge about the integrality gap for these LP-relaxations, which has been conjectured to be less than \(2\), but only at most \(2\) could be proved.
Better bounds are only known for special cases.
For example the result in \cite{FKOS16} implies an upper bound of \(\ln(4)\) for Steiner claw-free instances.
Furthermore, in \textit{quasi-bipartite} instances, a special case of Steiner claw-free instances, the current best bound is \(\frac{73}{60} \approx 1.217\) \cite{GORZ12}.
These are instances where no two non-required vertices are neighbouring, i.e.\ each edge is incident to at least one required vertex.
Note for these instances it is still NP-hard to approximate within a factor smaller than \(\frac{128}{127} \approx 1.007\) \cite{CC08}.
If there are no non-required vertices at all, i.e.\ one is in the minimum spanning tree case, then Edmonds \cite{Edmonds1967} shows that the LP-relaxations in \(\operatorname{\mathcal{BCR}}\) are exact.
In \cite{Goemans1994b} it is shown that \nameref{lp:BCR} is also exact in series-parallel graphs.
These are graphs which do not contain \(\operatorname{K_{4}}\) as minor where \(\operatorname{K_{4}}\) is the complete graph on \(4\) vertices.

For tackling the question about the integrality gap for the LP-relaxations in \(\operatorname{\mathcal{BCR}}\), it is useful to study the characteristics of instances with large integrality gaps.
The best lower bound so far is \(\frac{36}{31} \approx 1.161\) proved in \cite{BGRS13} by constructing a series of instances whose integrality gaps tend to \(\frac{36}{31}\).
Actually these instances can be used to lower bound the integrality gap of a stronger LP which is an extension of the bidirected cut relaxation.
This LP is obtained by adding so called flow-balance constraints on each Steiner vertex \cite{KM98} and induces the class of \textit{bidirected relaxations with Steiner vertex degree constraints} (\(\operatorname{\mathcal{BCR}^{+}}\)) (details are given in \cref{sec:steiner-vertex-degree-constraints}).
In \cite{PD03} it is shown that the optimum solution value of the LP-relaxations in \(\operatorname{\mathcal{BCR}^{+}}\) is bounded by the optimum solution value of the LP-relaxations in \(\operatorname{\mathcal{HYP}}\).
We denote by \(\operatorname{opt}_{\mathcal{P}}(I)\) the optimum solution value for given instance \(I\) of an optimization problem \(\mathcal{P}\).
Then we have the following general connection between the three classes \BCRclass, \BCRimprovedclass\ and \HYP, already stated in \cite[Sec.\ 2.3]{Pritchard2009}:
\begin{theorem}\label{thm:steiner-tree-lp-relaxation-chain}
	Let \((G, R, c)\) be a Steiner tree instance, then:
	\[\operatorname{opt}_{\operatorname{\mathcal{BCR}}}(G, R, c) \leq \operatorname{opt}_{\operatorname{\mathcal{BCR}^{+}}}(G, R, c) \leq \operatorname{opt}_{\operatorname{\mathcal{HYP}}}(G, R, c) \leq \operatorname{opt}_{\operatorname{STP}}(G, R, c)\]
\end{theorem}
\begin{proof}
	The last inequality holds as all formulations are LP-relaxations for \nameref{problem:STP}.
	The first inequality is clear as we only strengthen the formulation.
	The second inequality is proved in \cite{PD03}.
\end{proof}
Hence a natural question is, whether there is a major difference between \(\operatorname{\mathcal{BCR}}\) and \(\operatorname{\mathcal{BCR}^{+}}\), because the latter class is not more complex than the first one and in current research mainly \(\operatorname{\mathcal{BCR}}\) is considered.

Before the \((\ln(4) + \varepsilon)\)-approximation by Byrka et al.\ \cite{BGRS13} LP-relaxations were not successfully used in the design of algorithms beating \(2\)-approximations for \nameref{problem:STP}.
In \cite{RZ05} the previous best approximation algorithm is presented and other combinatorial approximation algorithms beating the ratio \(2\) are also noted.
An LP inducing a \(2\)-approximation is for example the undirected cut relaxation introduced in \cite{Aneja1980}.
In \cite{AKR95} this LP is generalized to the \textit{Steiner Forest Problem} and the authors obtain also a \(2\)-approximation.
More details can be for example found in \cite[Ch.\ 22]{Vaz01}.
The hypergraphic relaxations are actually the first theoretically efficiently approximable LP-relaxations for \nameref{problem:STP} for which the integrality gap is known to be less than \(2\).
A survey of other LP-relaxations for \nameref{problem:STP} can be found for example in \cite{PD01, PD03, Polzin2003, Daneshmand2004, Pritchard2009}.

\subsection{Our contributions}\label{sec:our-contributions}

Based on the work of Chakrabarty et al.\ \cite{CDV11} about embedding Steiner tree instances into simplices, we will show that there are instances for which the ratio of the optimum values between \(\operatorname{\mathcal{BCR}}\) and \(\operatorname{\mathcal{BCR}^{+}}\) tends at least to \(\frac{6}{5} = 1.2\).
This also implies the same lower bound on the worst case ratio between \(\operatorname{\mathcal{BCR}}\) and \(\operatorname{\mathcal{HYP}}\).
Moreover, this shows that the integrality gap for the bidirected cut relaxation is at least \(\frac{6}{5}\) improving the bound in \cite{BGRS13}.
The notion of worst case ratio and integrality gap can be easily combined in the following notion of a gap.
\begin{definition}\label{def:gap}
	Let \(\mathcal{P}\) and \(\mathcal{L}\) be both minimization problems on the same instance set \(\mathcal{I}\) such that for each \(I \in \mathcal{I}\) we have \(0 < \operatorname{opt}_{\mathcal{L}}(I) \leq \operatorname{opt}_{\mathcal{P}}(I)\).
	Then the \textbf{gap} of \(I\) for \(\mathcal{L}\) to \(\mathcal{P}\) is defined as \(\bm{{\operatorname{gap}}_{\mathcal{L}, \mathcal{P}}(I)} := \frac{\operatorname{opt}_{\mathcal{P}}(I)}{\operatorname{opt}_{\mathcal{L}}(I)}\).
	The \textbf{gap} of \(\mathcal{L}\) to \(\mathcal{P}\) is then
	\(\bm{{\operatorname{gap}}_{\mathcal{L}, \mathcal{P}}} := \sup_{I \in \mathcal{I}} \, \operatorname{gap}_{\mathcal{L}, \mathcal{P}}(I)\)
\end{definition}
Note that in our case it is sufficient to consider only instances with \(\operatorname{opt}_{\mathcal{L}}(I) > 0\) since \(\operatorname{opt}_{\mathcal{L}}(I) = 0 \Rightarrow \operatorname{opt}_{\mathcal{P}}(I) = 0\) holds for all problems \(\mathcal{P}\) and \(\mathcal{L}\) we are dealing with.

For \(\mathcal{P} \in \{\operatorname{\mathcal{BCR}}, \operatorname{\mathcal{BCR}^{+}}, \operatorname{\mathcal{HYP}}, \operatorname{STP}\}\) denote by \(\operatorname{\mathcal{P}_{\leq k}}\) the restriction of \(\mathcal{P}\) to instances with at most \(k\) required vertices.
Then our main theorem is simply:
\begin{theorem}\label{thm:refined-bidirected-relaxations-gap}
	\[\operatorname{gap}_{\operatorname{\mathcal{BCR}_{\leq k}}, \operatorname{STP_{\leq k}}} \geq \operatorname{gap}_{\operatorname{\mathcal{BCR}_{\leq k}}, \operatorname{\mathcal{HYP}_{\leq k}}} \geq \operatorname{gap}_{\operatorname{\mathcal{BCR}_{\leq k}}, \operatorname{\mathcal{BCR}_{\leq k}^{+}}} \geq \frac{6 \cdot (k - 1)}{5 \cdot (k - 1) + 1}\]
\end{theorem}
The first two inequalities follow directly by \cref{thm:steiner-tree-lp-relaxation-chain}.
Taking the limit, we obtain:
\begin{corollary}
	\[\operatorname{gap}_{\operatorname{\mathcal{BCR}}, \operatorname{STP}} \geq \operatorname{gap}_{\operatorname{\mathcal{BCR}}, \operatorname{\mathcal{HYP}}} \geq \operatorname{gap}_{\operatorname{\mathcal{BCR}}, \operatorname{\mathcal{BCR}^{+}}} \geq \frac{6}{5}\]
\end{corollary}
Hence we improve the previously best known lower bounds \(\operatorname{gap}_{\operatorname{\mathcal{BCR}}, \operatorname{\mathcal{HYP}}} \geq \frac{8}{7}\) and \(\operatorname{gap}_{\operatorname{\mathcal{BCR}}, \operatorname{STP}} \geq \frac{36}{31}\).
For a complete list of currently best known lower and upper bounds see \cref{table:gap-bounds}.
Hence even in a theoretical perspective it is worthwhile to explicitly consider \(\operatorname{\mathcal{BCR}^{+}}\) in addition to \(\operatorname{\mathcal{BCR}}\).

\begin{table}
	\centering
	%\hspace*{\fill}
	\begin{subtable}[t]{0.475\textwidth}
		\centering
		\begin{tabular}{l|ccc}
			\(\leq \operatorname{gap_{\mathcal{L}, \mathcal{P}}}\) & \(\mathcal{BCR}^{+}\) & \(\mathcal{HYP}\) & STP\\
			\hline
			\(\mathcal{BCR}\)\rule{0pt}{15pt} & \(\frac{6}{5}\) & \(\frac{6}{5}\) & \(\frac{6}{5}\)\\[3pt]
			\(\mathcal{BCR}^{+}\) & & \(\frac{8}{7}\) & \(\frac{36}{31}\) \\[3pt]
			\(\mathcal{HYP}\) &  &  & \(\frac{8}{7}\)
		\end{tabular}
		\caption{
			Note that \(\frac{6}{5} = 1.2\), \(\frac{36}{31} \approx 1.161\) and \(\frac{8}{7} \approx 1.142\).
			The lower bound for \(\operatorname{gap}_{\operatorname{\mathcal{BCR}^{+}}, \operatorname{\mathcal{HYP}}}\) is essentially proved in \cite{FKOS16}, for \(\operatorname{gap}_{\operatorname{\mathcal{BCR}^{+}}, \operatorname{STP}}\) essentially in \cite{BGRS13} and for \(\operatorname{gap}_{\operatorname{\mathcal{HYP}}, \operatorname{STP}}\) in \cite{KPT11}.
		}
		\label{table:gap-lower-bounds}
	\end{subtable}
	\hfill
	\begin{subtable}[t]{0.475\textwidth}
		\centering
		\begin{tabular}{l|ccc}
			\(\operatorname{gap_{\mathcal{L}, \mathcal{P}}} \leq \) & \(\mathcal{BCR}^{+}\) & \(\mathcal{HYP}\) & STP\\
			\hline
			\(\mathcal{BCR}\)\rule{0pt}{15pt} & \(2\) & \(2\) & \(2\)\\[3pt]
			\(\mathcal{BCR}^{+}\) &  & \(2\) & \(2\) \\[3pt]
			\(\mathcal{HYP}\) &  &  & \(\ln(4)\)
		\end{tabular}
		\caption{
			Note that \(\ln(4) \approx 1.387\).
			The upper bound for \(\operatorname{gap}_{\operatorname{\mathcal{HYP}}, \operatorname{STP}}\) is proved in \cite{GORZ12}.
			The remaining upper bounds are all due to the well known fact that \(\operatorname{gap}_{\operatorname{\mathcal{BCR}}, \operatorname{STP}} \leq 2\) (\cref{sec:bidirected-cut-relaxation}).
		}
		\label{table:gap-upper-bounds}
	\end{subtable}
	%\hspace*{\fill}
	\caption{
		% \hyperref[table:gap-lower-bounds]{a}
		% \hyperref[table:gap-upper-bounds]{b}
		After this work, to the best of our knowledge, the currently best known lower (\cref{table:gap-lower-bounds}) and upper (\cref{table:gap-upper-bounds}) bounds on the gaps for each pair of the relaxation chain \(\operatorname{\mathcal{BCR}} \leq \operatorname{\mathcal{BCR}^{+}} \leq \operatorname{\mathcal{HYP}} \leq \operatorname{STP}\).
	}
	\label{table:gap-bounds}
\end{table}

Before we define in \cref{sec:simplex-based-steiner-tree-instances-yielding-large-gaps} our new instances yielding that new lower bound, we will present in \cref{sec:bidirected-relaxations} different formulations in \(\operatorname{\mathcal{BCR}}\) and \(\operatorname{\mathcal{BCR}^{+}}\).
In \cite{GM93} various formulations has been proven to be equivalent to \nameref{lp:BCR} regarding the dominant: For a polyhedron \(P\) one can define the \textit{dominant} by \(\operatorname{dom}(P) := \{x : \exists x' \in P \ x' \leq x\}\).
Since we only work with non-negative edge-costs and we are dealing with minimization problems, it would be sufficient to require equivalence to \(\operatorname{dom}(\operatorname{proj}_{x}(P))\), where \(\operatorname{proj}_{x}(P) := \{x : (x, y) \in P\}\) denotes the projection of a polyhedron \(P\) to common variables \(x\).

We will slightly improve the results in \cite{GM93} by stating reduced formulations of their main relaxations which are all equivalent to the projection to the undirected edge variables.
Considering in addition a slightly changed formulation given in \cite{CDV11}, we are also able to simplify the line of arguments.
Moreover we show in \cref{sec:steiner-vertex-degree-constraints} for each stated formulation in \(\operatorname{\mathcal{BCR}}\) the respective equivalent Steiner vertex degree constraints obtaining the formulation in \(\operatorname{\mathcal{BCR}^{+}}\) and prove exactness for instances with at most \(3\) required vertices, which is in general not the case for \(\operatorname{\mathcal{BCR}}\).

\begin{theorem}\label{thm:improved-bidirected-relaxation-exact-for-3}
	Let \((G, R, c)\) be a Steiner tree instance.\\
	If \(|R| \leq 3\) then \(\operatorname{opt}_{\operatorname{\mathcal{BCR}^{+}}}(G, R, c) = \operatorname{opt}_{\operatorname{\mathcal{HYP}}}(G, R, c) = \operatorname{opt}_{\operatorname{STP}}(G, R, c)\).
\end{theorem}

Finally, in \cref{sec:simplex-based-steiner-tree-instances-yielding-large-gaps} we review the results about embedding \nameref{problem:STP} instances into simplices \cite{CDV11} (\cref{sec:simplex-embedding-steiner-tree-instances}), propose our instances (\cref{sec:simplex-based-steiner-tree-instances}) and prove \cref{thm:refined-bidirected-relaxations-gap} (\cref{sec:simplex-instances-gap-lower-bound}).
Additionally, we remark on a connection between a subclass of our instances to the Multiway Cut Problem in the case of \(3\) required vertices (\cref{sec:relation-to-multiway-cut-problem}).

In \cref{sec:set-cover-based-steiner-tree-instances} we generalize the instances in \cite{BGRS13} to arbitrary set cover instances and adapt the proofs in \cite{BGRS13} to \(\operatorname{\mathcal{BCR}^{+}}\), which is straightforward.

\subsection{Preliminaries}\label{sec:preliminaries}

We will use the common notation and terms for graph theory and linear programming from \cite{KV18}.
Further useful notation is as follows.
For \(n \in \mathbb{N}\) we define the index set \([n] := \{1, \dotsc, n\}\).
Let \(f : A \to \mathbb{R}\) be a function, then for a subset \(A' \subseteq A\) we define \(f(A') := \sum_{a \in A'} f(a)\).
Furthermore, we define the so called Iverson bracket \(\llbracket \cdot \rrbracket : \{\text{boolean expressions}\} \to \{0, 1\}\) by \(\llbracket P \rrbracket = 1\) if and only \(P\) is true.
%\[
%	\llbracket \text{boolean expression} \rrbracket :=
%	\begin{cases}
%		1 & \mbox{if boolean expression is true}\\
%		0 & \mbox{if boolean expression is false}\\
%	\end{cases}
%\]

For shorter notation we introduce the convention \(f(a, b) = f_{a}(b)\) for a function \(f : A \times B \to \mathbb{R}\) as well as \(f_{a}(B') = \sum_{b \in B'} f_{a}(b)\) for \(B' \subseteq B\).
Furthermore, if we apply a directed edge \((v, w)\) to a function \(f\) we omit the brackets, i.e.\ \(f(v, w) = f((v, w))\).

For a given graph \(G\) we denote by \(V(G)\) the vertices and \(E(G)\) the edges.
For a vertex set \(X \subseteq V(G)\) of an undirected graph \(G\) we define \(G[X] := (X, \{e \in E(G) : e \subseteq X\})\).
We will mainly be interested in undirected graphs \(G = (V, E)\) and their bidirected versions, i.e.\ we replace each undirected edge by the two edges directed in opposite directions.
Therefore, we define \(E_{\leftrightarrow} := \{(v, w), (w, v) : \{v, w\} \in E\}\).
Moreover, we refine the notation of incident edges \(\delta_{G}(v) := \{e \in E : v \in e\}\) by \(\delta_{G}^{+}(v) := \{(v, w) \in E_{\leftrightarrow} : \{v, w\} \in E\}\) and \(\delta_{G}^{-}(v) := \{(w, v) \in E_{\leftrightarrow} : \{v, w\} \in E\}\).
Analogously, we refine for a vertex set \(X \subseteq V\) \(\delta_{G}(X) := \{e \in E : |e \cap X| = 1\}\) by \(\delta_{G}^{+}(X) := \{(v, w) \in E_{\leftrightarrow} : v \in X \land w \notin X\}\) and \(\delta_{G}^{-}(X) := \{(v, w) \in E_{\leftrightarrow} : v \notin X \land w \in X\}\).

Let \(T\) be a tree and \(r\) be a leaf of \(T\).
We denote the arborescence rooted at \(r\) as \(T^{r\to}\), i.e.\ \(T^{r\to}\) is obtained from \(T\) by directing all edges from \(r\) to the other leafs.
For \(v, w \in V(T)\) we denote by \(T_{[v, w]}^{r\to}\) the directed path from \(v\) to \(w\) in \(T^{r\to}\) given that this path exists.

An important notion, when talking about bidirected relaxations, are balance flows.

\begin{definition}
	Let \(G = (V, E)\) be a directed graph, \(u : E \to \mathbb{R}_{\geq 0}\) edge-usage bounds and \(b : V \to \mathbb{R}\) vertex-balances.
	We call \(f : E \to \mathbb{R}_{\geq 0}\) a \textbf{balance flow} for \((G, u, b)\) if and only if the following two conditions hold:
	\[
		\begin{aligned}
			\forall e & \in E & f(e) & \leq u(e)\\
			\forall v & \in V & f(\delta_{G}^{+}(v)) - f(\delta_{G}^{-}(v)) & = b(v)
		\end{aligned}
	\]
\end{definition}

A trivially necessary condition for the existence of a balance-flow is \(b(V) = 0\).
This can easily be seen by summing up the edge flow values over all vertices.
However it is not sufficient as Gale's well-known theorem shows.

\begin{theorem}[\cite{Gale1957}]\label{thm:gale}
	Let \(G = (V, E)\) be a directed graph, \(u : E \to \mathbb{R}_{\geq 0}\) edge-usage bounds and \(b : V \to \mathbb{R}\) vertex-balances such that \(b(V) = 0\).
	Then there exists a balance-flow for \((G, u, b)\) if and only if \(u(\delta_{G}^{+}(X)) \geq b(X)\) for all \(X \subseteq V\).
\end{theorem}
A proof, as well as further results on balance-flows, can also be found in \cite{KV18}.

We will mainly be interested in balance-flows on bidirected graphs respecting the edge-usage bounds of the corresponding undirected graph.
Therefore we refine the notion of balance-flows for this special case.

\begin{definition}
	Let \(G = (V, E)\) be an undirected graph, \(u : E \to \mathbb{R}_{\geq 0}\) edge-usage bounds and \(b : V \to \mathbb{R}\) vertex-balances.
	We call \(f : E_{\leftrightarrow} \to \mathbb{R}_{\geq 0}\) a \textbf{bidirected balance flow} for \((G, u, b)\) if and only if the following two conditions hold:
	\[
		\begin{aligned}
			\forall \{v, w\} & \in E & f(v, w) + f(w, v) & \leq u(\{v, w\})\\
			\forall v & \in V & f(\delta_{G}^{+}(v)) - f(\delta_{G}^{-}(v)) & = b(v)
		\end{aligned}
	\]
\end{definition}

\hyperref[thm:gale]{Gale's theorem} can be adapted as follows.

\begin{theorem}\label{thm:gale-bidirected}
	Let \(G = (V, E)\) be an undirected graph, \(u : E \to \mathbb{R}_{\geq 0}\) be edge-usage bounds and \(b : V \to \mathbb{R}\) be vertex-balances such that \(b(V) = 0\).
	Then there exists a bidirected balance-flow for \((G, u, b)\) if and only if \(u(\delta_{G}(X)) \geq b(X)\) for all \(X \subseteq V\).
\end{theorem}
\begin{proof}
	Define \(u_{\leftrightarrow} : E_{\leftrightarrow} \to \mathbb{R}_{\geq 0}\) by \(u_{\leftrightarrow}(v, w) = u(\{v, w\})\).
	Then every bidirected balance-flow for \((G, u, b)\) is a balance flow for \(((V, E_{\leftrightarrow}), u_{\leftrightarrow}, b)\) and therefore for every \(X \subseteq V\) we have \(u(\delta_{G}(X)) = u_{\leftrightarrow}(\delta_{G}^{+}(X)) \geq b(X)\).\\
	So assume \(u(\delta_{G}(X)) \geq b(X)\) holds for every \(X \subseteq V\).
	By \cref{thm:gale} we know there exists a balance-flow \(f' : E_{\leftrightarrow} \to \mathbb{R}_{\geq 0}\) for \(((V, E_{\leftrightarrow}), u_{\leftrightarrow}, b)\).
	We define \(f : E_{\leftrightarrow} \to \mathbb{R}_{\geq 0}\) as \(f(v, w) = \max \{ f'(v, w) - f'(w, v), \ 0 \}\).
	Then \(f\) is a bidirected balance-flow for \((G, u, b)\) since \(f(v, w) + f(w, v) = |f'(v, w) - f'(w, v)| \leq \max \{f'(v, w), f'(w, v)\} \leq u(\{v, w\})\) and \(f(\delta_{G}^{+}(v)) - f(\delta_{G}^{-}(v)) = \sum_{(v, w) \in \delta_{G}^{+}(v)} \left( \ f'(v, w) - f'(w, v) \ \right) = f'(\delta_{G}^{+}(v)) - f'(\delta_{G}^{-}(v)) = b(v)\)
\end{proof}

	\pagebreak

\section{Bidirected relaxations with and without\texorpdfstring{\\}{ }Steiner vertex degree constraints}\label{sec:bidirected-relaxations}

We are interested in two classes of equivalent LP-relaxations for \nameref{problem:STP}.
The \textit{bidirected relaxations}, which we denote by \label{lp-class:BCR}\(\operatorname{\mathcal{BCR}}\), with the classical representative of the \textit{bidirected cut relaxation} and the \textit{bidirected relaxations with Steiner vertex degree constraints}, which we denote by \label{lp-class:BCR+SVDC}\(\operatorname{\mathcal{BCR}^{+}}\), consisting of the bidirected relaxations enhanced by certain degree constraints for each non-required vertex.

The first bidirected relaxation is stated in \cite{Edmonds1967} for the special case of spanning trees, i.e. all vertices are required, and shown to be exact.
In \cite{Wong1984} this approach is generalized for Steiner trees resulting in the bidirected cut relaxation (\nameref{lp:BCR}).
In \cite{KM98} additional constraints, which the authors call \textit{flow-balance constraints}, are added to \nameref{lp:BCR} to strengthen the relaxation in practice resulting in the bidirected cut relaxation with Steiner vertex degree constraints (\BCRimproved).

For this whole section we will fix a Steiner tree instance \((G = (V, E), R, c)\).
The idea of the bidirected relaxations is to consider in addition to the undirected edges of the Steiner tree \(T\), in contrast to the \textit{undirected cut relaxation} (\(\operatorname{UCR}\)\label{text:undirected-cut-relaxation}) introduced in \cite{Aneja1980}, also the directed edges of the arborescences \(T^{r\to}\) rooted in the required vertices \(r \in R\).
With this additional information one is able to define LPs stronger than \(\operatorname{UCR}\).

\subsection{Bidirected Cut Relaxation}\label{sec:bidirected-cut-relaxation}

For this formulation we only interested in one arborescence rooted in a fixed required vertex \(r \in R\).
We will later see that \(\operatorname{opt}_{\operatorname{BCR}}(G, R, c, r)\) is independent of the choice of \(r\).
This was first shown in \cite{GM93}.
\begin{linearprogram}[BCR]
	\label{lp:BCR}
	\(\min\) & \(\displaystyle \sum_{e \in E} c(e) \cdot u(e)\)\\[20pt]
	s.t. & \(u : E \to \mathbb{R}_{\geq 0}\, , \ f_{r} : E_{\leftrightarrow} \to \mathbb{R}_{\geq 0}\)\\[10pt]
	& \(
		\begin{aligned}[t]
			f_{r}(v, w) + f_{r}(w, v) & \leq u(\{v, w\}) & \{v, w\} & \in E\\
			f_{r}(\delta_{G}^{+}(X)) & \geq \llbracket r \in X \land R \setminus X \neq \emptyset \rrbracket & X & \subseteq V
		\end{aligned}
	\)
\end{linearprogram}
It is easy to verify that every solution of \nameref{lp:BCR} provides a solution of \(\operatorname{UCR}\), more precisely \(\operatorname{proj}_{u}(\operatorname{BCR}) \subseteq \operatorname{UCR}\).
In \cite{GD93} it is essentially proven that \(\operatorname{gap}_{\operatorname{UCR_{\leq k}}, \operatorname{BCR_{\leq k}}} = \operatorname{gap}_{\operatorname{UCR_{\leq k}}, \operatorname{STP}} = 2 \cdot ( 1 - \frac{1}{k} )\).
Therefore we have \(\operatorname{gap}_{\operatorname{\mathcal{BCR}}, \operatorname{STP}} \leq 2\) and, in addition to the case \(R = V\), \nameref{lp:BCR} is also exact for \(2\) required vertices.
A Steiner tree \(T\) with respect to \(G\) and \(R\) can be associated with the following integral solution:
\[
	\begin{aligned}
		u(\{v, w\}) & = \llbracket \{v, w\} \in E(T) \rrbracket\\
		f_{r}(v, w) & = \llbracket (v, w) \in E(T^{r\to}) \rrbracket
	\end{aligned}
\]

\subsection{Multi Commodity Flow Relaxation}\label{sec:multi-commodity-flow-relaxation}

\hyperref[thm:gale]{Gale's theorem} suggests the first equivalent LP-relaxation, which is introduced in \cite{Wong1984}.
The cut-constraints in \nameref{lp:BCR} can be seen as balance-flow existence conditions for \(|R| - 1\) different flows, namely a unit-flow for each \(s \in R \setminus \{r\}\) from \(r\) to \(s\).
We obtain the following LP-relaxation commonly known as the \textit{multi commodity flow relaxation}.
\begin{linearprogram}[MCFR]
	\label{lp:MCFR}
	\(\min\) & \(\displaystyle \sum_{e \in E} c(e) \cdot u(e)\)\\[20pt]
	s.t. & \(u : E \to \mathbb{R}_{\geq 0}\, , \ f_{r} : E_{\leftrightarrow} \to \mathbb{R}_{\geq 0}\, , \ g : R \setminus \{r\} \times E_{\leftrightarrow} \to \mathbb{R}_{\geq 0}\)\\[10pt]
	& \(
	\begin{aligned}[t]
	f_{r}(v, w) + f_{r}(w, v) & \leq u(\{v, w\}) & \{v, w\} \in E\\
	g_{s}(e) & \leq f_{r}(e) & \quad\quad e \in E_{\leftrightarrow}, s \in R \setminus \{r\}\\
	g_{s}(\delta_{G}^{+}(v)) - g_{s}(\delta_{G}^{-}(v)) & = \llbracket v = r \rrbracket - \llbracket v = s \rrbracket & \qquad v \in V, s \in R \setminus \{r\}
	\end{aligned}
	\)
\end{linearprogram}

\nameref{lp:MCFR} also induces an integer program for \nameref{problem:STP}.
A Steiner tree \(T\) with respect to \(G\) and \(R\) can be associated with the following integral solution:
\[
	\begin{aligned}
		u(\{v, w\}) & = \llbracket \{v, w\} \in E(T) \rrbracket\\
		f_{r}(v, w) & = \llbracket (v, w) \in E(T^{r\to}) \rrbracket\\
		g_{s}(v, w) & = \llbracket (v, w) \in E(T_{[r, s]}^{r\to}) \rrbracket
	\end{aligned}
\]
Note that we obtained an extended formulation for \(\operatorname{proj}_{u}(\operatorname{BCR})\) of compact size, namely the number of variables and constraints is in \(\mathcal{O}(|R| \cdot |E|)\).

\vspace{6pt}

\begin{theorem}[\cite{Wong1984}]
	\(\operatorname{proj}_{u, f_{r}}(\operatorname{MCFR}) = \operatorname{proj}_{u, f_{r}}(\operatorname{BCR})\)
\end{theorem}
\begin{proof}
	Let \((u, f_{r}, g)\) be a solution of \nameref{lp:MCFR}.
	For \(s \in R \setminus \{r\}\) define vertex-balances \(b_{s} : V \to \mathbb{R}\) as \(b_{s}(v) = \llbracket v = r \rrbracket - \llbracket v = s \rrbracket\), which satisfy \(b_{s}(V) = 0\).
	Then \(g_{s}\) is a balance flow for \(((V, E_{\leftrightarrow}), f_{r}, b_{s})\) and therefore by \cref{thm:gale} we have \(f_{r}(\delta_{G}^{+}(X)) \geq \llbracket r \in X \land s \notin X \rrbracket\).
	Hence overall we have \(f_{r}(\delta_{G}^{+}(X)) \geq \llbracket r \in X \land R \setminus X \neq \emptyset \rrbracket\).
	On the other hand \cref{thm:gale} shows the existence of \(g\) if \((u, f_{r})\) is a solution of \nameref{lp:BCR}.
\end{proof}

\subsection{Multi Balance Flow Relaxation}\label{sec:multi-balance-flow-relaxation}

Instead of only considering one arborescence rooted at a fixed required vertex \(r \in R\), we now consider the arborescences rooted at all required vertices.
By connecting these \(|R|\) directed trees at each vertex through degree constraints, we obtain another equivalent formulation which does not depend on a previously chosen root.
It is based on a similar one stated in \cite{CDV11}.
We call it the \textit{multi balance flow relaxation}.
\begin{linearprogram}[MBFR]
	\label{lp:MBFR}
	\(\min\) & \(\displaystyle \sum_{e \in E} c(e) \cdot u(e)\)\\[20pt]
	s.t. & \(u : E \to \mathbb{R}_{\geq 0}\, , \ b : V \to \mathbb{R}\, , \ f : R \times E_{\leftrightarrow} \to \mathbb{R}_{\geq 0}\)\\[10pt]
	& \(
		\begin{aligned}[t]
			f_{r}(v, w) + f_{r}(w, v) & \leq u(\{v, w\}) & \{v, w\} \in E, r \in R\\
			f_{r}(\delta_{G}^{+}(v)) - f_{r}(\delta_{G}^{-}(v)) & = b(v) + 2 \cdot \llbracket v = r \rrbracket & \qquad v \in V, r \in R
		\end{aligned}
	\)
\end{linearprogram}

\nameref{lp:MBFR} also induces an integer program for \nameref{problem:STP}.
A Steiner tree \(T\) with respect to \(G\) and \(R\) can be associated with the following integral solution:
\[
	\begin{aligned}
		u(\{v, w\}) & = \llbracket \{v, w\} \in E(T) \rrbracket\\
		b(v) & = \begin{cases}
			|\delta_{T}(v)| - 2 & \text{if } v \in V(T)\\
			0 & \text{if } v \in V(G) \setminus V(T)
			\end{cases}\\
		f_{r}(v, w) & = \llbracket (v, w) \in E(T^{r\to}) \rrbracket
	\end{aligned}
\]
Hence, the \(b\) values of an integral solution encode the degrees within the Steiner tree \(T\).
The variables \(f_{r}\) encode the corresponding arborescence rooted in \(s\).

The new variables can roughly be interpreted as global balance values for each of the bidirected flows \(f_{r}\).
The only difference occurs at each required vertex \(r \in R\).
More precisely, if we define \(b_{r} : V \to \mathbb{R}\) by \(b_{r}(v) = b(v) + 2 \cdot \llbracket v = r \rrbracket\), then \(f_{r}\) is a bidirected balance flow for \((G, u, b_{r})\).
Therefore we have the necessary condition \(b(V) = -2\).

As already mentioned, \nameref{lp:MBFR} is a slightly different version of a formulation in \cite{CDV11}.
The authors proved equivalence to \nameref{lp:BCR} directly.
Our proof of \cref{thm:equivalence-mbfr-mcfr} is an adapted version of theirs.

\vspace{6pt}

\begin{theorem}\label{thm:equivalence-mbfr-mcfr}
	\(\operatorname{proj}_{u}(\operatorname{MBFR}) = \operatorname{proj}_{u}(\operatorname{MCFR})\)
\end{theorem}
\begin{proof}
	We will prove \(\supseteq\) and \(\subseteq\) separately by translating solutions into each other.
	In both directions \(u : E \to \mathbb{R}_{\geq 0}\) will remain the same.
	Hence, if we can construct a solution with \(u\) given from the respective other problem, we have shown the statement.
	
	\textbf{Claim} \(\operatorname{proj}_{u}(\operatorname{MBFR}) \supseteq \operatorname{proj}_{u}(\operatorname{MCFR})\):
	
	Let \((u, f_{r}, g)\) be a solution of \nameref{lp:MCFR}.
	We need to define a solution \((u, b', f')\) of \nameref{lp:MBFR}.\\
	We define \(b'\) and \(f'\) as follows:
	\[
		\begin{aligned}
			b'(v) & := f_{r}(\delta_{G}^{+}(v)) - f_{r}(\delta_{G}^{-}(v)) - 2 \cdot \llbracket v = r \rrbracket\\
			f'_{r}(v, w) & := f_{r}(v, w)\\
			f'_{s}(v, w) & := f_{r}(v, w) - g_{s}(v, w) + g_{s}(w, v) & s \in R \setminus \{r\}
		\end{aligned}
	\]
	By definition these values are non-negative and fulfil the edge-usage bounds:
	\[f'_{s}(v, w) + f'_{s}(w, v) \leq f_{r}(v, w) + f_{r}(w, v) \leq u(\{v, w\}) \quad s \in R\]
	Again by definition the balance constraints are fulfilled for \(r\).\\
	For \(s \in R \setminus \{r\}\) we can easily check that they are also fulfilled:
	\[
		\begin{aligned}
			f'_{s}(\delta_{G}^{+}(v)) - f'_{s}(\delta_{G}^{-}(v))
			& = f_{r}(\delta_{G}^{+}(v)) - f_{r}(\delta_{G}^{-}(v)) - 2 \cdot ( \ g_{s}(\delta_{G}^{+}(v)) - g_{s}(\delta_{G}^{-}(v)) \ )\\
			& = b'(v) + 2 \cdot \llbracket v = r \rrbracket - 2 \cdot ( \ \llbracket v = r \rrbracket - \llbracket v = s \rrbracket \ )\\
			& = b'(v) + 2 \cdot \llbracket v = s \rrbracket
		\end{aligned}
	\]
	
	\textbf{Claim} \(\operatorname{proj}_{u}(\operatorname{MBFR}) \subseteq \operatorname{proj}_{u}(\operatorname{MCFR})\):
	
	Let \((u, b, f)\) be a solution of \nameref{lp:MBFR}.
	We need to define a solution \((u, f'_{r}, g')\) of \nameref{lp:MCFR}.\\
	We define \(f'_{r}\) and \(g'\) as follows:
	\[
		\begin{aligned}
			f'_{r}(v, w) & := \frac{1}{2} \cdot \left( \ f_{r}(v, w) - f_{r}(w, v) + u(\{v, w\}) \ \right)\\
			g'_{s}(v, w) & := \frac{1}{2} \cdot \max \{ f_{r}(v, w) - f_{r}(w, v) + f_{s}(w, v) - f_{s}(v, w), \ 0 \} & \quad s \in R \setminus \{r\}
		\end{aligned}
	\]
	We immediately see that \(f'_{r}(v, w) + f'_{r}(w, v) = u(\{v, w\})\).
	The flow bound constraints are fulfilled as well since \(|f_{s}(v, w) - f_{s}(w, v)| \leq \max \{f_{s}(v, w), f_{s}(w, v)\} \leq u(\{v, w\})\):
	\[g'_{s}(v, w) \leq \frac{1}{2} \cdot \max \{ f_{r}(v, w) - f_{r}(w, v) + u(\{v, w\}), \ 0 \} = f'_{r}(v, w)\]
	For \(s \in R \setminus \{r\}\) the flow constraints are also fulfilled:
	%\[
		\begin{align*}
			g'_{s}(\delta_{G}^{+}(v)) - g'_{s}(\delta_{G}^{-}(v))
			& = \sum_{(v, w) \in \delta_{G}^{+}(v)} ( \ g'_{s}(v, w) - g'_{s}(w, v) \ )\\
			& = \frac{1}{2} \cdot \sum_{(v, w) \in \delta_{G}^{+}(v)} ( \ f_{r}(v, w) - f_{r}(w, v) + f_{s}(w, v) - f_{s}(v, w) \ )\\
			& = \frac{1}{2} \cdot \left( \left( f_{r}(\delta_{G}^{+}(v)) - f_{r}(\delta_{G}^{-}(v)) \right) - \left( f_{s}(\delta_{G}^{+}(v)) - f_{s}(\delta_{G}^{-}(v)) \right) \right)\\
			& = \frac{1}{2} \cdot ( \ ( \ b(v) + 2 \cdot \llbracket v = r \rrbracket \ ) - ( \ b(v) + 2 \cdot \llbracket v = s \rrbracket \ ) \ )\\
			& = \llbracket v = r \rrbracket - \llbracket v = s \rrbracket \qedhere
		\end{align*}
	%\]
\end{proof}
\vspace{6pt}
\begin{corollary}[{\cite[Thm.\ 9]{GM93}}]
	\(\operatorname{opt}_{\operatorname{BCR}}(G, R, c, r)\) does not depend on \(r \in R\).
\end{corollary}

\subsection{Multi Balance Cut Relaxation}\label{sec:multi-balance-cut-relaxation}

Remembering the translation process between \nameref{lp:BCR} and \nameref{lp:MCFR}, it is natural to apply \cref{thm:gale-bidirected} to \nameref{lp:MBFR}.
We obtain the \textit{multi balance cut relaxation}.
\begin{linearprogram}[MBCR]
	\label{lp:MBCR}
	\(\min\) & \(\displaystyle \sum_{e \in E} c(e) \cdot u(e)\)\\[20pt]
	s.t. & \(u : E \to \mathbb{R}_{\geq 0} \, , \ b : V \to \mathbb{R}\)\\[10pt]
	& \(
		\begin{aligned}[t]
			u(\delta_{G}(X)) & \geq b(X) + 2 \cdot \llbracket R \cap X \neq \emptyset \rrbracket & \qquad X \subseteq V\\
			b(V) & = -2
		\end{aligned}
	\)
\end{linearprogram}
\nameref{lp:MBCR} is a reduced formulation of `\(R'_{xz}\)' in \cite{GM93}.
The additional constraints are presented in \cref{sec:valid-constraints}.
A Steiner tree \(T\) with respect to \(G\) and \(R\) can be associated with the following integral solution:
\[
	\begin{aligned}
		u(\{v, w\}) & = \llbracket \{v, w\} \in E(T) \rrbracket\\
		b(v) & = \begin{cases}
			|\delta_{T}(v)| - 2 & \text{if } v \in V(T)\\
			0 & \text{if } v \in V(G) \setminus V(T)
		\end{cases}
	\end{aligned}
\]

\begin{theorem}
	\(\operatorname{proj}_{u, b}(\operatorname{MBCR}) = \operatorname{proj}_{u, b}(\operatorname{MBFR})\)
\end{theorem}
\begin{proof}
	Let \((u, b, f)\) be a solution of \(\operatorname{MBFR}\).
	Then \(f_{s}\) for \(s \in R\) is a bidirected balance-flow for \((G, u, b_{s})\) where \(b_{s}(v) = b(v) + 2 \cdot \llbracket v = s \rrbracket\).
	With \cref{thm:gale-bidirected} we obtain for \(X \subseteq V\):
	\[u(\delta_{G}(X)) \geq b_{s}(X) = b(X) + 2 \cdot \llbracket s \in X \rrbracket\]
	This implies overall for \(X \subseteq V\):
	\[u(\delta_{G}(X)) \geq b(X) + 2 \cdot \llbracket R \cap X \neq \emptyset \rrbracket\]
	
	Let \((u, b)\) be a solution of \(\operatorname{MBCR}\).
	Then \cref{thm:gale-bidirected} shows the existence of \(f\) such that \((u, b, f)\) is a solution of \(\operatorname{MBFR}(G, R, c)\) since we have:
	\[u(\delta_{G}(X)) \geq b(X) + 2 \cdot \llbracket R \cap X \neq \emptyset \rrbracket \geq b(X) + 2 \cdot \llbracket s \in X \rrbracket \qedhere\]
\end{proof}

\subsection{Subtour Elimination Relaxation}\label{sec:subtour-elimination-relaxation}

For the next equivalent LP-relaxation we redefine the meaning of the auxiliary vertex-variables.
In an integral solution of \nameref{lp:MBCR} associated to a Steiner tree \(T\) we have for \(v \in V(T)\) that \(b(v) = |\delta_{T}(v)| - 2 = u(\delta_{G}(v)) - 2\).
Therefore the value \(\frac{1}{2} \cdot (u(\delta_{G}(v)) - b(v)) \in \{0, 1\}\) indicates whether the vertex \(v\) is contained in \(T\) or not.
Hence by introducing the variables \(y : V \to \mathbb{R}_{\geq 0}\), which enforce these values, we obtain the following LP-relaxation, which we call \textit{subtour elimination relaxation}.
\begin{linearprogram}[STER]
	\label{lp:STER}
	\(\min\) & \(\displaystyle \sum_{e \in E} c(e) \cdot u(e)\)\\[20pt]
	s.t. & \(u : E \to \mathbb{R}_{\geq 0} \, , \ y : V \to \mathbb{R}_{\geq 0}\)\\[10pt]
	& \(
		\begin{aligned}[t]
			u(E) & = y(V) - 1\\
			u(E(G[X])) & \leq y(X) - \llbracket R \cap X \neq \emptyset \rrbracket & \qquad X \subseteq V
		\end{aligned}
	\)
\end{linearprogram}
\nameref{lp:STER} is a reduced formulation of `\(P'_{xy}\)' in \cite{GM93}.
The additional constraints are presented in \cref{sec:valid-constraints}.
The constraints \(u(E(G[X])) \leq y(X) - \llbracket R \cap X \neq \emptyset \rrbracket\) forbid cycles involving required vertices within the graph associated to an integral solution and therefore they eliminate subtours.
A Steiner tree \(T\) with respect to \(G\) and \(R\) can be associated with the following integral solution:
\[
	\begin{aligned}
		u(\{v, w\}) & = \llbracket \{v, w\} \in E(T) \rrbracket\\
		y(v) & = \llbracket v \in V(T) \rrbracket
	\end{aligned}
\]

\begin{theorem}
	\(\operatorname{proj}_{u}(\operatorname{STER}) = \operatorname{proj}_{u}(\operatorname{MBCR})\)
\end{theorem}
\begin{proof}
	If we assume \(b(v) = u(\delta_{G}(v)) - 2 \cdot y(v)\), we have the following two equivalences, which prove the claimed result:
	\[
		\begin{aligned}
			\phantom{\Leftrightarrow} \quad & b(V) = -2\\
			\Leftrightarrow \quad & \sum_{v \in V} u(\delta_{G}(v)) - 2 \cdot y(V) = -2\\
			\Leftrightarrow \quad & 2 \cdot u(E) - 2 \cdot y(V) = -2\\
			\Leftrightarrow \quad & u(E) = y(V) - 1
		\end{aligned}
	\]
	%\[
		\begin{align*}
			\phantom{\Leftrightarrow} \quad & u(\delta_{G}(X)) \geq b(X) + 2 \cdot \llbracket R \cap X \neq \emptyset \rrbracket\\
			\Leftrightarrow \quad & u(\delta_{G}(X)) \geq \sum_{v \in X} u(\delta_{G}(v)) - 2 \cdot y(X) + 2 \cdot \llbracket R \cap X \neq \emptyset \rrbracket\\
			\Leftrightarrow \quad & u(\delta_{G}(X)) \geq u(\delta_{G}(X)) + 2 \cdot u(E(G[X])) - 2 \cdot ( \ y(X) - \llbracket R \cap X \neq \emptyset \rrbracket \ )\\
			\Leftrightarrow \quad & u(E(G[X])) \leq y(X) - \llbracket R \cap X \neq \emptyset \rrbracket \qedhere
		\end{align*}
	%\]
\end{proof}

\subsection{Valid constraints}\label{sec:valid-constraints}

Using \nameref{lp:MBCR} it is very easy to verify that \(\operatorname{proj}_{u}(\operatorname{BCR})\) fulfills the well-known \textit{Steiner partition} condition \cite{CR94} proven in \cite{Goemans1994b}:
Let \(S_{1} \dot{\cup} \dotsc \dot{\cup} S_{q} = V\) be a partition of the vertices and define \(E_{S} := \{\{v, w\} \in E : \exists i, j \in [q] \ i \neq j \land v \in S_{i} \land w \in S_{j}\}\), i.e.\ \(E_{S}\) contains all edges which connect different partition classes.
Then for a Steiner tree \(T\) we have the necessary condition that \(|E(T) \cap E_{S}| \geq |\{i \in [q] : S_{i} \cap R \neq \emptyset\}| - 1\).
For a solution \((u, b)\) of \(\operatorname{MBCR}\) we can compute:
\[u(E_{S}) = \frac{1}{2} \cdot \sum_{i \in [q]} u(\delta_{G}(S_{i})) \geq \frac{1}{2} \cdot \sum_{i \in [q]} ( \ 2 \cdot \llbracket S_{i} \cap R \neq \emptyset \rrbracket + b(S_{i}) \ ) = |\{i \in [q] : S_{i} \cap R \neq \emptyset\}| - 1\]
%\[
%	\begin{aligned}
%		2 \cdot u(E_{S}) & = \sum_{i \in [q]} u(\delta_{G}(S_{i}))\\
%		& \geq \sum_{i \in [q]} ( \ 2 \cdot \llbracket S_{i} \cap R \neq \emptyset \rrbracket + b(S_{i}) \ )\\
%		& = 2 \cdot |\{i \in [q] : S_{i} \cap R \neq \emptyset\}| + b(V)\\
%		& = 2 \cdot |\{i \in [q] : S_{i} \cap R \neq \emptyset\}| - 2
%	\end{aligned}
%\]
%Hence, \(u(E_{S}) \geq |\{i \in [q] : S_{i} \cap R \neq \emptyset\}| - 1\).

As already mentioned in \cref{sec:our-contributions}, in \cite{GM93} the formulations are compared with respect to \(\operatorname{dom}(\operatorname{proj}_{u}(\operatorname{BCR}))\), which is sufficient as we are only considering non-negative edge-costs.
However, this is also necessary as the authors state relaxations that are tighter in the sense that they bound the variables from above.
A trivial upper bound is for example \(u(\{v, w\}) \leq 1\).
In \cite[Thm.\ 6]{GM93} the following constraints are shown to be valid regarding the optimum value for the respective equivalent formulations in \BCRclass, i.e.\ not only are they valid for Steiner trees, they also do not change the optimum solution value of the respective LP-relaxations.
The first group of equivalent valid constraints is as follows:
\[
	\begin{aligned}
		f_{r}(\delta_{G}^{-}(v)) & \leq \llbracket v \neq r \rrbracket & v \in V\\
		u(\delta_{G}(v)) & \leq b(v) + 2 & v \in V\\
		y(v) & \leq 1 & v \in V
	\end{aligned}
\]
The second group is:
\[
	\begin{aligned}
		f_{r}(\delta_{G}^{-}(v)) & \leq f_{r}(\delta_{G}^{-}(X)) & v \in X \subseteq V \setminus R\\
		u(\delta_{G}(X)) - u(\delta_{G}(v)) & \geq b(X) - b(v) & v \in X \subseteq V \setminus R\\
		u(E(G[X])) & \leq y(X) - y(v) & v \in X \subseteq V \setminus R
	\end{aligned}
\]
Note that one has to prove only one constraint of each group, as the remaining ones can be translated similarly like the constraints in our equivalence proofs.

\subsection{Steiner vertex degree constraints}\label{sec:steiner-vertex-degree-constraints}

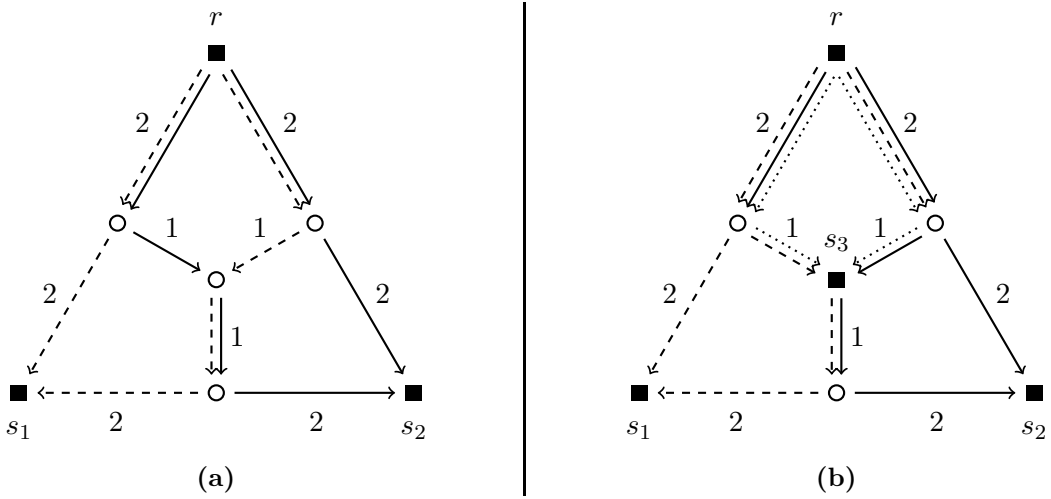
\begin{figure}
	\centering
	\begin{subfigure}[]{0.49\textwidth}
		\centering
		\begin{tikzpicture}[scale=1]
			\tikzset{vertex/.style={circle, thick, draw=black, fill=white, minimum size=0pt, inner sep=2pt, outer sep=4pt}}
			\tikzset{terminal/.style={rectangle, fill=black, minimum size=0pt, inner sep=3pt, outer sep=4pt}}
			\tikzset{color-red/.style={red}}
			\tikzset{color-green/.style={green}}
			\tikzset{color-blue/.style={blue}}
			\tikzset{color-red/.style={solid}}
			\tikzset{color-green/.style={dashed}}
			\tikzset{color-blue/.style={dashdotted}}
			
			\node[terminal, label=above:\(r\)] (r) at (90:3) {};
			\node[terminal, label=below:\(s_1\)] (s1) at (210:3) {};
			\node[terminal, label=below:\(s_2\)] (s2) at (330:3) {};
			
			\node[vertex] (v1) at (150:1.5) {};
			\node[vertex] (v2) at (30:1.5) {};
			\node[vertex] (v3) at (270:1.5) {};
			\node[vertex] (v4) at (0:0) {};
			
			\node at ( 65:2.3) {\(2\)};
			\node at (115:2.3) {\(2\)};
			
			\node at (185:2.2) {\(2\)};
			\node at (355:2.2) {\(2\)};
			
			\node at (235:2.3) {\(2\)};
			\node at (305:2.3) {\(2\)};
			
			\node at ( 50:0.9) {\(1\)};
			\node at (130:0.9) {\(1\)};
			\node at (290:0.8) {\(1\)};
			
			\TikzMultiLine{r}{v1}{+1}{0.6}{->, thick, color-red}{}
			\TikzMultiLine{r}{v1}{-1}{0.6}{->, thick, color-green}{}
			
			\TikzMultiLine{r}{v2}{+1}{0.6}{->, thick, color-red}{}
			\TikzMultiLine{r}{v2}{-1}{0.6}{->, thick, color-green}{}
			
			\TikzMultiLine{v1}{v4}{+0}{0}{->, thick, color-red}{}
			\TikzMultiLine{v2}{v4}{+0}{0}{->, thick, color-green}{}
			
			\TikzMultiLine{v4}{v3}{+1}{0.6}{->, thick, color-red}{}
			\TikzMultiLine{v4}{v3}{-1}{0.6}{->, thick, color-green}{}
			
			\TikzMultiLine{v1}{s1}{+0}{0}{->, thick, color-green}{}
			\TikzMultiLine{v3}{s1}{+0}{0}{->, thick, color-green}{}
			
			\TikzMultiLine{v2}{s2}{+0}{0}{->, thick, color-red}{}
			\TikzMultiLine{v3}{s2}{+0}{0}{->, thick, color-red}{}
		\end{tikzpicture}
		\caption{}
		\label{fig:bcr-small-integrality-gap-example}
	\end{subfigure}
	\vline
	\hfill
	\begin{subfigure}[]{0.49\textwidth}
		\centering
		\begin{tikzpicture}[scale=1]
			\tikzset{vertex/.style={circle, thick, draw=black, fill=white, minimum size=0pt, inner sep=2pt, outer sep=4pt}}
			\tikzset{terminal/.style={rectangle, fill=black, minimum size=0pt, inner sep=3pt, outer sep=4pt}}
			\tikzset{color-red/.style={red}}
			\tikzset{color-green/.style={green}}
			\tikzset{color-blue/.style={blue}}
			\tikzset{color-red/.style={solid}}
			\tikzset{color-green/.style={dashed}}
			\tikzset{color-blue/.style={dotted}}
			
			\node[terminal, label=above:\(r\)] (r) at (90:3) {};
			\node[terminal, label=below:\(s_1\)] (s1) at (210:3) {};
			\node[terminal, label=below:\(s_2\)] (s2) at (330:3) {};
			\node[terminal, label=above:\(s_3\)] (s3) at (0:0) {};
			
			\node[vertex] (v1) at (150:1.5) {};
			\node[vertex] (v2) at (30:1.5) {};
			\node[vertex] (v3) at (270:1.5) {};
			
			\node at ( 65:2.3) {\(2\)};
			\node at (115:2.3) {\(2\)};
			
			\node at (185:2.2) {\(2\)};
			\node at (355:2.2) {\(2\)};
			
			\node at (235:2.3) {\(2\)};
			\node at (305:2.3) {\(2\)};
			
			\node at ( 50:0.9) {\(1\)};
			\node at (130:0.9) {\(1\)};
			\node at (290:0.8) {\(1\)};
			
			\TikzMultiLine{r}{v1}{+1}{1.2}{->, thick, color-blue}{}
			\TikzMultiLine{r}{v1}{+0}{0}{->, thick, color-red}{}
			\TikzMultiLine{r}{v1}{-1}{1.2}{->, thick, color-green}{}
			
			\TikzMultiLine{r}{v2}{+1}{1.2}{->, thick, color-red}{}
			\TikzMultiLine{r}{v2}{+0}{0}{->, thick, color-green}{}
			\TikzMultiLine{r}{v2}{-1}{1.2}{->, thick, color-blue}{}
			
			\TikzMultiLine{v1}{s3}{+1}{0.6}{->, thick, color-blue}{}
			\TikzMultiLine{v1}{s3}{-1}{0.6}{->, thick, color-green}{}
			
			\TikzMultiLine{v2}{s3}{+1}{0.6}{->, thick, color-red}{}
			\TikzMultiLine{v2}{s3}{-1}{0.6}{->, thick, color-blue}{}
			
			\TikzMultiLine{s3}{v3}{+1}{0.6}{->, thick, color-red}{}
			\TikzMultiLine{s3}{v3}{-1}{0.6}{->, thick, color-green}{}
			
			\TikzMultiLine{v1}{s1}{+0}{0}{->, thick, color-green}{}
			\TikzMultiLine{v3}{s1}{+0}{0}{->, thick, color-green}{}
			
			\TikzMultiLine{v2}{s2}{+0}{0}{->, thick, color-red}{}
			\TikzMultiLine{v3}{s2}{+0}{0}{->, thick, color-red}{}
		\end{tikzpicture}
		\caption{}
		\label{fig:bcr-improved-small-integrality-gap-example}
	\end{subfigure}
	\caption{
		Small instances with integrality gaps larger than \(1\) for \BCRclass\ and \BCRimprovedclass.
		The required vertices are squared and the weight of the edges are given by the respective numbers.
		Optimum integral solutions for both instances have cost \(8\).
		Setting all shown directed edges to \(\frac{1}{2}\) leads to a feasible fractional solution of \nameref{lp:MCFR} with cost \(\frac{15}{2}\) and therefore to an integrality gap of at least \(\frac{16}{15}\).
		For \MCFRimproved\ this is only possible for the instance in \cref{fig:bcr-improved-small-integrality-gap-example}.
		In \cref{fig:bcr-small-integrality-gap-example} \MCFRimproved\ is exact as shown \cref{thm:improved-bcr-exactness}.
		For the instance in \cref{fig:bcr-improved-small-integrality-gap-example} one can use unit edge cost to obtain an integrality gap of at least \(\frac{10}{9}\) for both relaxations.
	}
	\label{fig:bcr-bcr-improved-small-integrality-gap-examples}
\end{figure}

A group of equivalent constraints which are not valid in general for optimum solutions for \BCRclass\ (an example is presented in \cref{fig:bcr-small-integrality-gap-example}) are the \textit{Steiner vertex degree constraints}:
\[
	\begin{aligned}
		f_{r}(\delta_{G}^{+}(v)) & \geq f_{r}(\delta_{G}^{-}(v)) & v \in V \setminus R\\
		b(v) & \geq 0 & v \in V \setminus R\\
		u(\delta_{G}(v)) & \geq 2 \cdot y(v) & v \in V \setminus R
	\end{aligned}
\]
A valid assumption for a minimal Steiner tree \(T\) is that all leaves are required vertices, as otherwise we can remove the respective part of the tree without increasing the cost.
Therefore each non-required vertex within the tree has at least degree \(2\).
Using this assumption it is possible to easily strengthen the bidirected relaxations.

The inequalities \(f(\delta_{G}^{+}(v)) \geq f(\delta_{G}^{-}(v))\) for non-required vertices were introduced in \cite{KM98}, which the authors call \textit{flow-balance constraints}, and are added to \nameref{lp:BCR} and \nameref{lp:MCFR} to strengthen the relaxation in practice.
We refer to the resulting formulations as \(\operatorname{BCR}^{+}\)\label{lp:BCRimproved} and \(\operatorname{MCFR}^{+}\)\label{lp:MCFRimproved}, respectively.

The inequalities \(b(v) \geq 0\) for non-required vertices were introduced in \cite{CDV11} as an improvement for \nameref{lp:MBFR} which also implies an improvement for \nameref{lp:MBCR}.
We denote the resulting formulations \(\operatorname{MBFR}^{+}\)\label{lp:MBFRimproved} and \(\operatorname{MBCR}^{+}\)\label{lp:MBCRimproved}, respectively.
For \(\operatorname{STER}^{+}\)\label{lp:STERimproved} the corresponding constraint is \(u(\delta_{G}(v)) \geq 2 \cdot y(v)\) for every \(v \in V \setminus R\).

In \cite[Sec.\ 2.2.2]{Pritchard2009} it is already noted that these improvements are equivalent.
We make this precise here.
It is easy to see that we obtain with the previous results:
\[
	\begin{aligned}
		\operatorname{proj}_{u, f_{r}}(\operatorname{BCR^{+}}) & = \operatorname{proj}_{u, f_{r}}(\operatorname{MCFR^{+}})\\
		\operatorname{proj}_{u, b}(\operatorname{MBCR}^{+}) & = \operatorname{proj}_{u, b}(\operatorname{MBFR}^{+})\\
		\operatorname{proj}_{u}(\operatorname{MBCR}^{+}) & = \operatorname{proj}_{u}(\operatorname{STER}^{+})
	\end{aligned}
\]

\begin{theorem}
	\(\operatorname{proj}_{u}(\operatorname{MBFR^{+}}) = \operatorname{proj}_{u}(\operatorname{MCFR^{+}})\)
\end{theorem}
\begin{proof}
	We only have to extend the proof of \cref{thm:equivalence-mbfr-mcfr}.
	
	For \(\operatorname{proj}_{u}(\operatorname{MBFR^{+}}) \supseteq \operatorname{proj}_{u}(\operatorname{MCFR^{+}})\) recall the definition of \(b'(v)\) for \(v \in V\):
	\[b'(v) := f_{r}(\delta_{G}^{+}(v)) - f_{r}(\delta_{G}^{-}(v)) - 2 \cdot \llbracket v = r \rrbracket\]
	Then for \(v \in V \setminus R\) we have \(b'(v) \geq 0\).
	
	For \(\operatorname{proj}_{u}(\operatorname{MBFR^{+}}) \subseteq \operatorname{proj}_{u}(\operatorname{MCFR^{+}})\) recall the definition of \(f'_{r}(v, w)\) for \((v, w) \in E_{\leftrightarrow}\):
	\[f'_{r}(v, w) := \frac{1}{2} \cdot \left( \ f_{r}(v, w) - f_{r}(w, v) + u(\{v, w\}) \ \right)\]
	Then we can compute for \(v \in V \setminus R\):
	\[f'_{r}(\delta_{G}^{+}(v)) - f'_{r}(\delta_{G}^{-}(v)) = f_{r}(\delta_{G}^{+}(v)) - f_{r}(\delta_{G}^{-}(v)) = b(v) + 2 \cdot \llbracket v = r \rrbracket = b(v) \geq 0 \qedhere\]
\end{proof}

Hence, analogously to \BCRclass, we denote the class of equivalent formulations for \BCRimproved\ by \(\operatorname{\mathcal{BCR}^{+}}\).\label{lp-class:BCRimproved}
\BCRimproved\ is, just like \nameref{lp:BCR}, efficiently solvable and independent of the chosen root \(r\).
We will see that there are instances for which the consideration of the additional constraints in \BCRimprovedclass\ compared to \BCRclass\ makes a significant difference regarding the optimum solution values for the respective LP-relaxations.

	
\pagebreak

\subsubsection{Exactness for 3 required vertices}\label{sec:exactness-3-terminals-bcr-improved}

We will now prove that the LP-relaxations in \BCRimprovedclass\ are exact for Steiner tree instances with at most \(3\) required vertices.
In particular, we only have to prove the case of exactly \(3\) required vertices as the LP-relaxations in \BCRclass\ are already exact for \(2\) required vertices.
This is also best possible as \cref{fig:bcr-bcr-improved-small-integrality-gap-examples} shows.

\begin{theorem}\label{thm:improved-bcr-exactness}
	Let \((G, R, c)\) be a Steiner tree instance.\\
	If \(|R| = 3\) then \(\operatorname{opt}_{\operatorname{\mathcal{BCR}^{+}}}(G, R, c) = \operatorname{opt}_{\operatorname{STP}}(G, R, c)\).
\end{theorem}

\begin{proof}
	We have to show \(\operatorname{opt}_{\operatorname{\mathcal{BCR}^{+}}}(G, R, c) \geq \operatorname{opt}_{\operatorname{STP}}(G, R, c)\).
	
	Let \((u, f_{r}, g)\) be an optimum solution of \MCFRimproved\ with respect to \((G, R, c, r)\).
	We can assume w.l.o.g.\ that for any edge \(e = \{v, w\} \in E(G)\) \(u(\{v, w\}) = f_{r}((v, w)) + f_{r}((w, v))\) and \(f_{r}(e) = \max_{s \in R \setminus \{r\}} g_{s}(e)\), as otherwise we can decrease \(u(e)\) and \(f_{r}(e)\) appropriately.
	This does not increase the objective value since \(c\) is non-negative.
	
	As \(|R| = 3\) we can write \(R = \{r, s_{1}, s_{2}\}\), where \(r\) is the chosen root.
	We define \(g_{s_1, s_2} : E_{\leftrightarrow} \to \mathbb{R}_{\geq 0}\), which will indicate the common flow on an edge \(e\), as follows:
	\[g_{s_1, s_2}(e) := \min \left\{ g_{s_1}(e),\ g_{s_2}(e) \right\}\]
	We see easily:
	\[g_{s_1}(e) + g_{s_2}(e) = \min\{ g_{s_1}(e),\ g_{s_2}(e)\} + \max\{ g_{s_1}(e),\ g_{s_2}(e)\} = g_{s_1, s_2}(e) + f_{r}(e)\]
	With that and using \(f_{r}(\delta_{G}^{+}(v)) \geq f_{r}(\delta_{G}^{-}(v))\) we obtain for \(v \in V \setminus \{r\}\):
	\[
		\begin{aligned}
			g_{s_{1}, s_{2}}(\delta_{G}^{-}(v)) - g_{s_{1}, s_{2}}(\delta_{G}^{+}(v))
			& = \sum_{i = 1}^{2} \left( g_{s_{i}}(\delta_{G}^{-}(v)) - g_{s_{i}}(\delta_{G}^{+}(v)) \right) - \left( f_{r}(\delta_{G}^{-}(v)) - f_{r}(\delta_{G}^{+}(v)) \right)\\
			& = \sum_{i = 1}^{2} \left(\ \llbracket v = s_{i} \rrbracket - \llbracket v = r \rrbracket \ \right) + f_{r}(\delta_{G}^{+}(v)) - f_{r}(\delta_{G}^{-}(v)) \geq 0
		\end{aligned}
	\]
	
	The idea now is to extract a Steiner tree.
	For \(3\) required vertices this is always a star.
	\[X_{f} := \left\{ v \in V \setminus \{r\} \ : \ g_{s_1, s_2}(\delta_{G}^{-}(v)) > g_{s_1, s_2}(\delta_{G}^{+}(v)) \right\}\]
	\(X_{f}\) contains possible branching vertices for a Steiner tree.
	
	Assume \(X_{f} \neq \emptyset\) and let \(v_{*} \in X_{f}\).
	Then there exists \(e_{s_1, s_2} \in \delta_{G}^{-}(v_{*})\) such that \(g_{s_1, s_2}(e_{s_1, s_2}) > 0\) and therefore there is a directed path from \(r\) to \(v_{*}\) in \((V, \{e \in E_{\leftrightarrow} : g_{s_1, s_2}(e) > 0\})\), since \(g_{s_1, s_2}(\delta_{G}^{-}(v)) \geq g_{s_1, s_2}(\delta_{G}^{+}(v))\) holds for all \(v \in V \setminus \{r\}\).
	Denote this path by \(P_{[r, v_{*}]}\).
	Furthermore, \(g_{s_1, s_2}(\delta_{G}^+(v_{*})) < g_{s_1, s_2}(\delta_{G}^-(v_{*})) \leq g_{s_{i}}(\delta_{G}^-(v_{*})) = g_{s_{i}}(\delta_{G}^+(v_{*}))\) for \(i \in \{1, 2\}\).
	
	Assume \(X_{f} = \emptyset\), i.e.\ \(g_{s_1, s_2}(\delta_{G}^{+}(v)) = g_{s_1, s_2}(\delta_{G}^{-}(v))\) for \(v \in V \setminus \{r\}\), and let \(v_{*} = r\).
	Then \(g_{s_1, s_2}(\delta_{G}^{+}(v_{*})) < g_{s_{i}}(\delta_{G}^{+}(v_{*}))\) for \(i \in \{1, 2\}\), as otherwise \(g_{s_{i}}(\delta_{G}^{+}(s_{i})) = g_{s_{i}}(\delta_{G}^{-}(s_{i}))\).
	
	Hence, in both cases, there exists \(e_{s_{i}} \in \delta_{G}^{+}(v_{*})\) such that \(g_{s_{i}}(e_{s_{i}}) > g_{s_1, s_2}(e_{s_{i}})\) for \(i \in \{1, 2\}\).
	So there is a path from \(v_{*}\) to \(s_{i}\) in \((V, \{e \in E_{\leftrightarrow} : g_{s_{i}}(e) > g_{s_{1}, s_{2}}(e)\})\), as \(g_{s_{1}, s_{2}}(\delta_{G}^{-}(v)) \geq g_{s_{1}, s_{2}}(\delta_{G}^{+}(v))\) for all \(v \in V \setminus \{r\}\).
	Denote these directed paths by \(P_{[v_{*}, s_{1}]}\) and \(P_{[v_{*}, s_{2}]}\).
	
	We obtain a directed Steiner tree \(T^{r\to} := P_{[r, v_{*}]} + P_{[v_{*}, s_1]} + P_{[v_{*}, s_2]}\).
	Denote by \(T\) the corresponding undirected one.
	We define further:
	\[
		\begin{aligned}
			\varepsilon_{v_{*}} & := g_{s_1, s_2}(\delta_{G}^{-}(v_{*})) - g_{s_1, s_2}(\delta_{G}^{+}(v_{*}))\\
			\varepsilon_{[r, v_{*}]} & := \min_{e \in E(P_{[r, v_{*}]})} g_{s_1, s_2}(e)\\
			\varepsilon_{[v_{*}, s_{i}]} & := \min_{e \in E(P_{[v_{*}, s_{i}]})} (\ g_{s_{i}}(e) - g_{s_1, s_2}(e) \ )
		\end{aligned}
	\]
	\[\varepsilon := \begin{cases}
	\min \left\{ \varepsilon_{v_{*}} \ , \ \varepsilon_{[r, v_{*}]} \ , \ \min_{i \in [2]} \varepsilon_{[v_{*}, s_{i}]} \right\} & \mbox{if } v_{*} \neq r\\
	\min_{i \in [2]} \varepsilon_{[v_{*}, s_{i}]} & \mbox{if } v_{*} = r
	\end{cases}\]
	By definition \(\varepsilon > 0\) and we can assume \(\varepsilon < 1\) as otherwise \(\operatorname{opt}_{\operatorname{STP}}(G, R, c) \leq \varepsilon \cdot \operatorname{opt}_{\operatorname{STP}}(G, R, c) \leq \varepsilon \cdot c(E(T)) \leq \operatorname{opt}_{\operatorname{\mathcal{BCR}^{+}}}(G, R, c)\).
	We define another solution \((u', f'_{r}, g')\) of \MCFRimproved\ as follows:
	\[
		\begin{aligned}
			g'_{s_{i}}(e) & := \frac{g_{s_{i}}(e) - \varepsilon \cdot \llbracket e \in E(T_{[r, s_{i}]}^{r}) \rrbracket}{1 - \varepsilon}\\[3pt]
			f'_{r}(e) & := \frac{f_{r}(e) - \varepsilon \cdot \llbracket e \in E(T^{r}) \rrbracket}{1 - \varepsilon}\\[3pt]
			u'(\{v, w\}) & := f'_{r}(v, w) + f'_{r}(w, v)
		\end{aligned}
	\]
	We will now prove that \((u', f'_{r}, g')\) is also a solution of \MCFRimproved.
	It is easy to see that \(f'\) is non-negative by the definition of \(\varepsilon\), in particular since \(\varepsilon\) is not larger than \(\varepsilon_{[r, v_{*}]}\) and \(\varepsilon_{[v_{*}, s_{i}]}\), and therefore \(u'\) is also non-negative.
	Hence, it suffices to prove the following three statements:
	\[
		\begin{aligned}
			\text{a)} \quad & \forall e \in E_{\leftrightarrow} & \max_{s \in R \setminus \{r\}} g'_{s}(e) & = f'_{r}(e)\\
			\text{b)} \quad & \forall v \in V & g'_{s_{i}}(\delta_{G}^{+}(v)) - g'_{s_{i}}(\delta_{G}^{-}(v)) & = \llbracket v = r \rrbracket - \llbracket v = s_{i} \rrbracket\\
			\text{c)} \quad & \forall v \in V \setminus R & f'_{r}(\delta_{G}^{+}(v)) & \geq f'_{r}(\delta_{G}^{-}(v))
		\end{aligned}
	\]
	
	a)
	By the definition of \(f'\) it suffices to show \(g'_{s_{i}}(e) \leq f'_{r}(e)\).
	We assume w.l.o.g.\ \(i = 1\).
	Only the case \(e \in E(T^{r}) \setminus E(T_{[r, s_{1}]}^{r}) = E(T_{[v_{*}, s_{2}]}^{r})\) is unclear as otherwise we reduce the values in the same way.
	For this case we compute:
	\[
		\begin{aligned}
			(1 - \varepsilon) \cdot (f'_{r}(e) - g'_{s_{1}}(e)) & = (f_{r}(e) - \varepsilon) - g_{s_{1}}(e)\\
			& = f_{r}(e) - g_{s_{1}}(e) - \varepsilon\\
			& = g_{s_{2}}(e) - g_{s_{1}, s_{2}}(e) - \varepsilon\\
			& \geq \varepsilon_{[v_{*}, s_{2}]} - \varepsilon\\
			& \geq 0
		\end{aligned}
	\]
	
	b)
	We compute for \(v \in V\):
	\[
		\begin{aligned}
			& (1 - \varepsilon) \cdot \left( g'_{s_{i}}(\delta_{G}^{+}(v)) - g'_{s_{i}}(\delta_{G}^{-}(v)) \right)\\
			& = g_{s_{i}}(\delta_{G}^{+}(v)) - g_{s_{i}}(\delta_{G}^{-}(v)) - \varepsilon \cdot \llbracket v = r \rrbracket + \varepsilon \cdot \llbracket v = s_{i} \rrbracket\\
			& = \llbracket v = r \rrbracket - \llbracket v = s_{i} \rrbracket - \varepsilon \cdot ( \llbracket v = r \rrbracket - \llbracket v = s_{i} \rrbracket )\\
			& = (1 - \varepsilon) \cdot ( \llbracket v = r \rrbracket - \llbracket v = s_{i} \rrbracket )
		\end{aligned}
	\]
	
	c)
	Using \(g_{s_1}(e) + g_{s_2}(e) = f_{r}(e) + g_{s_1, s_2}(e)\) we compute for \(v \in V \setminus R\):
	\[
		\begin{aligned}
			& (1 - \varepsilon) \cdot \left( f'_{r}(\delta_{G}^{+}(v)) - f'_{r}(\delta_{G}^{-}(v)) \right)\\
			& = f_{r}(\delta_{G}^{+}(v)) - f_{r}(\delta_{G}^{-}(v)) - \varepsilon \cdot \llbracket v = r \rrbracket + \varepsilon \cdot \sum_{i \in \{1, 2\}} \llbracket v = s_{i} \rrbracket - \varepsilon \cdot \llbracket v = v_{*} \rrbracket\\
			& = g_{s_{1}, s_{2}}(\delta_{G}^{-}(v)) - g_{s_{1}, s_{2}}(\delta_{G}^{+}(v)) - \varepsilon \cdot \llbracket v = v_{*} \rrbracket\\
			& \geq \varepsilon_{v_{*}} \cdot \llbracket v = v_{*} \rrbracket - \varepsilon \cdot \llbracket v = v_{*} \rrbracket\\
			& \geq 0
		\end{aligned}
	\]
	
	Since \((u', f'_{r}, g')\) is a solution of \MCFRimproved\ we obtain:
	\[
		\begin{aligned}
			\operatorname{opt}_{\operatorname{\mathcal{BCR}}^{+}}(G, R, c)
			& = \sum_{e \in E(G)} c(e) \cdot u(e)\\
			& = \sum_{e \in E(G)} c(e) \cdot \left( (1 - \varepsilon) \cdot u'(e) + \varepsilon \cdot \llbracket e \in E(T) \rrbracket \right)\\
			& = (1 - \varepsilon) \cdot \sum_{e \in E(G)} c(e) \cdot u'(e) + \varepsilon \cdot c(E(T))\\
			& \geq (1 - \varepsilon) \cdot \operatorname{opt}_{\operatorname{\mathcal{BCR}}^{+}}(G, R, c) + \varepsilon \cdot \operatorname{opt}_{\operatorname{STP}}(G, R, c)
		\end{aligned}
	\]
	And therefore:
	\[\operatorname{opt}_{\operatorname{\mathcal{BCR}}^{+}}(G, R, c) \geq \operatorname{opt}_{\operatorname{STP}}(G, R, c) \qedhere\]
\end{proof}

As already explained, \cref{thm:improved-bcr-exactness} with \cref{thm:steiner-tree-lp-relaxation-chain} immediately proves \cref{thm:improved-bidirected-relaxation-exact-for-3}.

\paragraph{Remark.}
If one were to iterate the procedure of extracting Steiner trees, one would obtain a finite convex combination of Steiner trees for the solution \((u, f)\).
For that, consider:
\begin{align*}
\phi(f) & := \left| \left\{ v \in V \setminus \{r\} : g_{s_1, s_2}(\delta_{G}^{-}(v)) > g_{s_1, s_2}(\delta_{G}^{+}(v)) \right\} \right|\\
& \quad\ + \left| \left\{ e \in E_{\leftrightarrow} : g_{s_1, s_2}(e) > 0 \right\} \right|\\
& \quad\ + \sum_{i \in [2]} \left| \left\{ e \in E_{\leftrightarrow} : g_{s_{i}}(e) > g_{s_1, s_2}(e) \right\} \right|
\end{align*}
\(\phi(f')\) is defined analogously.
By the definition of \(\varepsilon\) we have \(\phi(f') < \phi(f)\).
Therefore, by induction we know \((u', f')\) has a convex combination \(\sum_{i = 1}^{k} \lambda_i \cdot T_{i}\).
Then \((1 - \varepsilon) \cdot \sum_{i = 1}^{k} \lambda_i \cdot T_{i} + \varepsilon \cdot T\) is a convex combination of \((u, f)\).

	\pagebreak

\section{Simplex based Steiner tree instances yielding large gaps}\label{sec:simplex-based-steiner-tree-instances-yielding-large-gaps}

In \cite{BGRS13} it is proven that \(\operatorname{gap}_{\operatorname{\mathcal{BCR}}, \operatorname{STP}} \geq \frac{36}{31} \approx 1.161\).
The authors construct a series of instances which are based on an instance constructed by Skutella \cite{BGRS13, KPT11}, which itself can be expressed via a set cover instance.
Skutella's instance is actually a special case of the former instances and achieves an integrality gap of \(\frac{8}{7} \approx 1.142\).
The approach can easily be generalized to arbitrary set cover instances obtaining \textit{set cover based Steiner tree instances}, the proofs in \cite{BGRS13} can be adapted naturally and it is easy to see that the respective solutions are also valid for \BCRimprovedclass\ (\cref{sec:set-cover-based-steiner-tree-instances}).
In particular choosing again the set cover instance for Skutella's instance we obtain \(\operatorname{gap}_{\operatorname{\mathcal{BCR}^{+}}, \operatorname{STP}} \geq \frac{36}{31}\).

Hence, an interesting question is whether there is another class of instances which improves this bound by exploiting the weaknesses of \BCRclass\ towards \BCRimprovedclass.
Therefore, we are particularly interested in instances \((G, R, c)\) such that \(\operatorname{gap}_{\operatorname{\mathcal{BCR}}, \operatorname{\mathcal{BCR}^{+}}}(G, R, c)\) is large.
Based on the work of Chakrabarty et al.\ \cite{CDV11} about embedding Steiner tree instances into simplices, we will show that there is a class of instances yielding \(\operatorname{gap}_{\operatorname{\mathcal{BCR}}, \operatorname{\mathcal{BCR}^{+}}} \geq \frac{6}{5} = 1.2\).
Note that this also implies \(\operatorname{gap}_{\operatorname{\mathcal{BCR}}, \operatorname{STP}} \geq \frac{6}{5}\), improving the previous best known lower bound of \(\frac{36}{31} \approx 1.161\).

	
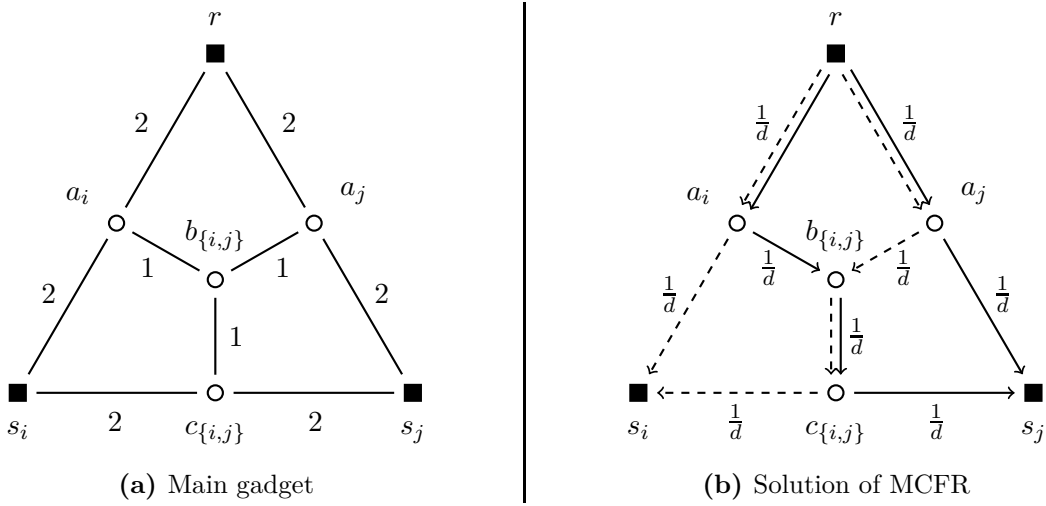
\begin{figure}
	\centering
	
	\begin{subfigure}[]{0.49\textwidth}
		\centering
		
		\begin{tikzpicture}[scale=1]
			\tikzset{vertex/.style={circle, thick, draw=black, fill=white, minimum size=0pt, inner sep=2pt, outer sep=4pt}}
			\tikzset{terminal/.style={rectangle, fill=black, minimum size=0pt, inner sep=3pt, outer sep=4pt}}
			\tikzset{color-red/.style={red}}
			\tikzset{color-green/.style={green}}
			\tikzset{color-blue/.style={blue}}
			\tikzset{color-red/.style={solid}}
			\tikzset{color-green/.style={dashed}}
			\tikzset{color-blue/.style={dashdotted}}
			
			\node[vertex, terminal, label={+90:\(r\)}] (r) at (90:3) {};
			\node[vertex, terminal, label={-90:\(s_{i}\)}] (si) at (210:3) {};
			\node[vertex, terminal, label={-90:\(s_{j}\)}] (sj) at (330:3) {};
			
			\node[vertex, label={+150:\(a_{i}\)}] (ai) at (150:1.5) {};
			\node[vertex, label={+30:\(a_{j}\)}] (aj) at (30:1.5) {};
			\node[vertex, label={+90:\(b_{\{i, j\}}\)}] (bij) at (0:0) {};
			\node[vertex, label={-90:\(c_{\{i, j\}}\)}] (cij) at (270:1.5) {};
			
			\node at ( 65:2.3) {\(2\)};
			\node at (115:2.3) {\(2\)};
			
			\node at (185:2.2) {\(2\)};
			\node at (355:2.2) {\(2\)};
			
			\node at (235:2.3) {\(2\)};
			\node at (305:2.3) {\(2\)};
			
			\node at ( 10:0.9) {\(1\)};
			\node at (170:0.9) {\(1\)};
			\node at (290:0.8) {\(1\)};
			
			\draw[thick] (r) -- (ai);
			\draw[thick] (r) -- (aj);
			
			\draw[thick] (ai) -- (si);
			\draw[thick] (aj) -- (sj);
			
			\draw[thick] (ai) -- (bij);
			\draw[thick] (aj) -- (bij);
			\draw[thick] (bij) -- (cij);
			
			\draw[thick] (cij) -- (si);
			\draw[thick] (cij) -- (sj);
		\end{tikzpicture}
		
		\caption{
			Main gadget
		}
		
		\label{fig:goemans-instance-series-graph-gadget}
	\end{subfigure}
	\vline
	\hfill
	\begin{subfigure}[]{0.49\textwidth}
		\centering
		
		\begin{tikzpicture}[scale=1]
			\tikzset{vertex/.style={circle, thick, draw=black, fill=white, minimum size=0pt, inner sep=2pt, outer sep=4pt}}
			\tikzset{terminal/.style={rectangle, fill=black, minimum size=0pt, inner sep=3pt, outer sep=4pt}}
			\tikzset{color-red/.style={red}}
			\tikzset{color-green/.style={green}}
			\tikzset{color-blue/.style={blue}}
			\tikzset{color-red/.style={solid}}
			\tikzset{color-green/.style={dashed}}
			\tikzset{color-blue/.style={dashdotted}}
			
			\node[vertex, terminal, label={+90:\(r\)}] (r) at (90:3) {};
			\node[vertex, terminal, label={-90:\(s_{i}\)}] (si) at (210:3) {};
			\node[vertex, terminal, label={-90:\(s_{j}\)}] (sj) at (330:3) {};
			
			\node[vertex, label={+150:\(a_{i}\)}] (ai) at (150:1.5) {};
			\node[vertex, label={+30:\(a_{j}\)}] (aj) at (30:1.5) {};
			\node[vertex, label={+90:\(b_{\{i, j\}}\)}] (bij) at (0:0) {};
			\node[vertex, label={-90:\(c_{\{i, j\}}\)}] (cij) at (270:1.5) {};
			
			\node at ( 65:2.3) {\(\frac{1}{d}\)};
			\node at (115:2.3) {\(\frac{1}{d}\)};
			
			\node at (185:2.2) {\(\frac{1}{d}\)};
			\node at (355:2.2) {\(\frac{1}{d}\)};
			
			\node at (235:2.3) {\(\frac{1}{d}\)};
			\node at (305:2.3) {\(\frac{1}{d}\)};
			
			\node at ( 10:0.9) {\(\frac{1}{d}\)};
			\node at (170:0.9) {\(\frac{1}{d}\)};
			\node at (290:0.8) {\(\frac{1}{d}\)};
			
%			\draw[thick, ->] (r) -- (ai);
%			\draw[thick, ->] (r) -- (aj);
%			
%			\draw[thick, ->] (ai) -- (si);
%			\draw[thick, ->] (aj) -- (sj);
%			
%			\draw[thick, ->] (ai) -- (bij);
%			\draw[thick, ->] (aj) -- (bij);
%			\draw[thick, ->] (bij) -- (cij);
%			
%			\draw[thick, ->] (cij) -- (si);
%			\draw[thick, ->] (cij) -- (sj);
			
			\TikzMultiLine{r}{ai}{+1}{0.6}{->, thick, color-red}{}
			\TikzMultiLine{r}{ai}{-1}{0.6}{->, thick, color-green}{}
			
			\TikzMultiLine{r}{aj}{+1}{0.6}{->, thick, color-red}{}
			\TikzMultiLine{r}{aj}{-1}{0.6}{->, thick, color-green}{}
			
			\TikzMultiLine{ai}{bij}{+0}{0}{->, thick, color-red}{}
			\TikzMultiLine{aj}{bij}{+0}{0}{->, thick, color-green}{}
			
			\TikzMultiLine{bij}{cij}{+1}{0.6}{->, thick, color-red}{}
			\TikzMultiLine{bij}{cij}{-1}{0.6}{->, thick, color-green}{}
			
			\TikzMultiLine{ai}{si}{+0}{0}{->, thick, color-green}{}
			\TikzMultiLine{cij}{si}{+0}{0}{->, thick, color-green}{}
			
			\TikzMultiLine{aj}{sj}{+0}{0}{->, thick, color-red}{}
			\TikzMultiLine{cij}{sj}{+0}{0}{->, thick, color-red}{}
		\end{tikzpicture}
		
		\caption{
			Solution of \nameref{lp:MCFR}
		}
		
		\label{fig:goemans-instance-series-flow}
	\end{subfigure}
	
	\caption{
		Goemans' instance series
	}

	\label{fig:goemans-instance-series}
\end{figure}

The first notable lower bound on the integrality gap for \nameref{lp:BCR} was proved by Goemans \cite[Exercise 22.11]{Vaz01}.
He constructed a series of instances also yielding an integrality gap of \(\frac{8}{7}\).
The main difference to the set cover based Steiner tree instances is, that on Goemans' instances \BCRimprovedclass\ is exact and therefore \(\operatorname{gap}_{\operatorname{\mathcal{BCR}}, \operatorname{\mathcal{BCR}^{+}}} \geq \frac{8}{7}\).
An instance consists of a graph with \(d + 1\) required vertices, namely \(s_{1}, \dotsc, s_{d}\) and a distinguished required vertex \(r\).
These required vertices are embedded in gadgets as depicted in \cref{fig:goemans-instance-series-graph-gadget}.
Note that we have already encountered these gadgets in \cref{fig:bcr-small-integrality-gap-example}.
The whole graph is given by the composition of these gadgets where corresponding vertices are identified.
That means, the graph has \(d + 2 \cdot \binom{d}{2}\) Steiner vertices, namely \(a_{i}\) for \(i \in [d]\) and \(b_{i, j}\), \(c_{i, j}\) for \(i, j \in [d]\) with \(i \neq j\) and \(2 \cdot d + (2 + 3) \cdot \binom{d}{2}\) undirected edges which each have cost as depicted.
Let \(\operatorname{GI}_{d}\) denote Goemans' instance for a certain \(d\).

It is easy to see that for each \(i \in [d]\) there are \(d\) edge-disjoint paths from \(r\) to \(s_{i}\) using only the directed edges in each gadget as indicated in \cref{fig:goemans-instance-series-flow}.
Therefore setting \(u \equiv \frac{1}{d}\) and the respective flow values to \(\frac{1}{d}\) or \(0\), we can easily verify that this is a solution of \nameref{lp:MCFR}.
The cost computes to
\(\frac{1}{d} \left( 2 \cdot \left( 2 \cdot d + 2 \cdot \binom{d}{2} \right) + 1 \cdot \left( 3 \cdot \binom{d}{2} \right) \right) = \frac{1}{d} \left( 7 \cdot \binom{d}{2} + 4 \cdot d \right) = \frac{7 \cdot d + 1}{2}\).
Hence, \(\operatorname{opt}_{\operatorname{\mathcal{BCR}}}(\operatorname{GI}_{d}) \leq \frac{7 \cdot d + 1}{2}\).
On the other hand, it is easy to see that \(\operatorname{opt}_{\operatorname{\mathcal{BCR}^{+}}}(\operatorname{GI}_{d}) \leq 4 \cdot d\), by connecting each \(s_{i}\) to \(r\) along \(a_{i}\).
In \cref{thm:goemans-instance-series-integral-optimum} we will show that \(\operatorname{opt}_{\operatorname{\mathcal{BCR}^{+}}}(\operatorname{GI}_{d}) = 4 \cdot d\) holds.
We obtain \(\operatorname{gap}_{\operatorname{\mathcal{BCR}}, \operatorname{\mathcal{BCR}^{+}}}(\operatorname{GI}_{d}) \geq \frac{8 \cdot d}{7 \cdot d + 1}\) and therefore \(\operatorname{gap}_{\operatorname{\mathcal{BCR}}, \operatorname{\mathcal{BCR}^{+}}} \geq \frac{8}{7}\).

\subsection{Simplex Embedding of Steiner tree instances}\label{sec:simplex-embedding-steiner-tree-instances}

Chakrabarty et al.\ \cite{CDV11} propose two lower bounds on the optimal cost of a Steiner tree for a given instance \((G, R, c)\).
The idea is to embed the vertices \(V(G)\) into a simplex of a certain size while respecting the \(\operatorname{L}^{1}\)-distance between neighbouring vertices given by the respective edge cost.
The lower bound is then given by spreading the required vertices in their assigned dimensions as far as possible and summing over their respective coordinates.
For a precise definition we need the following notion of simplices.
%An example is presented in \cref{fig:simplex}.

\begin{definition}
	For \(d \in \mathbb{N}\), \(\mathbb{K} \in \{\mathbb{R}, \mathbb{Z}\}\) and \(s \in \mathbb{K}_{\geq 0}\) we define:
	\[\triangle_{d, s}^{\mathbb{K}} := \left\{x \in \mathbb{K}_{\geq 0}^{d + 1} \ :\ \sum_{i = 1}^{d + 1} x_{i} = s\right\}\]
	We call \(\triangle_{d, s}^{\mathbb{K}}\) the \textbf{\(\bm{d}\)-dimensional continuous/discrete simplex of size \(\bm{s}\)}.
\end{definition}

For the rest of \cref{sec:simplex-based-steiner-tree-instances-yielding-large-gaps} we denote by \(e_{i}\) the \(i\)-th unit vector in the respective space.
\(\triangle_{d, s}^{\mathbb{R}}\) is the convex hull of the points \(\{s \cdot e_{i} \ : \ i \in [d + 1]\}\).

%\subfile{figures/figure-discrete-simplex}

For a Steiner tree instance \((G = (V, E), R, c)\) let \(d \in \mathbb{N}\) be such that \(|R| = d + 1\) and assume \(R = \{r_{1}, \dotsc, r_{d + 1}\}\).
In this way each required vertex is made to correspond with a dimension of \(\mathbb{R}^{d + 1}\).
Then we obtain the following two maximization problems from \cite{CDV11}.
Recall that \(\|x\|_{1} = \sum_{i = 1}^{n} |x_{i}|\) denotes the \(\operatorname{L}^{1}\)-distance for a vector \(x \in \mathbb{R}^{n}\).

\begin{linearprogram}[SE]
	\label{lp:SE}
	\(\max\) & \(\displaystyle 2 \cdot \left( \sum_{i = 1}^{d + 1} y(r_{i})_{i} - s \right)\)\\[20pt]
	\(\text{s.t.}\) & \(y : V \to \mathbb{R}_{\geq 0}^{d + 1}\),\quad \(s \in \mathbb{R}_{\geq 0}\)\\[10pt]
	& \(
		\begin{aligned}[t]
			\|y(v) - y(w)\|_{1} & \leq c(\{v, w\}) & \quad \{v, w\} & \in E\\[3pt]
			y(v) & \in \triangle_{d, s}^{\mathbb{R}} & \quad v & \in V
		\end{aligned}
	\)
\end{linearprogram}

\begin{linearprogram}[SE^{+}]
	\label{lp:SE-above}
	\(\max\) & \(\displaystyle 2 \cdot \left( \sum_{i = 1}^{d + 1} y(r_{i})_{i} - s \right)\)\\[20pt]
	\(\text{s.t.}\) & \(y : V \to \mathbb{R}_{\geq 0}^{d + 1}\),\quad \(s \in \mathbb{R}_{\geq 0}\)\\[10pt]
	& \(
		\begin{aligned}[t]
			\|y(v) - y(w)\|_{1} & \leq c(\{v, w\}) & \quad \{v, w\} & \in E\\[3pt]
			y(r) & \in \triangle_{d, s}^{\mathbb{R}} & \quad r & \in R\\
			y(v) & \in \bigcup_{s \leq s'} \triangle^{\mathbb{R}}_{d, s'} & \quad v & \in V \setminus R
		\end{aligned}
	\)
\end{linearprogram}

\begin{theorem}[\cite{CDV11}]\label{thm:simplex-embedding-steiner-tree-instances}
	Let \((G, R, c)\) be a Steiner tree instance, then
	\[
		\begin{aligned}
			\operatorname{opt}_{\operatorname{SE}}(G, R, c) & = \operatorname{opt}_{\operatorname{\mathcal{BCR}}}(G, R, c)\\[6pt]
			\operatorname{opt}_{\operatorname{SE^{+}}}(G, R, c) & = \operatorname{opt}_{\operatorname{\mathcal{BCR}^{+}}}(G, R, c)
		\end{aligned}
	\]
\end{theorem}

Initially Chakrabarty et al.\ \cite{CDV11} prove combinatorially that \nameref{lp:SE} and \nameref{lp:SE-above} is a lower bound for \nameref{problem:STP}.
But they also show, that these two maximization problems are actually dual formulations of a slightly different but equivalent version of \nameref{lp:MBFR}.\\
The constraints \(\|y(v) - y(w)\|_{1} \leq c(\{v, w\})\) can be implemented using auxiliary variables \(z_{i}(\{v, w\})\) for each dimension \(i \in [d + 1]\) as we want to maximize \(y(r_{i})_{i}\), namely \(|y(v)_{i} - y(w)_{i}| \leq z_{i}(\{v, w\})\) and \(\sum_{i \in [d + 1]} z_{i}(\{v, w\}) \leq c(\{v, w\})\).
These dual variables correspond to the primal constraints \(f_{r}(v, w) + f_{r}(w, v) \leq u(\{v, w\})\).

In \cite{CDV11} \nameref{lp:SE} is called \textit{simplex embedding} since \(y\) is actually a function which maps to \(\triangle_{d, s}^{\mathbb{R}}\).
For \nameref{lp:SE-above} we allow Steiner vertices to be mapped to \(\bigcup_{s \leq s'} \triangle_{d, s'}^{\mathbb{R}}\), we adopt the term from \cite{CDV11} and say the Steiner vertices are mapped \textit{above the simplex} \(\triangle_{d, s}^{\mathbb{R}}\).
With this geometrical relation the authors construct a \(\frac{4}{3}\)-approximation algorithm for quasi-bipartite instances and also upper bound the integrality gaps for these instances by the same value.
Note that the current best factor of an approximation algorithm for these instances as well as the upper bound on their integrality gaps is \(\frac{73}{60} \approx 1.217\) \cite{GORZ12}.

Since our main motivation is to construct instances \((G, R, c)\) such that \(\operatorname{gap}_{\operatorname{\mathcal{BCR}}, \operatorname{\mathcal{BCR}^{+}}}(G, R, c)\) is large, we will investigate these dual formulations further.

The optimization goal in \nameref{lp:SE} and \nameref{lp:SE-above} is to spread the required vertices in their corresponding dimensions as far as the respective graph structure with the given edge-costs admits.
Therefore, we want to construct instances which for given \(s \in \mathbb{R}_{\geq 0}\) are easy to embed above \(\triangle_{d, s}^{\mathbb{R}}\), but need a much more compact embedding into the simplex itself, i.e.\ it is much more complicated to spread the \(d + 1\) required vertices in their dimensions.

%Consider the following maximization problem which only depends on \(d\):
%
%\begin{linearprogram}[SL]
%	\label{lp:SL}
%	\(\max\) & \(\displaystyle 2 \cdot \left( \sum_{i = 1}^{d + 1} y(e_{i})_{i} - 1 \right)\)\\[20pt]
%	\(\text{s.t.}\) & \(y : \bigcup_{1 \leq s'} \triangle^{\mathbb{R}}_{d, s'} \to \triangle_{d, 1}^{\mathbb{R}}\)\\[10pt]
%	& \(
%	\begin{aligned}[t]
%		\|y(v) - y(w)\|_{1} & \leq \|v - w\|_{1} & \quad v, w & \in {\textstyle\bigcup_{1 \leq s'}} \triangle^{\mathbb{R}}_{d, s'}\\[3pt]
%		y(e_{i}) & \in \triangle_{d, 1}^{\mathbb{R}} & \quad i & \in [d + 1]
%	\end{aligned}
%	\)
%\end{linearprogram}
%
%\begin{theorem}
%	Denote by \(\operatorname{\mathcal{BCR}_{\leq d + 1}}\) and \(\operatorname{\mathcal{BCR}_{\leq d + 1}^{+}}\) the restrictions of \BCRclass\ and \BCRimprovedclass\ to instances with at most \(d + 1\) required vertices.
%	Then:
%	\[\operatorname{gap}_{\operatorname{\mathcal{BCR}_{\leq d + 1}}, \operatorname{\mathcal{BCR}_{\leq d + 1}^{+}}} \geq \frac{2\cdot d}{\operatorname{opt}_{\operatorname{SL}}(d)}\]
%\end{theorem}
%\begin{proof}
%	Consider the infinite complete graph on the vertex set \(\bigcup_{s \leq s'} \triangle^{\mathbb{R}}_{d, s'}\) with the required vertices \(\{s \cdot e_{1}, \dotsc, s \cdot e_{d + 1}\}\) and edge-costs \(\|v - w\|_{1}\) for \(v, w \in \bigcup_{s \leq s'} \triangle^{\mathbb{R}}_{d, s'}\).
%	Denote this instance by \(\operatorname{SI}_{d, s}^{\mathbb{R}}\).
%	We directly obtain \(\operatorname{opt}_{\operatorname{SE^{+}}}(\operatorname{SI}_{d, s}^{\mathbb{R}}) \geq 2 \cdot d \cdot s\).
%\end{proof}

\subsection{Simplex based Steiner tree instances}\label{sec:simplex-based-steiner-tree-instances}

%\begin{wrapfigure}{r}{0.3\textwidth}
%	\centering
%	\begin{tikzpicture}[scale=0.5]
%	\coordinate (s1-1) at (-3.,  1.73205081);
%	\coordinate (s1-2) at ( 3.,  1.73205081);
%	\coordinate (s1-3) at ( 0., -3.46410162);
%	
%	\coordinate (s0-1) at (-1.5,  2.59807621);
%	\coordinate (s0-2) at ( 1.5,  2.59807621);
%	\coordinate (s0-3) at ( 3.0,  0.);
%	\coordinate (s0-4) at ( 1.5, -2.59807621);
%	\coordinate (s0-5) at (-1.5, -2.59807621);
%	\coordinate (s0-6) at (-3.0,  0.);
%	
%	\draw[thick, draw=orange] (s1-1) -- (s1-2) -- (s1-3) -- cycle;
%	\draw[thick, draw=blue] (s0-1) -- (s0-2) -- (s0-3) -- (s0-4) -- (s0-5) -- (s0-6) -- cycle;
%	\end{tikzpicture}
%	\caption{
%		unit balls
%	}
%	\label{fig:2d-simplex-unit-balls}
%\end{wrapfigure}

For given \(d \in \mathbb{N}\) and \(s \in \mathbb{R}_{\geq 0}\) our idea is to create a graph on the vertex set \(\bigcup_{s \leq s'} \triangle^{\mathbb{R}}_{d, s'}\) with \(\{s \cdot e_{i} \ : \ i \in [d + 1]\}\) as required vertices and the \(\operatorname{L}^{1}\)-distances between neighbouring vertices as edge-costs.
One could consider the infinite complete graph on the vertex set \(\bigcup_{s \leq s'} \triangle^{\mathbb{R}}_{d, s'}\), but there might be many vertices and edges which do not contribute to a more compact embedding into \(\triangle_{d, s}^{\mathbb{R}}\).

Therefore we start with a discretization of \(\triangle_{d, s}^{\mathbb{R}}\), namely \(\triangle_{d, s}^{\mathbb{Z}}\), and try to figure out which additional points in \(\bigcup_{s \leq s'} \triangle^{\mathbb{R}}_{d, s'}\) with appropriate edges contribute to a more compact embedding.
The following observation suggests a possible construction:

A useful decomposition of \(\triangle_{d, s}^{\mathbb{Z}}\) is given by the notion of support.
We define for a point \(v \in \triangle_{d, s}^{\mathbb{Z}}\):
\[
	\begin{aligned}
		\operatorname{support}_{d}(v) & := \{i \in [d + 1] : v_{i} > 0\}\\
		\operatorname{level}_{d}(v) & := |\operatorname{support}_{d}(v)| - 1
	\end{aligned}
\]
\(\triangle_{d, s}^{\mathbb{Z}}\) contains a point on level \(l\) if and only if \(l + 1 \leq s\).
Let \(d, s \in \mathbb{N}\) and \(v \in \triangle_{d, s + 1}^{\mathbb{Z}}\).
Consider the following star graph:
\[\triangle_{d}^{*}(v) := ( \{ v \} \cup \{ v - e_{i} : i \in \operatorname{support}_{d}(v) \},\ \{ \{ v, v - e_{i} \} : i \in i \in \operatorname{support}_{d}(v) \} )\]
\(\triangle_{d}^{*}(v)\) can easily be embed above \(\triangle_{d, s}^{\mathbb{R}}\) with unit edge-costs, since \(\left\| v - (v - e_{i}) \right\|_{1} = 1\) and \(v - e_{i} \in \triangle_{d, s}^{\mathbb{R}}\).
If we want to embed \(\triangle_{d}^{*}(v)\) into \(\triangle_{d, s}^{\mathbb{R}}\) we need to project \(v\) on \(\triangle_{d, s}^{\mathbb{R}}\), which could be done by a orthogonal projection on the point \(v - \frac{1}{\operatorname{level}_{d}(v) + 1} \sum_{j \in \operatorname{support}_{d}(v)} e_{j}\), which is the centre of the points \(v - e_{i}\) and therefore within \(\triangle_{d, s}^{\mathbb{R}}\).
But now we need a much more compact embedding since:
\[
	\begin{aligned}
		& \left\| \left( v - \frac{1}{\operatorname{level}_{d}(v) + 1} \sum_{j \in \operatorname{support}_{d}(v)} e_{j} \right) - (v - e_{i}) \right\|_{1}\\
		& = \left( 1 - \frac{1}{\operatorname{level}_{d}(v) + 1} \right) + \operatorname{level}_{d}(v) \cdot \frac{1}{\operatorname{level}_{d}(v) + 1}\\
		& = 2 \cdot \frac{\operatorname{level}_{d}(v)}{\operatorname{level}_{d}(v) + 1}
	\end{aligned}
\]

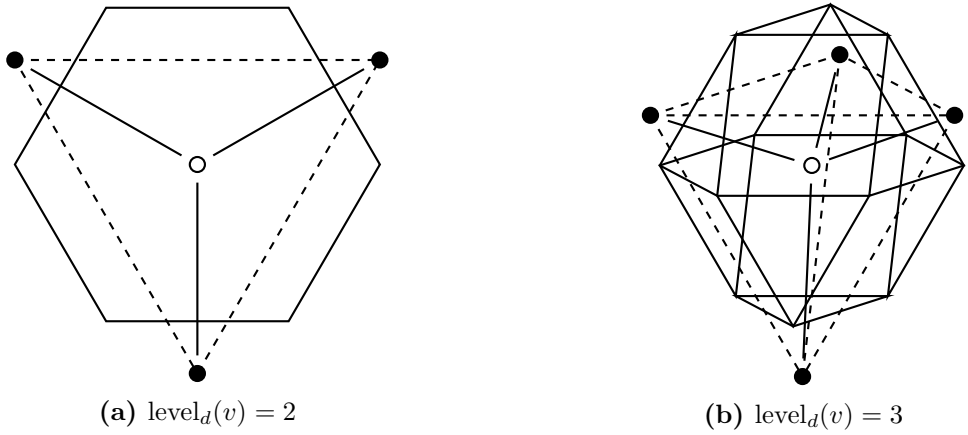
\begin{figure}
	\centering
	\begin{subfigure}[c]{0.49\textwidth}
		\centering
		\begin{tikzpicture}[scale=0.8]
			\tikzset{vertex/.style={circle, thick, minimum size=0pt, inner sep=2pt, outer sep=2pt}}
			\tikzset{vertexinternalzero/.style={vertex, draw=blue, fill=blue}}
			\tikzset{vertexinternalone/.style={vertex, draw=orange, fill=orange}}
			
			\tikzset{vertex/.style={circle, thick, draw=black, fill=white, minimum size=0pt, inner sep=2pt, outer sep=4pt}}
			\tikzset{terminal/.style={rectangle, fill=black, minimum size=0pt, inner sep=3pt, outer sep=4pt}}
			\tikzset{vertexinternalzero/.style={vertex, draw=black, fill=black}}
			\tikzset{vertexinternalone/.style={vertex, draw=black, fill=white}}
			\tikzset{color-blue/.style={solid}}
			\tikzset{color-orange/.style={dashed}}
			
			\coordinate (s1-0) at ( 0.,  0.);
			
			\coordinate (s1-1) at (-3.,  1.73205081);
			\coordinate (s1-2) at ( 3.,  1.73205081);
			\coordinate (s1-3) at ( 0., -3.46410162);
			
			\coordinate (s0-1) at (-1.5,  2.59807621);
			\coordinate (s0-2) at ( 1.5,  2.59807621);
			\coordinate (s0-3) at ( 3.0,  0.);
			\coordinate (s0-4) at ( 1.5, -2.59807621);
			\coordinate (s0-5) at (-1.5, -2.59807621);
			\coordinate (s0-6) at (-3.0,  0.);
			
			\draw[thick, color-orange] (s1-1) -- (s1-2) -- (s1-3) -- cycle;
			\draw[thick, color-blue] (s0-1) -- (s0-2) -- (s0-3) -- (s0-4) -- (s0-5) -- (s0-6) -- cycle;
			
			\node[vertexinternalone] (v1-0) at (s1-0) {};
			
			\node[vertexinternalzero] (v1-1) at (s1-1) {};
			\node[vertexinternalzero] (v1-2) at (s1-2) {};
			\node[vertexinternalzero] (v1-3) at (s1-3) {};
			
			\draw[thick, -] (v1-1) -- (v1-0);
			\draw[thick, -] (v1-2) -- (v1-0);
			\draw[thick, -] (v1-0) -- (v1-3);
		\end{tikzpicture}
		\caption{
			\(\operatorname{level}_{d}(v) = 2\)
		}
	\end{subfigure}
	\begin{subfigure}[c]{0.49\textwidth}
		\centering
		\begin{tikzpicture}[scale=1.0]
			\pgfsetxvec{\pgfpoint{+1.0cm}{+0.0cm}}
			\pgfsetyvec{\pgfpoint{+0.0cm}{+1.0cm}}
			\pgfsetzvec{\pgfpoint{-0.15cm}{-0.6cm}}
			
			\tikzset{vertex/.style={circle, thick, minimum size=0pt, inner sep=2pt, outer sep=2pt}}
			\tikzset{vertexinternalzero/.style={vertex, draw=blue, fill=blue}}
			\tikzset{vertexinternalone/.style={vertex, draw=orange, fill=orange}}
			
			\tikzset{vertex/.style={circle, thick, draw=black, fill=white, minimum size=0pt, inner sep=2pt, outer sep=4pt}}
			\tikzset{terminal/.style={rectangle, fill=black, minimum size=0pt, inner sep=3pt, outer sep=4pt}}
			\tikzset{vertexinternalzero/.style={vertex, draw=black, fill=black}}
			\tikzset{vertexinternalone/.style={vertex, draw=black, fill=white}}
			\tikzset{color-blue/.style={solid}}
			\tikzset{color-orange/.style={dashed}}
			
			\coordinate (s1-0) at ( 0.,  0.,  0.);
			
			\coordinate (s1-1) at (-2.,  1.15470054,  0.81649658);
			\coordinate (s1-2) at ( 2.,  1.15470054,  0.81649658);
			\coordinate (s1-3) at ( 0., -2.30940108,  0.81649658);
			\coordinate (s1-4) at ( 0.,  0.,         -2.44948974);
			
			\coordinate (s0-1) at ( 0., -1.15470054,  1.63299316);
			\coordinate (s0-2) at ( 1.,  0.57735027,  1.63299316);
			\coordinate (s0-3) at (-1.,  0.57735027,  1.63299316);
			
			\coordinate (s0-4) at ( 0.,  1.15470054, -1.63299316);
			\coordinate (s0-5) at ( 1.,  1.73205081,  0.);
			\coordinate (s0-6) at (-1.,  1.73205081,  0.);
			
			\coordinate (s0-7) at (-1., -0.57735027, -1.63299316);
			\coordinate (s0-8) at (-1., -1.73205081,  0.);
			\coordinate (s0-9) at (-2.,  0.,          0.);
			
			\coordinate (s0-10) at ( 1., -0.57735027, -1.63299316);
			\coordinate (s0-11) at ( 1., -1.73205081,  0.);
			\coordinate (s0-12) at ( 2.,  0.,          0.);
			
			\coordinate (as0-1) at ( 0., -1.15470054,  1.63299316);
			\coordinate (as0-2) at ( 0.,  1.15470054, -1.63299316);
			\coordinate (as0-3) at (-1., -0.57735027, -1.63299316);
			\coordinate (as0-4) at ( 1.,  0.57735027,  1.63299316);
			\coordinate (as0-5) at (-1.,  0.57735027,  1.63299316);
			\coordinate (as0-6) at ( 1., -0.57735027, -1.63299316);
			\coordinate (as0-7) at (-1., -1.73205081,  0.);
			\coordinate (as0-8) at ( 1.,  1.73205081,  0.);
			\coordinate (as0-9) at (-1.,  1.73205081,  0.);
			\coordinate (as0-10) at ( 1., -1.73205081, 0.);
			\coordinate (as0-11) at ( 2., 0., 0.);
			\coordinate (as0-12) at (-2., 0., 0.);
			
			\draw[thick, color-orange] (s1-1) -- (s1-2);
			\draw[thick, color-orange] (s1-1) -- (s1-3);
			\draw[thick, color-orange] (s1-1) -- (s1-4);
			\draw[thick, color-orange] (s1-2) -- (s1-3);
			\draw[thick, color-orange] (s1-2) -- (s1-4);
			\draw[thick, color-orange] (s1-3) -- (s1-4);
			
			\draw[thick, color-blue] (s0-1) -- (s0-2) -- (s0-3) -- cycle;
			\draw[thick, color-blue] (s0-4) -- (s0-5) -- (s0-6) -- cycle;
			\draw[thick, color-blue] (s0-7) -- (s0-8) -- (s0-9) -- cycle;
			\draw[thick, color-blue] (s0-10) -- (s0-11) -- (s0-12) -- cycle;
			
			\draw[thick, color-blue] (s0-4) -- (s0-7) -- (s0-10) -- cycle;
			\draw[thick, color-blue] (s0-1) -- (s0-8) -- (s0-11) -- cycle;
			\draw[thick, color-blue] (s0-2) -- (s0-5) -- (s0-12) -- cycle;
			\draw[thick, color-blue] (s0-3) -- (s0-6) -- (s0-9) -- cycle;
			
			\node[vertexinternalone] (v1-0) at (s1-0) {};
			
			\node[vertexinternalzero] (v1-1) at (s1-1) {};
			\node[vertexinternalzero] (v1-2) at (s1-2) {};
			\node[vertexinternalzero] (v1-3) at (s1-3) {};
			\node[vertexinternalzero] (v1-4) at (s1-4) {};
			
			\draw[thick, -] (v1-1) -- (v1-0);
			\draw[thick, -] (v1-2) -- (v1-0);
			\draw[thick, -] (v1-4) -- (v1-0);
			\draw[thick, -] (v1-0) -- (v1-3);
		\end{tikzpicture}
		\caption{
			\(\operatorname{level}_{d}(v) = 3\)
		}
	\end{subfigure}
	\caption{
		In both pictures we see a vertex \(v \in \triangle_{d, s + 1}^{\mathbb{Z}}\) (white point) and the respective star graph \(\triangle_{d}^{*}(v)\).
		Therefore the black points are the vertices \(v - e_{i}\) for \(i \in \operatorname{support}_{d}(v)\).
		The dashed polygon shows the intersection of the unit ball with \(\triangle_{d, s}^{\mathbb{Z}}\) with respect to \(\operatorname{L}^{1}\) around \(v\) and the solid polygon shows the intersection of the unit ball with \(\triangle_{d, s}^{\mathbb{Z}}\) with respect to \(\operatorname{L}^{1}\) around the orthogonal projection \(v - \frac{1}{\operatorname{level}_{d}(v) + 1} \sum_{j \in \operatorname{support}_{d}(v)} e_{j}\) onto \(\triangle_{d, s}^{\mathbb{R}}\).
	}
	\label{fig:simplex-unit-balls}
\end{figure}

Therefore this approach seems quite promising.
Another way to illustrate this property is to consider the unit ball on \(\triangle_{d, s}^{\mathbb{R}}\) of \(v\) and its projection.
Examples for \(\operatorname{level}_{d}(v) \in \{2, 3\}\) are presented in \cref{fig:simplex-unit-balls}.
Using these gadgets \(\triangle_{d}^{*}(v)\) to connect the points in \(\triangle_{d, s}^{\mathbb{Z}}\) results in the following simplex based Steiner tree instance.

\begin{definition}
	Let \(d, s \in \mathbb{N}\).
	We define the \textbf{\(\bm{d}\)-dimensional simplex graph of size \(\bm{s}\)} to be the undirected graph \(\bm{{\operatorname{SG}}_{d, s} = (V_{d, s}, E_{d, s})}\)\label{graph:SGds} given by:
	\[
		\begin{aligned}
			V_{d, s} & := \triangle_{d, s}^{\mathbb{Z}} \cup \left\{ v \in \triangle_{d, s + 1}^{\mathbb{Z}} : \max_{i \in [d + 1]} v_{i} \leq s \right\}\\
			E_{d, s} & := \{\{v, w\} : v, w \in V_{d, s} \land \|v - w\|_{1} = 1\}
		\end{aligned}
	\]
	Furthermore, we define the \textbf{\(\bm{d}\)-dimensional simplex instance of size \(\bm{s}\)} to be the Steiner tree instance \(\bm{{\operatorname{SI}}_{d, s} = ({\operatorname{SG}}_{d, s}, R_{d, s}, c_{\operatorname{L}^{1}})}\)\label{instance:SIds} given by \(R_{d, s} := \{r_{i} : i \in [d + 1]\}\) where \(r_{i} := s \cdot e_{i} \in \triangle_{d, s}^{\mathbb{Z}}\), i.e.\ \((r_{i})_{j} = s \cdot \llbracket i = j \rrbracket\), and \(c_{\operatorname{L}^{1}}(\{v, w\}) := \|v - w\|_{1}\).
\end{definition}

For an edge \(\{v, w\} \in E_{d, s}\) by definition we have either \(v \in \triangle_{d, s}^{\mathbb{Z}} \land w \in \triangle_{d, s + 1}^{\mathbb{Z}}\) or vice versa, since points within each respective simplex always have \(\operatorname{L}^{1}\)-distance at least \(2\).
The \(d + 1\) vertices \(\{v \in \triangle_{d, s + 1}^{\mathbb{Z}} : \max_{i \in [d + 1]} v_{i} = s + 1\}\) are of no interest as they are just antennas to the required vertices \(R_{d, s}\).
\(R_{d, s}\) consists of the \(d + 1\) outermost vertices in \(\triangle_{d, s}^{\mathbb{Z}}\).
\(R_{d, s}\) is defined in view of \nameref{lp:SE} and \nameref{lp:SE-above}.
An example is shown in \cref{fig:simplex-graph-dim-2-3}.
\SGds\ is actually a composition of the star graphs \(\triangle_{d}^{*}(v)\) from our motivation at each \(v \in \triangle_{d, s + 1}^{\mathbb{Z}}\).

\subfile{figure-simplex-graph-dim-2-3}

We can easily prove the optimum solution values for \BCRimprovedclass, \HYP\ and \nameref{problem:STP}:

\begin{theorem}\label{thm:simplex-terminal-instance-optimum-solution}
	Let \(\operatorname{SI}_{d, s}\) be a simplex instance, then:
	\[\operatorname{opt}_{\operatorname{\mathcal{BCR}^{+}}}(\operatorname{SI}_{d, s}) = \operatorname{opt}_{\operatorname{\mathcal{HYP}}}(\operatorname{SI}_{d, s}) = \operatorname{opt}_{\operatorname{STP}}(\operatorname{SI}_{d, s}) = 2 \cdot s \cdot d\]
\end{theorem}
\begin{proof}
	The distance between two required vertices with respect to the edge-costs is exactly \(2 \cdot s\).
	As we have \(d + 1\) required vertices, a minimum spanning tree induces a Steiner tree of length \(2 \cdot s \cdot d\).
	This implies \(\operatorname{opt}_{\operatorname{STP}}(\operatorname{SI}_{d, s}) \leq 2 \cdot s \cdot d\).
	By \cref{thm:simplex-embedding-steiner-tree-instances} we also know that \(\operatorname{opt}_{\operatorname{\mathcal{BCR}^{+}}}(\operatorname{SI}_{d, s}) \geq 2 \cdot s \cdot d\).
	The vertices themselves induce the dual solution \(y : V_{d, s} \to \mathbb{R}^{d + 1}\) with the required properties.
	Since \((r_{i})_{i} = s\) this implies an objective value of \(2 \cdot \sum_{i \in [d + 1]} y(r_{i})_{i} - 2 \cdot s = 2 \cdot s \cdot d\).
	The claimed equalities then follow immediately with \cref{thm:steiner-tree-lp-relaxation-chain}.
\end{proof}

Note that this implies that \nameref{problem:STP} is easy on these instances, as already a minimum spanning tree induced solution suffices.

Looking back at Goemans' instance \(\operatorname{GI}_{d}\), we see that the corresponding graph is a minor of \(\operatorname{SG}_{d, 2}\) respecting the cost of the edges.
Therefore we obtain with the previous upper bound:

\begin{corollary}\label{thm:goemans-instance-series-integral-optimum}
	\(\operatorname{opt}_{\operatorname{\mathcal{BCR}^{+}}}(\operatorname{GI}_{d}) = \operatorname{opt}_{\operatorname{\mathcal{HYP}}}(\operatorname{GI}_{d}) = \operatorname{opt}_{\operatorname{STP}}(\operatorname{GI}_{d}) = 4 \cdot d\)
\end{corollary}

With \cref{thm:goemans-instance-series-integral-optimum} we obtain \(\operatorname{gap}_{\operatorname{\mathcal{BCR}}, \operatorname{\mathcal{BCR}^{+}}}(\operatorname{SI}_{d, 2}) \geq \operatorname{gap}_{\operatorname{\mathcal{BCR}}, \operatorname{\mathcal{BCR}^{+}}}(\operatorname{GI}_{d}) \geq \frac{8}{7}\) and therefore \(\operatorname{gap}_{\operatorname{\mathcal{BCR}}, \operatorname{\mathcal{BCR}^{+}}} \geq \frac{8}{7} \approx 1.142\).
We want to improve the lower bound for \(\operatorname{gap}_{\operatorname{\mathcal{BCR}}, \operatorname{\mathcal{BCR}^{+}}}\) and for that we have to determine a good upper bound on the optimum value of \SIds\ for arbitrary \(s\) regarding \BCRclass.

\pagebreak

\subsection{A lower bound on the gap of simplex based instances}\label{sec:simplex-instances-gap-lower-bound}

\begin{wraptable}{r}{0.32\textwidth}
	\centering
	\begin{tabular}{cc}
		d & \(\operatorname{gap}_{\operatorname{\mathcal{BCR}}, \operatorname{\mathcal{BCR}^{+}}}(\operatorname{SI}_{d, d})\)\\[6pt]
		1 & 1\\
		2 & 1.06666\\
		3 & 1.09459\\
		4 & 1.12116\\
		5 & 1.13939\\
		6 & 1.15042\\
		7 & 1.16094\\
		8 & 1.16883\\
		9 & 1.17340
	\end{tabular}
	\caption{}
	\label{table:simplex-terminal-instance-integrality-gaps}
\end{wraptable}

Exemplary LP-solver solutions stated in \cref{table:simplex-terminal-instance-integrality-gaps} confirm that the instances \SIds\ yield large integrality gaps (the values are all truncated after the fifth decimal place).
We beat \(\frac{8}{7} \approx 1.142\) at \(d = s = 6\) and even pass the current best lower bound of \(\frac{36}{31} \approx 1.161\) on the integrability gap of \BCRclass\ at \(d = s = 8\).
Note that the number of variables grows exponentially in the dimension \(d\).
For this reason we were only able to compute solution values up to \(d = s = 9\).

It turns out that it is rather difficult to define (optimum) solutions in general that are provably good enough to improve on the lower bounds for the respective gaps, even with the help of exemplary LP-solver solutions.
Even though the simplex based instances are quite symmetrical, optimum LP-solver solutions show that the variable values are complicated and miscellaneous in contrast to set cover based instances (\cref{sec:set-cover-based-steiner-tree-instances}).
This characteristic is understandable if one remembers that the gadgets \(\triangle_{d}^{*}(v)\) have different factors of influence on a more compact embedding in \(\triangle_{d, s}^{\mathbb{R}}\) with respect to the \(\operatorname{L}^{1}\)-norm.
At least it seems that there are quantitative patterns for optimum solutions \((u, b)\) of \nameref{lp:MBFR}:

Our gadgets \(\triangle_{d}^{*}(v)\) at vertices \(v \in \triangle_{d, s + 1}^{\mathbb{Z}}\) within a certain \(\operatorname{L}^{1}\)-distance to the required vertices do not seem to contribute to a more compact embedding.
Therefore we only consider a specific centre of \SGds\ and connect at the border directly to the required vertices while still respecting the costs.
This also avoids unnecessarily difficult notation in the end.

\begin{definition}\label{instance:SGdsdelta}
	Let \(d, s, \delta \in \mathbb{N}\) be such that \(2 \cdot \delta \leq s\).\\
	We define the \textbf{\(\bm{d}\)-dimensional simplified simplex graph of size \(\bm{s}\)} to be the undirected graph \(\bm{{\operatorname{SG}}_{d, s, \delta} = (V_{d, s, \delta} \cup R_{d, s}, E_{d, s, \delta} \cup F_{d, s, \delta})}\), given as follows:\label{graph:SGdsdelta}
	\[
		\begin{aligned}
			V_{d, s, \delta} & := \left\{ v \in \triangle_{d, s}^{\mathbb{Z}} \cup \triangle_{d, s + 1}^{\mathbb{Z}} : \max_{i \in [d + 1]} v_{i} \leq s - \delta \right\}\\
			E_{d, s, \delta} & := \left\{ \{v, w\} : v, w \in V_{d, s, \delta} \land \|v - w\|_1 = 1 \right\}\\
			F_{d, s, \delta} & := \left\{ \{r_{i}, v\} : r_{i} \in R_{d, s, \delta}, v \in \triangle_{d, s}^{\mathbb{Z}} \cap V_{d, s, \delta} \land v_{i} = s - \delta \right\}
		\end{aligned}
	\]
	We define the \textbf{\(\bm{d}\)-dimensional simplified simplex instance of size \(\bm{s}\)} to be the Steiner tree instance \(\bm{{\operatorname{SI}}_{d, s, \delta} = ({\operatorname{SG}}_{d, s, \delta}, R_{d, s}, c_{{\operatorname{L}}^{1}})}\).\label{instance:SIdsdelta}
\end{definition}

By construction \(\operatorname{SI}_{d, s, 0} = \operatorname{SI}_{d, s}\), however, we will restrict ourselves to \(\delta \geq 1\) such that \(V_{d, s, \delta}\) and \(R_{d, s}\) are disjoint.
It is easy to see that \(\operatorname{opt}_{\operatorname{\mathcal{BCR}}}(\operatorname{SI_{d, s}}) \leq \operatorname{opt}_{\operatorname{\mathcal{BCR}}}(\operatorname{SI_{d, s, \delta}})\) and \(\operatorname{opt}_{\operatorname{\mathcal{BCR}^{+}}}(\operatorname{SI_{d, s}}) = \operatorname{opt}_{\operatorname{\mathcal{BCR}^{+}}}(\operatorname{SI_{d, s, \delta}})\).
We have the following vertex degree structure in \SIdsdelta:
\[
	\begin{aligned}
		|\delta_{SG_{d, s, \delta}}(
		%\textcolor{green!40!black}
		{r_{i} \in R_{d, s, \delta}})| & = \left| \left\{ v \in \triangle_{d, s}^{\mathbb{Z}} : v_{i} = s - \delta \right\} \right| = \left| \triangle_{d - 1, \delta}^{\mathbb{Z}} \right|\\
		|\delta_{SG_{d, s}}(
		%\textcolor{blue}
		{v \in \triangle_{d, s}^{\mathbb{Z}} \cap V_{d, s, \delta}})| & = \underbrace{(d + 1) - |\operatorname{support_{d}}(v)|}_{\substack{\text{neighbour } 
				%\textcolor{orange}
				{w \in \triangle_{d, s + 1}^{\mathbb{Z}}}\\\operatorname{level_{d}}(w) = \operatorname{level_{d}}(v) + 1}} \ \, + \underbrace{|\operatorname{support_{d}}(v)|}_{\substack{\text{neighbour } 
				%\textcolor{orange}
				{w \in \triangle_{d, s + 1}^{\mathbb{Z}}}\\\operatorname{level_{d}}(w) = \operatorname{level_{d}}(v)\\\text{or neighbour } 
				%\textcolor{green!40!black}
				{r_{i} \in R_{d, s, \delta}}}} = d + 1\\
		|\delta_{SG_{d, s}}(
		%\textcolor{orange}
		{v \in \triangle_{d, s + 1}^{\mathbb{Z}} \cap V_{d, s, \delta}})| & = \underbrace{|\operatorname{support_{d}}(v)|}_{\substack{\text{neighbour } 
				%\textcolor{blue}
				{w \in \triangle_{d, s}^{\mathbb{Z}}}}} = \operatorname{level_{d}}(v) + 1
	\end{aligned}
\]
%An example is given in \cref{fig:simplified-simplex-graph-dim-2}\todo{undefined}.
%\subfile{figures/figure-simplified-simplex-graph-dim-2}

With our notion of level we have already partitioned the vertices of \SGds.
This can also be done for the edges by defining \(\operatorname{level}_{d}(\{v, w\}) := \max \{\operatorname{level}_{d}(v), \operatorname{level}_{d}(w)\}\).
Moreover, if we divide the vertex balances \(b\) into \(b_{i}\) for vertices in \(\triangle_{d, s + i}^{\mathbb{Z}}\) for \(i \in \{0, 1\}\), then, apart from vertices and edges within a certain \(\operatorname{L}^{1}\)-distance to the required vertices, it seems that the edge usages \(u\) and vertex balances \(b_{i}\) are homogeneous on their respective level \(l\)  and connected in the following way:
\[
	\begin{aligned}
		b_{0}(l) & = + (l - 1) \cdot u(l) + (d - l) \cdot u(l + 1)\\
		b_{1}(l) & = - (l - 1) \cdot u(l)
	\end{aligned}
\]

Restricting ourselves to the above solution structure, we were able to define and prove solutions on \SIdsdelta\ for certain \(\delta\) but only in the case \(u(l) = 0\) for \(l \geq 3\).
Hence, we are only using a small subset of the edges, namely edges on level \(1\) and \(2\), even though LP-solver solutions suggest that the remaining edges generally contribute to the optimum solution as well (see \cref{table:simplex-instances-integrality-gaps-comparison} in \cref{sec:conclusion}).
It is still an open question whether this observation above is in general the truth.
Although the partitioning into \(b_{i}(l)\) and \(u(l)\) seems quite promising, we think the graph on which this holds is much more complicated than \(\operatorname{SG}_{d, s, \delta}\).
We think this is mostly due to the fact that the unit balls within \(\triangle_{l, s}^{\mathbb{R}}\) are much more complicated in higher dimensions \(l\).
It is likely that a different approach is needed to find even better bounds on the optimum value for this class of instances.

%\pagebreak

Before we prove an upper bound in \cref{thm:existence-solution-simplified-simplex-instance} on the optimum solution value of \BCRclass\ for \SIds, in particular \SIdsdelta, we will first propose our underlying solution of \nameref{lp:MBFR} only using edges on level \(1\) and \(2\) in \cref{thm:existence-bidirected-balance-flow-simplified-simplex-graph}.
For that, we will need two lemmas, which show how to handle the flow on specific subgraphs.
\cref{thm:flow-level-1} will be needed for distributing flow along edges on level \(1\) and \cref{thm:flow-level-2} will be needed for distributing flow along edges on level \(2\).
For illustrations see \cref{fig:simplified-simplex-terminal-instance-flow-subdim-2}, in particular \cref{fig:simplified-simplex-terminal-instance-flow-subdim-2-k-not-in-support}.

\begin{lemma}\label{thm:flow-level-1}
	Let \(P\) be an undirected path of length \(p\), i.e.\ \(V(P) = \{v_{1}, \dotsc, v_{p + 1}\}\) and \(E(P) = \{\{v_{i}, v_{i + 1}\} : i \in [p]\}\).
	Let \(\gamma \in \mathbb{R}_{\geq 0}\) and \(b : V(P) \to \mathbb{R}\) be given by
	\[
		b(v_{i}) = \begin{cases}
			- (p - 1) \cdot \gamma & \text{if } i \in \{1, p + 1\}\\
			2 \cdot \gamma & \text{otherwise}
		\end{cases}
	\]
	Then there exists a bidirected balance-flow \(f : E_{\leftrightarrow}(P) \to \mathbb{R}_{\geq 0}\) on \(P\) with respect to \(b\) such that for all \(\{v, w\} \in E(P)\) we have \(f(v, w) + f(w, v) \leq (p - 1) \cdot \gamma\).
\end{lemma}

\begin{proof}
	Note that \(b(V(P)) = 2 \cdot (- (p - 1) \cdot \gamma) + (p - 1) \cdot (2 \cdot \gamma) = 0\).
	
	We define \(f : E_{\leftrightarrow}(P) \to \mathbb{R}_{\geq 0}\) by
	\[
		\begin{aligned}
			f(v_{i}, v_{i + 1}) & = (i - 1) \cdot \gamma & \quad i & \in [p]\\
			f(v_{i + 1}, v_{i}) & = (p - i) \cdot \gamma & \quad i & \in [p]
		\end{aligned}
	\]
	We immediately see that \(f(v_{i}, v_{i + 1}) + f(v_{i + 1}, v_{i}) \leq (p - 1) \cdot \gamma\).\\
	Hence, we only have to check the flow-balance conditions:
	\[
		\begin{aligned}
			f(\delta_{P}^{+}(v_{1})) - f(\delta_{P}^{-}(v_{1})) & = f(v_{1}, v_{2}) - f(v_{2}, v_{1}) = - (p - 1) \cdot \gamma\\
			f(\delta_{P}^{+}(v_{p + 1})) - f(\delta_{P}^{-}(v_{p + 1})) & = f(v_{p + 1}, v_{p}) - f(v_{p}, v_{p + 1}) = - (p - 1) \cdot \gamma
		\end{aligned}
	\]
	For \(i \in [p + 1] \setminus \{1, p + 1\}\) we have:
	%\[
		\begin{align*}
			f(\delta_{P}^{+}(v_{i})) - f(\delta_{P}^{-}(v_{i}))
			& = f(v_{i}, v_{i + 1}) + f(v_{i}, v_{i - 1}) - f(v_{i + 1}, v_{i}) - f(v_{i - 1}, v_{i})\\
			& = (i - 1) \cdot \gamma + (p - (i - 1)) \cdot \gamma - (p - i) \cdot \gamma - ((i - 1) - 1) \cdot \gamma\\
			& = 2 \cdot \gamma \qedhere
		\end{align*}
	%\]
\end{proof}

\begin{lemma}\label{thm:flow-level-2}
	Let \(s, \delta \in \mathbb{N}\) be such that \(s = 3 \cdot \delta - 2\).
	
	Let \(G = (V, E)\) be given by
	\[
		\begin{aligned}
			V & := \left\{ v \in \triangle_{2, s}^{\mathbb{Z}} : \forall i \in [3] \ \ 0 \leq v_{i} \leq s - \delta \right\} \cup \left\{ v \in \triangle_{2, s + 1}^{\mathbb{Z}} : \forall i \in [3] \ \ 1 \leq v_{i} \leq s - \delta + 1 \right\}\\
			E & := \left\{ \{v, w\} : v, w \in V \land \|v - w\|_1 = 1 \right\}
		\end{aligned}
	\]
	Let \(\gamma \in \mathbb{R}_{\geq 0}\) and \(b : V \to \mathbb{R}\) be given by
	\[
		b(v) = \begin{cases}
			+ \gamma & \text{if } v \in \triangle_{2, s}^{\mathbb{Z}} \cap V\\
			- \gamma & \text{if } v \in \triangle_{2, s + 1}^{\mathbb{Z}} \cap V
		\end{cases}
	\]
	Then there exists a bidirected balance-flow \(f : E_{\leftrightarrow} \to \mathbb{R}_{\geq 0}\) on \(G\) with respect to \(b\) such that for all \(\{v, w\} \in E\) we have \(f(v, w) + f(w, v) \leq \gamma\).
\end{lemma}

\begin{proof}
	To obtain a bidirected balance-flow we will use a perfect matching in \(G\) between the vertices in \(\triangle_{2, s}^{\mathbb{Z}} \cap V\) and \(\triangle_{2, s + 1}^{\mathbb{Z}} \cap V\), motivated by the definition of \(b\).
	
	We define for \(i \in [3]\) the following vertex-sets:
	\[
		\begin{aligned}
			A_{i} & := \left\{ v \in V \ : \ v_{i} \leq \frac{s + 2}{3} - 1 \ \land \ \min_{j \in [3] \setminus \{i\}} v_{j} \geq \frac{s + 2}{3} \right\}\\
			B_{i} & := \left\{ v \in V \ : \ v_{i} \geq \frac{s + 2}{3} \ \land \ \max_{j \in [3] \setminus \{i\}} v_{j} \leq \frac{s + 2}{3} - 1 \right\}
		\end{aligned}
	\]
	The sets \(A_{i}\) and \(B_{i}\) form a partition of \(V\):
	Since \(s \leq \sum_{i \in [3]} v_{i} \leq s + 1\), there exist \(i_{1}, i_{2} \in [3]\) with \(i_{1} \neq i_{2}\) such that \(v_{i_{1}} \leq \frac{s + 2}{3} - 1\) and \(v_{i_{2}} \geq \frac{s + 2}{3}\).
	Therefore, either \(v \in A_{i_{1}}\) or \(v \in B_{i_{2}}\).
	We will define the perfect matching within each set \(A_{i}\) and \(B_{i}\).
	Note that \(\frac{s + 2}{3} = \delta \geq 1\) as \(s = 3 \cdot \delta - 2\).
	
	If for \(v \in \triangle_{2, s}^{\mathbb{Z}} \cap V\) there exists \(i \in [3]\) such that \(v_{i} = 0\), then \(v \in A_{i}\) as otherwise \(\max_{j \in [3]} v_{j} \geq s - (0 + (\frac{s + 2}{3} - 1)) = s - \delta + 1 > s - \delta\).
	If for \(v \in \triangle_{2, s + 1}^{\mathbb{Z}} \cap V\) there exists \(i \in [3]\) such that \(v_{i} = s - \delta + 1\), then \(v \in B_{i}\) as otherwise \(\min_{j \in [3]} v_{j} \leq (s + 1) - ((s - \delta + 1) + \frac{s + 2}{3}) = 0\).
	
	For \(v \in \triangle_{2, s}^{\mathbb{Z}} \cap V\) we have \(|\{i \in [3] : v_{i} = 0\}| \in \{0, 1\}\) as otherwise we obtain the contradiction \(\sum_{i \in [3]} v_{i} \leq s - \delta < s\) as \(\delta \geq 1\).
	For \(v \in \triangle_{2, s + 1}^{\mathbb{Z}} \cap V\) we have \(|\{i \in [3] : v_{i} = s - \delta + 1\}| \in \{0, 1\}\) as otherwise we obtain the contradiction \(\sum_{i \in [3]} v_{i} \geq 2 \cdot (s - \delta + 1) + 1 = s + 1 + \delta > s + 1\) as \(\delta \geq 1\).
	
	Therefore, if \(v \in A_{i} \cap \triangle_{2, s}^{\mathbb{Z}}\), then \(v + e_{i} \in A_{i} \cap \triangle_{2, s + 1}^{\mathbb{Z}}\) as otherwise \(\max_{j \in [3]} v_{j} \leq \frac{s + 2}{3} - 1\) contradicting \(\sum_{j \in [3]} v_{j} = s\), and if \(v \in B_{i} \cap \triangle_{2, s + 1}^{\mathbb{Z}}\), then \(v - e_{i} \in B_{i} \cap \triangle_{2, s}^{\mathbb{Z}}\) as otherwise \(\min_{j \in [3]} v_{j} \geq \frac{s + 2}{3}\) contradicting \(\sum_{j \in [3]} v_{j} = s + 1\).
	On the other hand, if \(v - e_{i} \in B_{i} \cap \triangle_{2, s}^{\mathbb{Z}}\), then \(v \in B_{i} \cap \triangle_{2, s + 1}^{\mathbb{Z}}\), and if \(v \in A_{i} \cap \triangle_{2, s + 1}^{\mathbb{Z}}\), then \(v - e_{i} \in A_{i} \cap \triangle_{2, s}^{\mathbb{Z}}\).
	
	Finally, we obtain for \(i \in [3]\) and \(v \in \triangle_{2, s}^{\mathbb{Z}} \cap V\) that \(v + e_{i} \in A_{i} \Leftrightarrow v \in A_{i}\) and \(v + e_{i} \in B_{i} \Leftrightarrow v \in B_{i}\)
	and for \(v \in \triangle_{2, s + 1}^{\mathbb{Z}} \cap V\) we have \(v - e_{i} \in A_{i} \Leftrightarrow v \in A_{i}\) and \(v - e_{i} \in B_{i} \Leftrightarrow v \in B_{i}\).
	
	This induces a perfect matching in \(G\) between the vertices in \(\triangle_{2, s}^{\mathbb{Z}} \cap V\) and \(\triangle_{2, s + 1}^{\mathbb{Z}} \cap V\) within each set \(A_{i}\) and \(B_{i}\).
	We define for \(v \in \triangle_{2, s + 1}^{\mathbb{Z}} \cap V\) and \(i \in [3]\):
	\[f(v - e_{i}, v) := \gamma \cdot \llbracket v \in A_{i} \lor v \in B_{i} \rrbracket\]
	\(f\) respects the edge-usage bounds and is a bidirected balance-flow with respect to \(b\).
\end{proof}

\pagebreak

\subfile{figure-simplified-simplex-terminal-instance-flow-subdim-2}

We are now able to propose our underlying solution of \nameref{lp:MBFR} for \SIdsdelta.
For that we will consider a slightly different Graph \(\operatorname{SG}'_{d, s, \delta}\), in which we split up the required vertices for easier notation.
In \cref{thm:existence-solution-simplified-simplex-instance} we will see that \cref{thm:existence-bidirected-balance-flow-simplified-simplex-graph} implies a solution of \nameref{lp:MBFR} for \SIdsdelta, by contracting these auxiliary vertices in \(\operatorname{SG}'_{d, s, \delta}\) back to the respective required vertices.

\begin{lemma}\label{thm:existence-bidirected-balance-flow-simplified-simplex-graph}
	Let \(d \in \mathbb{N}\) and \(k \in [d + 1]\).
	Let \(s, \delta \in \mathbb{N}\) be such that \(2 \cdot \delta \leq s\).\\
	Let \(R_{i} := \left\{ v \in \triangle_{d, s + 1}^{\mathbb{Z}} : v_{i} = s - \delta + 1 \right\}\) be auxiliary vertices for each required vertex \(r_{i} \in R_{d, s}\) (\(R_{i}\) are disjoint as \(2 \cdot \delta \leq s\)).
	Let \(\operatorname{SG}'_{d, s, \delta} := (V_{d, s, \delta}', E_{d, s, \delta}')\) be given by
	\[
		\begin{aligned}
			V_{d, s, \delta}' & := V_{d, s, \delta} \cup \bigcup_{i \in [d + 1]} R_{i}\\
			E_{d, s, \delta}' & := \left\{ \{v, w\} : v, w \in V_{d, s, \delta}' \land \|v - w\|_1 = 1 \right\}
		\end{aligned}
	\]
	
	Let \(u' : [d] \to \mathbb{R}_{\geq 0}\) be given by
	\[
		u'(l) = \frac{1}{\binom{d + 1}{2} \cdot (2 \cdot s - 3 \cdot \delta + 1)} \cdot \begin{cases}
			\hfil (d + 1) \cdot s - d \cdot (3 \cdot \delta - 2) - 1 & \text{if } l = 1\\
			\hfil 3 & \text{if } l = 2\\
			\hfil 0 & \text{if } l \geq 3\\
		\end{cases}
	\]
	
	Let \(b_{k} : V_{d, s, \delta}' \to \mathbb{R}\) be given as follows.
	Let \(v \in V_{d, s, \delta}'\) and \(l := \operatorname{level}_{d}(v)\).\\
	If \(v \in V_{d, s, \delta}\), i.e.\ \(\max_{i \in [d + 1]} v_{i} \leq s - \delta\):
	\[
		b_{k}(v) = \begin{cases}
			+ (l - 1) \cdot u'(l) + (d - l) \cdot u'(l + 1) & \text{if } v \in \triangle_{d, s}^{\mathbb{Z}}\\
			- (l - 1) \cdot u'(l) & \text{if } v \in \triangle_{d, s + 1}^{\mathbb{Z}}
		\end{cases}
	\]
	If \(v \in R_{i}\):
	\[b_{k}(v) = (\llbracket i = k \rrbracket - \llbracket i \neq k \rrbracket) \cdot u'(l)\]
	
	If \(d = 2\) or \(d \geq 3 \land s = 3 \cdot \delta - 2\), then there exists for \(\operatorname{SG}'_{d, s, \delta}\) a bidirected balance flow \(f_{k} : E_{\leftrightarrow}(\operatorname{SG}'_{d, s, \delta}) \to \mathbb{R}_{\geq 0}\)  with respect to \(b_{k}\) and \(u : E_{d, s, \delta}' \to \mathbb{R}_{\geq 0}\) defined by \(u(e) = u'(\operatorname{level}_{d}(e))\).
\end{lemma}

\begin{proof}
	Note that we only have to deal with vertices on level \(1\) and \(2\), as \(b_{k}(v) = 0\) for \(v \in V_{d, s, \delta}'\) with \(\operatorname{level}_{d}(v) \geq 3\).
	Also the edges on levels higher than \(2\) have value \(0\) under \(f_{k}\) induced by \(u\).
	Moreover, edges that we do not specify \(f_{k}\) for in the following have value \(0\) under \(f_{k}\) as well.
	
	We will first define \(f_{k}\) for edges incident to vertices which contain \(k\) in their support.
	After this we proceed with the definition of \(f_{k}\) for edges between vertices without \(k\) in their support.
	
	So we now consider the edges incident to the edge-level representatives \(v \in \triangle_{d, s + 1}^{\mathbb{Z}} \cap V_{d, s, \delta}'\) with \(k \in \operatorname{support}_{d}(v)\).
	For \(i \in \operatorname{support}_{d}(v)\) with \(v - e_{i} \in \triangle_{d, s}^{\mathbb{Z}} \cap V_{d, s, \delta}'\) we define:
	\[
		\begin{aligned}
			f_{k}(v - e_{i}, v) & := u'(\operatorname{level}_{d}(v)) \cdot \llbracket i \neq k \rrbracket\\
			f_{k}(v, v - e_{i}) & := u'(\operatorname{level}_{d}(v)) \cdot \llbracket i = k \rrbracket
		\end{aligned}
	\]
	For an illustration see \cref{fig:simplified-simplex-terminal-instance-flow-subdim-2-k-in-support}.
	
	We easily see that \(f_{k}(v - e_{i}, v) + f_{k}(v, v - e_{i}) \leq u'(\operatorname{level}_{d}(v)) = u(\{v, v - e_{i}\})\).
	
	Furthermore, the balance conditions hold:
	Let \(v \in V_{d, s, \delta}'\) be such that \(k \in \operatorname{support}_{d}(v)\).
	It is clear for \(v \in R_{i}\) as \(\{v, v - e_{i}\} \in E_{d, s, \delta}'\) is the only incident edge and \(b_{k}(v) = (\llbracket i = k \rrbracket - \llbracket i \neq k \rrbracket) \cdot u'(\operatorname{level}_{d}(v))\).
	So assume \(v \in V_{d, s, \delta}\), i.e.\ \(\max_{i \in [d + 1]} v_{i} \leq s - \delta\):
	
	\vspace{12pt}
	
	\begin{minipage}[t]{0.2\textwidth}
		If \(v \in \triangle_{d, s + 1}^{\mathbb{Z}}\):
	\end{minipage}
	\begin{minipage}[t]{0.8\textwidth}
		\vspace{-10pt}
		\[
			\begin{aligned}
				& f_{k}(\delta_{\operatorname{SG}'_{d, s, \delta}}^{+}(v)) - f_{k}(\delta_{\operatorname{SG}'_{d, s, \delta}}^{-}(v))\\
				& = \sum_{i \in \operatorname{support}_{d}(v)} ( \ f_{k}(v, v - e_{i}) - f_{k}(v - e_{i}, v) \ )\\
				& = u'(\operatorname{level}_{d}(v)) - (|\operatorname{support}_{d}(v)| - 1) \cdot u'(\operatorname{level}_{d}(v))\\
				& = - (\operatorname{level}_{d}(v) - 1) \cdot u'(\operatorname{level}_{d}(v)) = b_{k}(v)
			\end{aligned}
		\]
	\end{minipage}

	\vspace{12pt}
	
	\begin{minipage}[t]{0.2\textwidth}
		If \(v \in \triangle_{d, s}^{\mathbb{Z}}\):
	\end{minipage}
	\begin{minipage}[t]{0.8\textwidth}
	\vspace{-22pt}
		\[
			\begin{aligned}
				& f_{k}(\delta_{\operatorname{SG}'_{d, s, \delta}}^{+}(v)) - f_{k}(\delta_{\operatorname{SG}'_{d, s, \delta}}^{-}(v))\\
				& = \sum_{i \in [d + 1]} ( \ f_{k}(v, v + e_{i}) - f_{k}(v + e_{i}, v) \ )\\
				& = - u'(\operatorname{level}_{d}(v + e_{k})) + \sum_{i \in [d + 1] \setminus \{k\}} u'(\operatorname{level}_{d}(v + e_{i}))\\
				& = - u'(\operatorname{level}_{d}(v)) + (|\operatorname{support}_{d}(v)| - 1) \cdot u'(\operatorname{level}_{d}(v))\\
				& \qquad + (d + 1 - |\operatorname{support}_{d}(v)|) \cdot u'(\operatorname{level}_{d}(v) + 1)\\
				& = (\operatorname{level}_{d}(v) - 1) \cdot u'(\operatorname{level}_{d}(v)) + (d - \operatorname{level}_{d}(v)) \cdot u'(\operatorname{level}_{d}(v) + 1) = b_{k}(v)
			\end{aligned}
		\]
	\end{minipage}

	\vspace{6pt}
	
	This proves all required properties for vertices which contain \(k\) in their support.
	
	Now we deal with edges between vertices without \(k\) in their support.
	From the construction of \(f_{k}\) so far, we only have incoming flow at vertices \(v \in \triangle_{d, s}^{\mathbb{Z}} \cap V_{d, s, \delta}\) with \(v_{k} = 0\) and it is of value \(u'(\operatorname{level}_{d}(v) + 1)\) as \(\operatorname{level}_{d}(v + e_{k}) = \operatorname{level}_{d}(v) + 1\) and \(f_{k}(v + e_{k}, v) = u'(\operatorname{level}_{d}(v) + 1)\).
	
	Note that we actually only have positive flow on edges on level \(1\) and \(2\) and therefore only vertices in \(\triangle_{d, s}^{\mathbb{Z}} \cap V_{d, s, \delta}\) on level \(1\) have incoming flow and it is of value \(u'(2)\).
	For an illustration see again \cref{fig:simplified-simplex-terminal-instance-flow-subdim-2-k-in-support}.
	
	Let \(v \in \triangle_{d, s}^{\mathbb{Z}} \cap V_{d, s, \delta}\) be such that \(v_{k} = 0\) and \(\operatorname{level}_{d}(v) = 1\).
	As \(b_{k}(v) = (d - 1) \cdot u'(2)\) we have to accommodate \(d \cdot u'(2)\) units of flow.
	Further note that for \(v' \in \triangle_{d, s + 1}^{\mathbb{Z}} \cap V_{d, s, \delta}'\) with \(\operatorname{level}_{d}(v') = 1\) we have
	\[
	b_{k}(v') = \begin{cases}
	0 & \mbox{if } \max_{i \in [d + 1]} v'_{i} \leq s - \delta\\
	- u'(1) & \mbox{if } \max_{i \in [d + 1]} v'_{i} = s - \delta + 1
	\end{cases}
	\]
	This allows us to use \cref{thm:flow-level-1} with \(\gamma = u'(2)\) and \(p = s - 2 \cdot \delta + 2\) to move \(2 \cdot u'(2)\) units of outgoing flow at \(v\) to the two vertices \(v^{i} \in R_{i}\) with \(i \in \operatorname{support}_{d}(v)\), i.e.\ \((v^{i})_{i} = s - \delta + 1\), along the \(1\)-dimensional subgraph with \(2 \cdot (s - 2 \cdot \delta) + 2\) edges, which can be interpreted as a path of length \(s - 2 \cdot \delta + 2\) since \(b_{k}(v') = 0\) for \(v' \in \triangle_{d, s + 1}^{\mathbb{Z}} \cap V_{d, s, \delta}\) with \(\operatorname{level}_{d}(v') = 1\).
	To see that this is possible we need the case distinction \(d = 2\) and \(d \geq 3\).
	
	First consider \(d = 2\).
	The values of \(u'\) simplify to:
	\[
		u'(l) = \frac{1}{2 \cdot s - 3 \cdot \delta + 1} \cdot \begin{cases}
			\hfil s - 2 \cdot \delta + 1 & \text{if } l = 1\\
			\hfil 1 & \text{if } l = 2
		\end{cases}
	\]
%	\[
%		\begin{aligned}
%			u'(1) & = \frac{(d + 1) \cdot s - d \cdot (3 \cdot \delta - 2) - 1}{\binom{d + 1}{2} \cdot (2 \cdot s - 3 \cdot \delta + 1)} = \frac{s - 2 \cdot \delta + 1}{2 \cdot s - 3 \cdot \delta + 1}\\
%			u'(2) & = \frac{3}{\binom{d + 1}{2} \cdot (2 \cdot s - 3 \cdot \delta + 1)} = \frac{1}{2 \cdot s - 3 \cdot \delta + 1}
%		\end{aligned}
%	\]
	Then \(b_{k}(v^{i}) = - u'(1) = - (s - 2 \cdot \delta + 1) \cdot u'(2) = - (p - 1) \cdot \gamma\) and \(b_{k}(v) = 2 \cdot \gamma\).
	Hence we can apply \cref{thm:flow-level-1} with the chosen \(\gamma\) and \(p\).
	Note that this concludes the proof for \(d = 2\).
	
	Now consider \(d \geq 3\).
	We have the restriction \(s = 3 \cdot \delta - 2\).
%	\footnote{
%		We were able to define flows for \(s = 3 \cdot \delta - \alpha\) with \(\alpha \in \{0, 1, 2\}\), so \(s\) could be arbitrary.
%		But the cases \(\alpha \in \{0, 1\}\) are more complicated than \(\alpha = 2\) and, as in the end we are only interested in the asymptotic behaviour in \(s\), it is enough to deal with a specific \(\alpha\).
%	}
	The values of \(u'\) simplify to:
	\[
	u'(l) = \frac{1}{\binom{d + 1}{2}} \cdot \begin{cases}
	1 & \mbox{if } l = 1\\
	\frac{3}{s - 1} & \mbox{if } l = 2\\
	0 & \mbox{if } l \geq 3
	\end{cases}
	\]
	Then \(b_{k}(v^{i}) = - u'(1) = - \frac{s - 1}{3} \cdot u'(2) = - (\delta - 1) \cdot u'(2) = - (s - 2 \cdot \delta + 1) \cdot u'(2) = - (p - 1) \cdot \gamma\) and \(b_{k}(v) = 2 \cdot \gamma\).
	Hence we can also apply \cref{thm:flow-level-1} with the chosen \(\gamma\) and \(p\).
	
	Now, all flow balance conditions on level \(1\) are fulfilled, except the \((d - 2) \cdot u'(2)\) units of flow left at the vertices \(v \in \triangle_{d, s}^{\mathbb{Z}} \cap V_{d, s, \delta}\) on level \(1\).
	Moreover, since \(d \geq 3\), we still have to deal with vertices on level \(2\) without \(k\) in their support.
	
	For \(v \in \triangle_{d, s}^{\mathbb{Z}} \cap V_{d, s, \delta}\) on level \(1\) with \(v_{k} = 0\) we have \(d - 2\) different supports \(I \subseteq [d + 1]\) of size \(3\) such that \(\operatorname{support}_{d}(v) \subset I\) and \(k \notin I\).
	On the other hand, for all supports \(I \subseteq [d + 1]\) of size \(3\) and \(k \notin I\) we still need to handle the vertex-balances of the respective vertices, which are \(u'(2)\) for vertices in \(v \in \triangle_{d, s}^{\mathbb{Z}} \cap V_{d, s, \delta}'\) and \(- u'(2)\) for vertices in \(v \in \triangle_{d, s + 1}^{\mathbb{Z}} \cap V_{d, s, \delta}'\).
	
	We will distribute the flow using \cref{thm:flow-level-2} with \(\gamma = u'(2)\).
	For an illustration see \cref{fig:simplified-simplex-terminal-instance-flow-subdim-2-k-not-in-support}.
	
	More precisely, for given \(I \subseteq [d + 1]\) such that \(|I| = 3\) and \(k \notin I\) we need to define the flow for the edges incident to the vertices \(v \in \triangle_{d, s + 1}^{\mathbb{Z}}\) with \(\operatorname{support}_{d}(v) = I\).
	The subgraph of \(\operatorname{SG}'_{d, s, \delta}\) induced by these edges is isomorphic to the one considered in \cref{thm:flow-level-2}.
	With the outgoing flow at the vertices \(v \in \triangle_{d, s}^{\mathbb{Z}} \cap V_{d, s, \delta}'\) on level \(1\), we obtain the following balance values:
	\[
		\tilde{b}(v) = \begin{cases}
			+ u'(2) & \text{if } v \in \triangle_{d, s}^{\mathbb{Z}} \cap V_{d, s, \delta}' \land \operatorname{support}_{d}(v) \subseteq I\\
			- u'(2) & \text{if } v \in \triangle_{d, s + 1}^{\mathbb{Z}} \cap V_{d, s, \delta}' \land \operatorname{support}_{d}(v) = I
		\end{cases}
	\]
	Hence with our chosen \(\gamma\) we obtain with \cref{thm:flow-level-2} a bidirected balance-flow on each subgraph induced by an \(I \subseteq [d + 1]\) such that \(|I| = 3\) and \(k \notin I\) as described above.
	
	In total we have accumulated a bidirected balance-flow \(f_{k}\) on \(\operatorname{SG}'_{d, s, \delta}\) for \(u\) and \(b_{k}\) and this concludes the proof.
\end{proof}

%\pagebreak

We will now provide combinatorial formulas, which help for counting certain vertices and edges in our simplified graph, to be able to compute the cost of our solution in \cref{thm:existence-bidirected-balance-flow-simplified-simplex-graph}.
An important formula is the number of points in a \(d\)-dimensional discrete simplex of size \(s\), i.e.\ \(|\triangle_{d, s}^{\mathbb{Z}}|\), as all following formulas are built on this information.
The numbers \(|\triangle_{d, s}^{\mathbb{Z}}|\) are also known as \textit{simplicial polytopic numbers}, but to the best of our knowledge there is no proof that fits our context for counting the points in \(\triangle_{d, s}^{\mathbb{Z}}\).
However, it is easy to prove:

\begin{lemma}
	\[|\triangle_{d, s}^{\mathbb{Z}}| = \binom{d + s}{d}\]
\end{lemma}
\begin{proof}
	We will prove this equation by constructing a bijection \(\varphi : \triangle_{d, s}^{\mathbb{Z}} \leftrightarrow \{A \subseteq [d + s] : |A| = d\}\).
	Since \(|\{A \subseteq [d + s] : |A| = d\}| = \binom{d + s}{d}\) this is sufficient.
	\(\varphi\) is defined by:
	\[\varphi(x) := \left\{\sum_{i = 1}^{k} x_{i} + k \ : \ k \in [d]\right\}\]
	The inverse \(\varphi^{-1}\) can be defined as follows:
	\[\varphi^{-1}(\{1 \leq a_1 < \dotsc < a_d \leq d + s\}) := \left(x_i = \begin{cases}
	a_{i} - 1 & \mbox{if } i = 1\\
	a_{i} - a_{i - 1} - 1 & \mbox{if } i \in [d] \setminus \{1\}\\
	s + d - a_{i - 1} & \mbox{if } i = d + 1
	\end{cases}\right)_{i \in [d + 1]}\]
	With \(\varphi^{-1}\) it can be easily verified that \(\varphi\) is indeed a bijection.
	Note that for \(x \in \triangle_{d, s}^{\mathbb{Z}}\) \(x_{d + 1}\) is already determined by \(x_{1}, \dotsc, x_{d}\).
\end{proof}

Often the binomial coefficient \(\binom{n}{k}\) is only defined for \(0 \leq k \leq n\).
A convenient convention allows arbitrary values by defining \(\binom{n}{k}\) to be \(0\) if \(k\) does not fulfill the above condition.
We will make use of this.

\begin{lemma}\label{thm:simplex-points-in-radius}
	Let \(d, s, k \in \mathbb{N}\) be such that \(2 \cdot k + 1 \geq s\).
	\[\left| \left\{ v \in \triangle_{d, s}^{\mathbb{Z}} : \max_{i \in [d + 1]} v_{i} \leq k \right\} \right| = \binom{d + s}{d} - (d + 1) \cdot \binom{d + s - (k + 1)}{d}\]
\end{lemma}
\begin{proof}
	Let \(v \in \triangle_{d, s}^{\mathbb{Z}}\).
	Then \(|\{i \in [d + 1] : v_{i} \geq k + 1\}| \in \{0, 1\}\) since \(2 \cdot k + 1 \geq s\) and \(\sum_{i \in [d + 1]} v_{i} = s\).
	With that in mind we can prove the claimed equality:
	\begin{align*}
		\left| \left\{ v \in \triangle_{d, s}^{\mathbb{Z}} : \max_{i \in [d + 1]} v_{i} \leq k \right\} \right|
		& = \left| \triangle_{d, s}^{\mathbb{Z}} \right| - \left| \left\{ v \in \triangle_{d, s}^{\mathbb{Z}} : \exists i \in [d + 1] \quad v_{i} \geq k + 1 \right\} \right|\\
		& = \left| \triangle_{d, s}^{\mathbb{Z}} \right| - \sum_{i \in [d + 1]} \left| \left\{ v \in \triangle_{d, s}^{\mathbb{Z}} : v_{i} \geq k + 1 \right\} \right|\\
		& = \left| \triangle_{d, s}^{\mathbb{Z}} \right| - \sum_{i \in [d + 1]} \left| \triangle_{d, s - (k + 1)}^{\mathbb{Z}} \right|\\
		& = \binom{d + s}{d} - (d + 1) \cdot \binom{d + s - (k + 1)}{d} \qedhere
	\end{align*}
\end{proof}

%\pagebreak

\begin{lemma}\label{thm:simplex-points-in-radius-on-level}
	Let \(d, s, k \in \mathbb{N}\) be such that \(2 \cdot k + 1 \geq s\) and \(l \in \mathbb{N}_{0}\).
	\[\left| \left\{ v \in \triangle_{d, s}^{\mathbb{Z}} : \max_{i \in [d + 1]} v_{i} \leq k \land \operatorname{level}_{d}(v) = l \right\} \right| = \binom{d + 1}{l + 1} \cdot \left( \binom{s - 1}{l} - (l + 1) \cdot \binom{s - 1 - k}{l} \right)\]
\end{lemma}
\begin{proof}
	\[
		\begin{aligned}
			& \left| \left\{ v \in \triangle_{d, s}^{\mathbb{Z}} : \max_{i \in [d + 1]} v_{i} \leq k \land \operatorname{level}_{d}(v) = l \right\} \right|\\
			& = \sum_{\substack{I \subseteq [d + 1]\\|I| = l + 1}} \left| \left\{ v \in \triangle_{d, s}^{\mathbb{Z}} : \max_{i \in [d + 1]} v_{i} \leq k \land \operatorname{support}_{d}(v) = I \right\} \right|\\
			& = \sum_{\substack{I \subseteq [d + 1]\\|I| = l + 1}} \left| \left\{ v \in \triangle_{l, s}^{\mathbb{Z}} : \max_{i \in [l + 1]} v_{i} \leq k \land \forall i \in [l + 1] \ v_{i} > 0 \right\} \right|\\
			& = \binom{d + 1}{l + 1} \cdot \left| \left\{ v \in \triangle_{l, s}^{\mathbb{Z}} : \max_{i \in [l + 1]} v_{i} \leq k \land \forall i \in [l + 1] \ v_{i} > 0 \right\} \right|\\
			& = \binom{d + 1}{l + 1} \cdot \left| \left\{ v \in \triangle_{l, s - (l + 1)}^{\mathbb{Z}} : \max_{i \in [l + 1]} v_{i} \leq k - 1 \right\} \right|\\
			& = \binom{d + 1}{l + 1} \cdot \left( \binom{s - 1}{l} - (l + 1) \cdot \binom{s - 1 - k}{l} \right)
		\end{aligned}
	\]
	
	Note for the last equality that \(2 \cdot (k - 1) + 1 \geq s - 1 \geq s - (l + 1)\) as \(l \geq 0\).
	This allows the application of \cref{thm:simplex-points-in-radius}.
\end{proof}

Note that in the case \(k = s\) \cref{thm:simplex-points-in-radius-on-level} implies \(\left| \left\{ v \in \triangle_{d, s}^{\mathbb{Z}} : \operatorname{level}_{d}(v) = l \right\} \right| = \binom{d + 1}{l + 1} \cdot \binom{s - 1}{l}\).

\pagebreak

We are now able to compute the cost of our solution implied by \cref{thm:existence-bidirected-balance-flow-simplified-simplex-graph}.

\begin{lemma}\label{thm:existence-solution-simplified-simplex-instance}
	Let \(d \in \mathbb{N}\) and \(s, \delta \in \mathbb{N}\) be such that \(2 \cdot \delta \leq s\).
	Let \(\operatorname{SI}_{d, s, \delta} = (\operatorname{SG}_{d, s, \delta}, R_{d, s}, c_{\operatorname{L}^{1}})\) be a simplified simplex instance.
	If \(d = 2\) or \(d \geq 3 \land s = 3 \cdot \delta - 2\), then there exists a solution \(u : E(\operatorname{SG}_{d, s, \delta}) \to \mathbb{R}_{\geq 0}\) of \nameref{lp:MBCR} for \(\operatorname{SI}_{d, s, \delta}\) of cost
	\[
		\sum_{e \in E(\operatorname{SG}_{d, s, \delta})} c_{\operatorname{L}^{1}}(e) \cdot u(e) = \begin{cases}
			\hfil 3 \cdot s + \frac{3}{2} \cdot \frac{(s - \delta) \cdot (s - \delta + 1)}{2 \cdot s - 3 \cdot \delta + 1} & \mbox{if } d = 2\\
			\hfil \frac{s}{3} \cdot \left( 5 \cdot d + 1 + \frac{d - 1}{s} \right) & \mbox{if } d \geq 3 \land s = 3 \cdot \delta - 2
		\end{cases}
	\]
\end{lemma}

\begin{proof}
	We will define a solution \((u, b, f)\) of \nameref{lp:MBFR} for \SIdsdelta\ with the respective solution cost.
	
	Let \(u' : [d] \to \mathbb{R}_{\geq 0}\) be given by
	\[
		u'(l) = \frac{1}{\binom{d + 1}{2} \cdot (2 \cdot s - 3 \cdot \delta + 1)} \cdot \begin{cases}
			\hfil (d + 1) \cdot s - d \cdot (3 \cdot \delta - 2) - 1 & \text{if } l = 1\\
			\hfil 3 & \text{if } l = 2\\
			\hfil 0 & \text{if } l \geq 3\\
		\end{cases}
	\]
	and \(u : E(\operatorname{SG}_{d, s, \delta}) \to \mathbb{R}_{\geq 0}\) be given by \(u(e) = u'(\operatorname{level}_{d}(e))\) as in \cref{thm:existence-bidirected-balance-flow-simplified-simplex-graph}.
	
	Let \(b : V(\operatorname{SG}_{d, s, \delta}) \to \mathbb{R}\) be given as follows:\\
	For \(r_{i} \in R_{d, s}\) we have \(b(r_{i}) = -1\) and for \(v \in V_{d, s, \delta}\) and \(l := \operatorname{level}_{d}(v)\):
	\[
		b(v) = \begin{cases}
			- (l - 1) \cdot u'(l) & \mbox{if } v \in \triangle_{d, s + 1}^{\mathbb{Z}}\\
			+ (l - 1) \cdot u'(l) + (d - l) \cdot u'(l + 1) & \mbox{if } v \in \triangle_{d, s}^{\mathbb{Z}}
		\end{cases}
	\]
	
	Now we need to define for each \(k \in [d + 1]\) a bidirected balance-flow for \(u\) and \(b_{r_{k}}\) where \(r_{k} \in R_{d, s}\) and \(b_{r_{k}}(v) = b(v) + 2 \cdot \llbracket v = r_{k} \rrbracket\).
	In particular that means \(b_{r_{k}}(r_{i}) = \llbracket i = k \rrbracket - \llbracket i \neq k \rrbracket\).
	\cref{thm:existence-bidirected-balance-flow-simplified-simplex-graph} shows the existence of such bidirected balance-flows \(f_{r_{k}}\).
	This is the case because, as remarked
	earlier, we only have to contract the auxiliary vertices \(R_{i}\) in \(\operatorname{SG}'_{d, s, \delta}\) to the respective required vertices \(r_{i}\) to obtain the corresponding bidirected balance-flow on \(\operatorname{SG}_{d, s, \delta}\).
	We have \(b_{r_{k}}(r_{i}) = b_{k}(R_{i}) = f_{r_{k}}(\delta_{\operatorname{SG}'_{d, s, \delta}}^{+}(R_{i})) - f_{r_{k}}(\delta_{\operatorname{SG}'_{d, s, \delta}}^{-}(R_{i})) = \left( \llbracket i = k \rrbracket - \llbracket i \neq k \rrbracket \right) \cdot u(\delta_{\operatorname{SG}_{d, s, \delta}}(R_{i}))\) and we further compute:
	\[
		\begin{aligned}
			u(\delta_{\operatorname{SG}_{d, s, \delta}}(R_{i}))
			& = \sum_{l \in [d]} \left| \left\{ v \in \triangle_{d, s + 1}^{\mathbb{Z}} : v_{i} = s - \delta + 1 \land \operatorname{level}_{d}(v) = l \right\} \right| \cdot u'(l)\\
			& = \sum_{l \in [d]} \left| \left\{ v \in \triangle_{d - 1, \delta}^{\mathbb{Z}} : \operatorname{level}_{d - 1}(v) = l - 1 \right\} \right| \cdot u'(l)\\
			& = \sum_{l \in [d]} \binom{d}{l} \cdot \binom{\delta - 1}{l - 1} \cdot u'(l)\\
			& = d \cdot u'(1) + \binom{d}{2} \cdot (\delta - 1) \cdot u'(2)\\
			& = \frac{d \cdot \left( (d + 1) \cdot s - d \cdot (3 \cdot \delta - 2) - 1 \right) + \binom{d}{2} \cdot (\delta - 1) \cdot 3}{\binom{d + 1}{2} \cdot (2 \cdot s - 3 \cdot \delta + 1)}
			= 1
		\end{aligned}
	\]
	
%	\[
%		\begin{aligned}
%			d \cdot u'(1)
%			& = d \cdot \frac{(d + 1) \cdot s - d \cdot (3 \cdot \delta - 2) - 1}{\binom{d + 1}{2} \cdot (2 \cdot s - 3 \cdot \delta + 1)}\\
%			& = \frac{\binom{d + 1}{2} \cdot (2 \cdot s - 3 \cdot \delta + 1) - 3 \cdot \binom{d}{2} \cdot (\delta - 1)}{\binom{d + 1}{2} \cdot (2 \cdot s - 3 \cdot \delta + 1)}\\
%			& = 1 - \frac{3 \cdot \binom{d}{2} \cdot (\delta - 1)}{\binom{d + 1}{2} \cdot (2 \cdot s - 3 \cdot \delta + 1)}
%			= 1 - \binom{d}{2} \cdot (\delta - 1) \cdot u'(2)
%		\end{aligned}
%	\]
%	\[
%		\begin{aligned}
%			u(\delta_{\operatorname{SG}_{d, s, \delta}}(r_{i}))
%			& = \sum_{l \in [d]} \left| \left\{ v \in \triangle_{d, s}^{\mathbb{Z}} : v_{i} = s - \delta \land \operatorname{level}_{d}(v) = l \right\} \right| \cdot u'(l)\\
%			& = \sum_{l \in [d]} \left| \left\{ v \in \triangle_{d - 1, \delta}^{\mathbb{Z}} : \operatorname{level}_{d - 1}(v) = l - 1 \right\} \right| \cdot u'(l)\\
%			& = \sum_{l \in [d]} \binom{d}{l} \cdot \binom{\delta - 1}{l - 1} \cdot u'(l)\\
%			& = d \cdot u'(1) + \binom{d}{2} \cdot (\delta - 1) \cdot u'(2)\\
%			& = \frac{d \cdot \left( (d + 1) \cdot s - d \cdot (3 \cdot \delta - 2) - 1 \right) + \binom{d}{2} \cdot (\delta - 1) \cdot 3}{\binom{d + 1}{2} \cdot (2 \cdot s - 3 \cdot \delta + 1)}
%			= 1
%		\end{aligned}
%	\]
	
	Hence \((u, b, f)\) is a solution of \nameref{lp:MBCR} for \SIdsdelta.
	
	With the formulas stated in \cref{thm:simplex-points-in-radius} and \cref{thm:simplex-points-in-radius-on-level} we can compute the solution cost.
	Note that \(2 \cdot (s - \delta) + 1 \geq s + 1\) if \(2 \cdot \delta \leq s\) and therefore we can use \cref{thm:simplex-points-in-radius-on-level} for \(k = s - \delta\).
	We first compute:
	\[
		\begin{aligned}
			|\{e \in E_{d, s, \delta} : \operatorname{level}_{d}(e) = l\}|
			& = (l + 1) \cdot \left| \left\{ v \in V_{d, s, \delta} \cap \triangle_{d, s + 1}^{\mathbb{Z}} : \operatorname{level}_{d}(v) = l \right\} \right|\\
			& = (l + 1) \cdot \left| \left\{ v \in \triangle_{d, s + 1}^{\mathbb{Z}} : \max_{i \in [d + 1]} v_{i} \leq s - \delta \land \operatorname{level}_{d}(v) = l \right\} \right|\\
			& = (l + 1) \cdot \binom{d + 1}{l + 1} \cdot \left( \binom{s}{l} - (l + 1) \cdot \binom{\delta}{l} \right)\\
			& = (d + 1) \cdot \binom{d}{l} \cdot \left( \binom{s}{l} - (l + 1) \cdot \binom{\delta}{l} \right)
		\end{aligned}
	\]
	\[
		\begin{aligned}
			|\{e \in F_{d, s, \delta} : \operatorname{level}_{d}(e) = l\}|
			& = \sum_{i \in [d + 1]} \left| \left\{ v \in \triangle_{d, s}^{\mathbb{Z}} : v_{i} = s - \delta \land \operatorname{level}_{d}(v) = l \right\} \right|\\
			& = (d + 1) \cdot \left| \left\{ v \in \triangle_{d - 1, \delta}^{\mathbb{Z}} : \operatorname{level}_{d - 1}(v) = l - 1 \right\} \right|\\
			& = (d + 1) \cdot \binom{d}{l} \cdot \binom{\delta - 1}{l - 1}
		\end{aligned}
	\]
	
	By definition we have \(c_{\operatorname{L}^{1}}(e) = 2 \cdot \delta\) for \(e \in F_{d, s, \delta}\) and \(c_{\operatorname{L}^{1}}(e) = 1\) for \(e \in E_{d, s, \delta}\).\\
	Therefore we compute:
	\[
		\begin{aligned}
			& \sum_{e \in E(\operatorname{SG}_{d, s, \delta})} c_{\operatorname{L}^{1}}(e) \cdot u(e)\\
			& = \sum_{l \in [d]} \left(\ |\{e \in E_{d, s, \delta} : \operatorname{level}_{d}(e) = l\}| + 2 \cdot \delta \cdot |\{e \in F_{d, s, \delta} : \operatorname{level}_{d}(e) = l\}| \ \right) \cdot u'(l)\\
			& = (d + 1) \cdot \sum_{l \in [d]} \binom{d}{l} \cdot \left( \binom{s}{l} - (l + 1) \cdot \binom{\delta}{l} + 2 \cdot \delta \cdot \binom{\delta - 1}{l - 1} \right) \cdot u'(l)\\
			%& = (d + 1) \cdot \sum_{l \in [d]} \binom{d}{l} \cdot \left( \binom{s}{l} - (l + 1) \cdot \binom{\delta}{l} + 2 \cdot l \cdot \binom{\delta}{l} \right) \cdot u'(l)\\
			& = (d + 1) \cdot \sum_{l \in [d]} \binom{d}{l} \cdot \left( \binom{s}{l} + (l - 1) \cdot \binom{\delta}{l} \right) \cdot u'(l)\\
			& = (d + 1) \cdot \left( s \cdot d \cdot u'(1) + \binom{d}{2} \cdot \left( \binom{s}{2} + \binom{\delta}{2} \right) \cdot u'(2) \right)\\
			& = (d + 1) \cdot \left( s - \binom{d}{2} \cdot \left( \binom{s}{2} + \binom{\delta}{2} - s \cdot (\delta - 1) \right) \cdot u'(2) \right)\\
			%& = (d + 1) \cdot \left( s \cdot \left( 1 - \frac{3 \cdot \binom{d}{2} \cdot (\delta - 1)}{\binom{d + 1}{2} \cdot (2 \cdot s - 3 \cdot \delta + 1)} \right) + \frac{\binom{d}{2} \cdot \left( \binom{s}{2} + \binom{\delta}{2} \right) \cdot 3}{\binom{d + 1}{2} \cdot (2 \cdot s - 3 \cdot \delta + 1)} \right)\\
			%& = (d + 1) \cdot \left( s  - 3 \cdot \frac{d - 1}{d + 1} \cdot \frac{\binom{s}{2} + \binom{\delta}{2} - s \cdot (\delta - 1)}{2 \cdot s - 3 \cdot \delta + 1} \right)\\
			%& = s \cdot (d + 1) + 3 \cdot (d - 1) \cdot \frac{\binom{s}{2} + \binom{\delta}{2} - s \cdot (\delta - 1)}{2 \cdot s - 3 \cdot \delta + 1}\\
			& = s \cdot (d + 1) + \frac{3}{2} \cdot \frac{(s - \delta) \cdot (s - \delta + 1)}{2 \cdot s - 3 \cdot \delta + 1} \cdot (d - 1)
		\end{aligned}
	\]
	
	If \(s = 3 \cdot \delta - 2\), we further compute:\footnote{
		One can check that the term above is minimized for \(0 \leq \delta \leq \frac{s}{2}\) if \(s = 3 \cdot \delta - \alpha\) for a constant \(\alpha\).
		%For the asymptotic behaviour in \(s\) the actual value of \(\alpha\) is irrelevant, as it is a constant.
	}
	\begin{align*}
		& s \cdot (d + 1) + \frac{3}{2} \cdot \frac{(s - \delta) \cdot (s - \delta + 1)}{2 \cdot s - 3 \cdot \delta + 1} \cdot (d - 1)\\
		& = s \cdot (d + 1) + \frac{1}{6} \cdot \frac{(3 \cdot s - 3 \cdot \delta) \cdot (3 \cdot s - 3 \cdot \delta + 3)}{2 \cdot s - 3 \cdot \delta + 1} \cdot (d - 1)\\
		& = s \cdot (d + 1) + \frac{1}{6} \cdot \frac{(2 \cdot s - 2) \cdot (2 \cdot s + 1)}{s - 1} \cdot (d - 1)\\
		& = s \cdot (d + 1) + \frac{1}{3} \cdot (2 \cdot s + 1) \cdot (d - 1)%\\
		%& 
		= \frac{s}{3} \cdot \left( 5 \cdot d + 1 + \frac{d - 1}{s} \right) \qedhere
	\end{align*}
\end{proof}

\begin{theorem}\label{thm:bounds-solution-simplified-simplex-instance}
	Let \(d, s \in \mathbb{N}\) such that \(s \geq 2\) and \(s \equiv 1 \mod 2\).
	Then
	\[\operatorname{opt}_{\operatorname{\mathcal{BCR}}}(\operatorname{SI}_{d, s}) \leq \frac{s}{3} \cdot \left( 5 \cdot d + 1 + \frac{d - 1}{s} \right) \quad\mathrm{and}\quad \operatorname{gap}_{\operatorname{\mathcal{BCR}}, \operatorname{\mathcal{BCR}^{+}}}(\operatorname{SI}_{d, s}) \geq \frac{6 \cdot d}{5 \cdot d + 1 + \frac{d - 1}{s}}\]
\end{theorem}
\begin{proof}
	This follows with \cref{thm:existence-solution-simplified-simplex-instance} and the fact that \(\operatorname{opt}_{\operatorname{\mathcal{BCR}}}(\operatorname{SI}_{d, s}) \leq \operatorname{opt}_{\operatorname{\mathcal{BCR}}}(\operatorname{SI}_{d, s, \delta})\).
	As \(s \geq 2\) and \(s \equiv 1 \mod 2\), there exists \(\delta \in \mathbb{N}\) such that \(s = 3 \cdot \delta - 2\) and \(1 \leq \delta \leq \frac{s}{2}\).
	Hence, with \cref{thm:existence-solution-simplified-simplex-instance} we obtain the claimed upper bound on the optimum value, since \(\operatorname{opt}_{\operatorname{\mathcal{BCR}}}(\operatorname{SI}_{d, s}) \leq \operatorname{opt}_{\operatorname{\mathcal{BCR}}}(\operatorname{SI}_{d, s, \delta})\).
	Moreover, we already know by \cref{thm:simplex-terminal-instance-optimum-solution} that \(\operatorname{opt}_{\operatorname{\mathcal{BCR}^{+}}}(\operatorname{SI}_{d, s}) = 2 \cdot s \cdot d\) and therefore the claimed lower bound on the gap is also true.
\end{proof}

\cref{thm:bounds-solution-simplified-simplex-instance} immediately proves \cref{thm:refined-bidirected-relaxations-gap} since we have:
\[\operatorname{gap}_{\operatorname{\mathcal{BCR}_{\leq d + 1}}, \operatorname{\mathcal{BCR}_{\leq d + 1}^{+}}} \geq \lim_{\delta \to \infty} \operatorname{gap}_{\operatorname{\mathcal{BCR}}, \operatorname{\mathcal{BCR}^{+}}}(\operatorname{SI}_{d, 3 \cdot \delta - 2}) \geq \lim_{\delta \to \infty} \frac{6 \cdot d}{5 \cdot d + 1 + \frac{d - 1}{3 \cdot \delta - 2}} = \frac{6 \cdot d}{5 \cdot d + 1} \qedhere\]

\vspace{30pt}

%\pagebreak

\subsection{Relation to the C\texorpdfstring{\u{a}}{a}linescu-Karloff-Rabani LP-relaxation\texorpdfstring{\\}{ }of the Multiway Cut Problem for three required vertices}\label{sec:relation-to-multiway-cut-problem}

For \SIdsdelta\ with \(d = 2\) there exists an interesting connection between the bidirected cut relaxation and the C{\u{a}}linescu-Karloff-Rabani LP-relaxation \cite{CKR98} for the \shortmathproblem[MCP]{Multiway Cut Problem}, where we are also given a Steiner tree instance \(((V, E), R, c)\), but the task is now to find an edge set \(F \subseteq E\) such that the required vertices \(R\) are in pairwise different connected components in \((V, E \setminus F)\) and \(c(F)\) is minimum.

%\begin{mathproblem}{Multiway Cut Problem\label{problem:MCP}}[MCP]
%	Instance & A terminal instance \(((V, E), R, c)\)\\[3pt]
%	Task & Find an edge set \(F \subseteq E\) such that the terminals \(R\) are in pairwise different connected components in \((V, E \setminus F)\) and \(c(F)\) is minimum.\\[3pt]
%\end{mathproblem}

\nameref{problem:MCP} is also known as Multi-Terminal Problem.
Notice that \nameref{problem:STP} can be formulated similarly:
Find an edge set \(F \subseteq E\) such that the required vertices \(R\) are all in the same connected component in \((V, F)\) and \(c(F)\) is minimum.
In this section we will work with this definition.
In contrast to \nameref{problem:STP}, \nameref{problem:MCP} is NP-hard even for fixed \(|R| \geq 3\).
Further information can be found for example in \cite[Ch.\ 4 and 19]{Vaz01}.

Let \(((V, E), R, c)\) be a Steiner tree instance and write \(R = \{r_{1}, \dotsc, r_{d + 1}\}\), then the C{\u{a}}linescu-Karloff-Rabani LP-relaxation for \nameref{problem:MCP} (\cite{CKR98}) can be formulated as follows:
\begin{linearprogram}[CKR]
	\label{lp:CKR}
	\(\max\) & \(\displaystyle \sum_{\{v, w\} \in E} c(\{v, w\}) \cdot \frac{\|x(v) - x(w)\|_{1}}{2}\)\\[20pt]
	s.t. & \(x : V \to \triangle_{d, 1}^{\mathbb{R}}\)\\[10pt]
	& \(x(r_{i})_{i} = 1 \qquad i \in [d + 1]\)
\end{linearprogram}
The relaxation embeds the vertices in the \(d\)-dimensional continuous unit simplex \(\triangle_{d, 1}^{\mathbb{R}}\) with the restriction that each required vertex is mapped to its respective corner of the simplex.
If we require integrality for \(x\) then it encodes a partition of the vertices, namely the connected components associated to the respective required vertices if one is contained.
Only edges between these connected components in the original graph are counted in the objective, which exactly encodes \nameref{problem:MCP}.

In \cite{CCT06}, \nameref{lp:CKR} restricted to Steiner tree instances with \(3\) required vertices is studied.
There it is shown that the following Steiner tree instances \((G_{q}, R_{q}, \cdot)\), defined for each \(q \in \mathbb{N}\), are worst case instances for \nameref{lp:CKR} regarding the integrality gap:
\begin{align*}
	G_{q} & = \left( \triangle_{2, q}^{\mathbb{Z}},\ \left\{ \{v, w\} \ :\ v, w \in \triangle_{2, q}^{\mathbb{Z}} \land \|v - w\|_{1} = 2 \right\} \right)\\
	R_{q} & = \left\{ v \in \triangle_{2, q}^{\mathbb{Z}} \ :\ \max_{i \in [3]} v_{i} = q \right\}
\end{align*}
Only the edge costs are unspecified.
So for every Steiner tree instance \((G, R, c)\) there exists a terminal instance \((G_{q}, R_{q}, c')\) such that \(\operatorname{gap}_{\operatorname{CKR}, \operatorname{MCP}}(G, R, c) \leq \operatorname{gap}_{\operatorname{CKR}, \operatorname{MCP}}(G_{q}, R_{q}, c')\).
Analyzing these instances, the authors prove for this class:
\[\operatorname{gap}_{\operatorname{CKR}, \operatorname{MCP}}(G_{q}, R_{q}, \cdot) = \begin{cases}
	\frac{12 \cdot q + 12}{11 \cdot q + 12} & \mbox{if } q \equiv 0 \mod 3\\[3pt]
	\frac{12 \cdot q}{11 \cdot q + 1} & \mbox{if } q \equiv 1 \mod 3\\[3pt]
	\frac{12 \cdot q^{2}}{11 \cdot q^{2} + q - 1} & \mbox{if } q \equiv 2 \mod 3
\end{cases}\]
This implies \(\operatorname{gap}_{\operatorname{CKR_{3}}, \operatorname{MCP_{3}}} = \frac{12}{11}\) restricted to Steiner tree instances with \(3\) required vertices.

%\subfile{figures/figure-planar-duality}\todo{reformat pictures}

\begin{wrapfigure}{r}{0.62\textwidth}
	\centering
	\captionsetup{justification=centering,margin=1cm}
	\begin{tikzpicture}[scale=0.75]
	\tikzset{gridline/.style={thick, draw=red}}
	\tikzset{vertex/.style={circle, thick, minimum size=5pt, inner sep=0pt, outer sep=0pt}}
	\tikzset{vertexinternalzero/.style={vertex, draw=blue, fill=blue}}
	\tikzset{vertexinternalone/.style={vertex, draw=orange, fill=orange}}
	\tikzset{vertexterminal/.style={vertex, rectangle, outer sep=3pt, draw=green!40!black, fill=green!40!black}}
	\tikzset{edgeinternal/.style={thick, draw=gray}}
	\tikzset{edgeterminal/.style={thick, draw=green}}
	
	\tikzset{vertexinternalzero/.style={vertex, draw=black, fill=black}}
	\tikzset{vertexinternalone/.style={vertex, draw=black, fill=white}}
	\tikzset{vertexterminal/.style={vertex, rectangle, inner sep=2pt, outer sep=3pt, draw=black, fill=black}}
	\tikzset{edgeinternal/.style={thick, solid}}
	\tikzset{edgeterminal/.style={thick, solid}}
	\tikzset{gridline/.style={thick, draw=black, dashed}}
	
	\coordinate (simplex0I0-0-9) at (10.0, 0.5773502691896257);
	\coordinate (simplex0I0-1-8) at (9.0, 0.5773502691896257);
	\coordinate (simplex0I0-2-7) at (8.0, 0.5773502691896257);
	\coordinate (simplex0I0-3-6) at (7.0, 0.5773502691896257);
	\coordinate (simplex0I0-4-5) at (6.0, 0.5773502691896257);
	\coordinate (simplex0I0-5-4) at (5.0, 0.5773502691896257);
	\coordinate (simplex0I0-6-3) at (4.0, 0.5773502691896257);
	\coordinate (simplex0I0-7-2) at (3.0, 0.5773502691896257);
	\coordinate (simplex0I0-8-1) at (2.0, 0.5773502691896257);
	\coordinate (simplex0I0-9-0) at (1.0, 0.5773502691896257);
	\coordinate (simplex0I1-0-8) at (9.5, 1.4433756729740643);
	\coordinate (simplex0I1-1-7) at (8.5, 1.4433756729740643);
	\coordinate (simplex0I1-2-6) at (7.5, 1.4433756729740643);
	\coordinate (simplex0I1-3-5) at (6.5, 1.4433756729740643);
	\coordinate (simplex0I1-4-4) at (5.5, 1.4433756729740643);
	\coordinate (simplex0I1-5-3) at (4.5, 1.4433756729740643);
	\coordinate (simplex0I1-6-2) at (3.5, 1.4433756729740643);
	\coordinate (simplex0I1-7-1) at (2.5, 1.4433756729740643);
	\coordinate (simplex0I1-8-0) at (1.5, 1.4433756729740643);
	\coordinate (simplex0I2-0-7) at (9.0, 2.309401076758503);
	\coordinate (simplex0I2-1-6) at (8.0, 2.309401076758503);
	\coordinate (simplex0I2-2-5) at (7.0, 2.309401076758503);
	\coordinate (simplex0I2-3-4) at (6.0, 2.309401076758503);
	\coordinate (simplex0I2-4-3) at (5.0, 2.309401076758503);
	\coordinate (simplex0I2-5-2) at (4.0, 2.309401076758503);
	\coordinate (simplex0I2-6-1) at (3.0, 2.309401076758503);
	\coordinate (simplex0I2-7-0) at (2.0, 2.309401076758503);
	\coordinate (simplex0I3-0-6) at (8.5, 3.1754264805429417);
	\coordinate (simplex0I3-1-5) at (7.5, 3.1754264805429417);
	\coordinate (simplex0I3-2-4) at (6.5, 3.1754264805429417);
	\coordinate (simplex0I3-3-3) at (5.5, 3.1754264805429417);
	\coordinate (simplex0I3-4-2) at (4.5, 3.1754264805429417);
	\coordinate (simplex0I3-5-1) at (3.5, 3.1754264805429417);
	\coordinate (simplex0I3-6-0) at (2.5, 3.1754264805429417);
	\coordinate (simplex0I4-0-5) at (8.0, 4.04145188432738);
	\coordinate (simplex0I4-1-4) at (7.0, 4.04145188432738);
	\coordinate (simplex0I4-2-3) at (6.0, 4.04145188432738);
	\coordinate (simplex0I4-3-2) at (5.0, 4.04145188432738);
	\coordinate (simplex0I4-4-1) at (4.0, 4.04145188432738);
	\coordinate (simplex0I4-5-0) at (3.0, 4.04145188432738);
	\coordinate (simplex0I5-0-4) at (7.5, 4.907477288111819);
	\coordinate (simplex0I5-1-3) at (6.5, 4.907477288111819);
	\coordinate (simplex0I5-2-2) at (5.5, 4.907477288111819);
	\coordinate (simplex0I5-3-1) at (4.5, 4.907477288111819);
	\coordinate (simplex0I5-4-0) at (3.5, 4.907477288111819);
	\coordinate (simplex0I6-0-3) at (7.0, 5.773502691896258);
	\coordinate (simplex0I6-1-2) at (6.0, 5.773502691896258);
	\coordinate (simplex0I6-2-1) at (5.0, 5.773502691896258);
	\coordinate (simplex0I6-3-0) at (4.0, 5.773502691896258);
	\coordinate (simplex0I7-0-2) at (6.5, 6.639528095680696);
	\coordinate (simplex0I7-1-1) at (5.5, 6.639528095680696);
	\coordinate (simplex0I7-2-0) at (4.5, 6.639528095680696);
	\coordinate (simplex0I8-0-1) at (6.0, 7.505553499465135);
	\coordinate (simplex0I8-1-0) at (5.0, 7.505553499465135);
	\coordinate (simplex0I9-0-0) at (5.5, 8.371578903249572);
	\coordinate (simplex1I0-0-10) at (10.5, 0.28867513459481287);
	\coordinate (simplex1I0-1-9) at (9.5, 0.28867513459481287);
	\coordinate (simplex1I0-2-8) at (8.5, 0.28867513459481287);
	\coordinate (simplex1I0-3-7) at (7.5, 0.28867513459481287);
	\coordinate (simplex1I0-4-6) at (6.5, 0.28867513459481287);
	\coordinate (simplex1I0-5-5) at (5.5, 0.28867513459481287);
	\coordinate (simplex1I0-6-4) at (4.5, 0.28867513459481287);
	\coordinate (simplex1I0-7-3) at (3.5, 0.28867513459481287);
	\coordinate (simplex1I0-8-2) at (2.5, 0.28867513459481287);
	\coordinate (simplex1I0-9-1) at (1.5, 0.28867513459481287);
	\coordinate (simplex1I0-10-0) at (0.5, 0.28867513459481287);
	\coordinate (simplex1I1-0-9) at (10.0, 1.1547005383792515);
	\coordinate (simplex1I1-1-8) at (9.0, 1.1547005383792515);
	\coordinate (simplex1I1-2-7) at (8.0, 1.1547005383792515);
	\coordinate (simplex1I1-3-6) at (7.0, 1.1547005383792515);
	\coordinate (simplex1I1-4-5) at (6.0, 1.1547005383792515);
	\coordinate (simplex1I1-5-4) at (5.0, 1.1547005383792515);
	\coordinate (simplex1I1-6-3) at (4.0, 1.1547005383792515);
	\coordinate (simplex1I1-7-2) at (3.0, 1.1547005383792515);
	\coordinate (simplex1I1-8-1) at (2.0, 1.1547005383792515);
	\coordinate (simplex1I1-9-0) at (1.0, 1.1547005383792515);
	\coordinate (simplex1I2-0-8) at (9.5, 2.02072594216369);
	\coordinate (simplex1I2-1-7) at (8.5, 2.02072594216369);
	\coordinate (simplex1I2-2-6) at (7.5, 2.02072594216369);
	\coordinate (simplex1I2-3-5) at (6.5, 2.02072594216369);
	\coordinate (simplex1I2-4-4) at (5.5, 2.02072594216369);
	\coordinate (simplex1I2-5-3) at (4.5, 2.02072594216369);
	\coordinate (simplex1I2-6-2) at (3.5, 2.02072594216369);
	\coordinate (simplex1I2-7-1) at (2.5, 2.02072594216369);
	\coordinate (simplex1I2-8-0) at (1.5, 2.02072594216369);
	\coordinate (simplex1I3-0-7) at (9.0, 2.886751345948129);
	\coordinate (simplex1I3-1-6) at (8.0, 2.886751345948129);
	\coordinate (simplex1I3-2-5) at (7.0, 2.886751345948129);
	\coordinate (simplex1I3-3-4) at (6.0, 2.886751345948129);
	\coordinate (simplex1I3-4-3) at (5.0, 2.886751345948129);
	\coordinate (simplex1I3-5-2) at (4.0, 2.886751345948129);
	\coordinate (simplex1I3-6-1) at (3.0, 2.886751345948129);
	\coordinate (simplex1I3-7-0) at (2.0, 2.886751345948129);
	\coordinate (simplex1I4-0-6) at (8.5, 3.7527767497325675);
	\coordinate (simplex1I4-1-5) at (7.5, 3.7527767497325675);
	\coordinate (simplex1I4-2-4) at (6.5, 3.7527767497325675);
	\coordinate (simplex1I4-3-3) at (5.5, 3.7527767497325675);
	\coordinate (simplex1I4-4-2) at (4.5, 3.7527767497325675);
	\coordinate (simplex1I4-5-1) at (3.5, 3.7527767497325675);
	\coordinate (simplex1I4-6-0) at (2.5, 3.7527767497325675);
	\coordinate (simplex1I5-0-5) at (8.0, 4.618802153517006);
	\coordinate (simplex1I5-1-4) at (7.0, 4.618802153517006);
	\coordinate (simplex1I5-2-3) at (6.0, 4.618802153517006);
	\coordinate (simplex1I5-3-2) at (5.0, 4.618802153517006);
	\coordinate (simplex1I5-4-1) at (4.0, 4.618802153517006);
	\coordinate (simplex1I5-5-0) at (3.0, 4.618802153517006);
	\coordinate (simplex1I6-0-4) at (7.5, 5.484827557301445);
	\coordinate (simplex1I6-1-3) at (6.5, 5.484827557301445);
	\coordinate (simplex1I6-2-2) at (5.5, 5.484827557301445);
	\coordinate (simplex1I6-3-1) at (4.5, 5.484827557301445);
	\coordinate (simplex1I6-4-0) at (3.5, 5.484827557301445);
	\coordinate (simplex1I7-0-3) at (7.0, 6.3508529610858835);
	\coordinate (simplex1I7-1-2) at (6.0, 6.3508529610858835);
	\coordinate (simplex1I7-2-1) at (5.0, 6.3508529610858835);
	\coordinate (simplex1I7-3-0) at (4.0, 6.3508529610858835);
	\coordinate (simplex1I8-0-2) at (6.5, 7.216878364870322);
	\coordinate (simplex1I8-1-1) at (5.5, 7.216878364870322);
	\coordinate (simplex1I8-2-0) at (4.5, 7.216878364870322);
	\coordinate (simplex1I9-0-1) at (6.0, 8.08290376865476);
	\coordinate (simplex1I9-1-0) at (5.0, 8.08290376865476);
	\coordinate (simplex1I10-0-0) at (5.5, 8.948929172439199);
	\coordinate (simplex2I0-0-11) at (11.0, 0.0);
	\coordinate (simplex2I0-1-10) at (10.0, 0.0);
	\coordinate (simplex2I0-2-9) at (9.0, 0.0);
	\coordinate (simplex2I0-3-8) at (8.0, 0.0);
	\coordinate (simplex2I0-4-7) at (7.0, 0.0);
	\coordinate (simplex2I0-5-6) at (6.0, 0.0);
	\coordinate (simplex2I0-6-5) at (5.0, 0.0);
	\coordinate (simplex2I0-7-4) at (4.0, 0.0);
	\coordinate (simplex2I0-8-3) at (3.0, 0.0);
	\coordinate (simplex2I0-9-2) at (2.0, 0.0);
	\coordinate (simplex2I0-10-1) at (1.0, 0.0);
	\coordinate (simplex2I0-11-0) at (0.0, 0.0);
	\coordinate (simplex2I1-0-10) at (10.5, 0.8660254037844386);
	\coordinate (simplex2I1-1-9) at (9.5, 0.8660254037844386);
	\coordinate (simplex2I1-2-8) at (8.5, 0.8660254037844386);
	\coordinate (simplex2I1-3-7) at (7.5, 0.8660254037844386);
	\coordinate (simplex2I1-4-6) at (6.5, 0.8660254037844386);
	\coordinate (simplex2I1-5-5) at (5.5, 0.8660254037844386);
	\coordinate (simplex2I1-6-4) at (4.5, 0.8660254037844386);
	\coordinate (simplex2I1-7-3) at (3.5, 0.8660254037844386);
	\coordinate (simplex2I1-8-2) at (2.5, 0.8660254037844386);
	\coordinate (simplex2I1-9-1) at (1.5, 0.8660254037844386);
	\coordinate (simplex2I1-10-0) at (0.5, 0.8660254037844386);
	\coordinate (simplex2I2-0-9) at (10.0, 1.7320508075688772);
	\coordinate (simplex2I2-1-8) at (9.0, 1.7320508075688772);
	\coordinate (simplex2I2-2-7) at (8.0, 1.7320508075688772);
	\coordinate (simplex2I2-3-6) at (7.0, 1.7320508075688772);
	\coordinate (simplex2I2-4-5) at (6.0, 1.7320508075688772);
	\coordinate (simplex2I2-5-4) at (5.0, 1.7320508075688772);
	\coordinate (simplex2I2-6-3) at (4.0, 1.7320508075688772);
	\coordinate (simplex2I2-7-2) at (3.0, 1.7320508075688772);
	\coordinate (simplex2I2-8-1) at (2.0, 1.7320508075688772);
	\coordinate (simplex2I2-9-0) at (1.0, 1.7320508075688772);
	\coordinate (simplex2I3-0-8) at (9.5, 2.598076211353316);
	\coordinate (simplex2I3-1-7) at (8.5, 2.598076211353316);
	\coordinate (simplex2I3-2-6) at (7.5, 2.598076211353316);
	\coordinate (simplex2I3-3-5) at (6.5, 2.598076211353316);
	\coordinate (simplex2I3-4-4) at (5.5, 2.598076211353316);
	\coordinate (simplex2I3-5-3) at (4.5, 2.598076211353316);
	\coordinate (simplex2I3-6-2) at (3.5, 2.598076211353316);
	\coordinate (simplex2I3-7-1) at (2.5, 2.598076211353316);
	\coordinate (simplex2I3-8-0) at (1.5, 2.598076211353316);
	\coordinate (simplex2I4-0-7) at (9.0, 3.4641016151377544);
	\coordinate (simplex2I4-1-6) at (8.0, 3.4641016151377544);
	\coordinate (simplex2I4-2-5) at (7.0, 3.4641016151377544);
	\coordinate (simplex2I4-3-4) at (6.0, 3.4641016151377544);
	\coordinate (simplex2I4-4-3) at (5.0, 3.4641016151377544);
	\coordinate (simplex2I4-5-2) at (4.0, 3.4641016151377544);
	\coordinate (simplex2I4-6-1) at (3.0, 3.4641016151377544);
	\coordinate (simplex2I4-7-0) at (2.0, 3.4641016151377544);
	\coordinate (simplex2I5-0-6) at (8.5, 4.330127018922193);
	\coordinate (simplex2I5-1-5) at (7.5, 4.330127018922193);
	\coordinate (simplex2I5-2-4) at (6.5, 4.330127018922193);
	\coordinate (simplex2I5-3-3) at (5.5, 4.330127018922193);
	\coordinate (simplex2I5-4-2) at (4.5, 4.330127018922193);
	\coordinate (simplex2I5-5-1) at (3.5, 4.330127018922193);
	\coordinate (simplex2I5-6-0) at (2.5, 4.330127018922193);
	\coordinate (simplex2I6-0-5) at (8.0, 5.196152422706632);
	\coordinate (simplex2I6-1-4) at (7.0, 5.196152422706632);
	\coordinate (simplex2I6-2-3) at (6.0, 5.196152422706632);
	\coordinate (simplex2I6-3-2) at (5.0, 5.196152422706632);
	\coordinate (simplex2I6-4-1) at (4.0, 5.196152422706632);
	\coordinate (simplex2I6-5-0) at (3.0, 5.196152422706632);
	\coordinate (simplex2I7-0-4) at (7.5, 6.06217782649107);
	\coordinate (simplex2I7-1-3) at (6.5, 6.06217782649107);
	\coordinate (simplex2I7-2-2) at (5.5, 6.06217782649107);
	\coordinate (simplex2I7-3-1) at (4.5, 6.06217782649107);
	\coordinate (simplex2I7-4-0) at (3.5, 6.06217782649107);
	\coordinate (simplex2I8-0-3) at (7.0, 6.928203230275509);
	\coordinate (simplex2I8-1-2) at (6.0, 6.928203230275509);
	\coordinate (simplex2I8-2-1) at (5.0, 6.928203230275509);
	\coordinate (simplex2I8-3-0) at (4.0, 6.928203230275509);
	\coordinate (simplex2I9-0-2) at (6.5, 7.794228634059947);
	\coordinate (simplex2I9-1-1) at (5.5, 7.794228634059947);
	\coordinate (simplex2I9-2-0) at (4.5, 7.794228634059947);
	\coordinate (simplex2I10-0-1) at (6.0, 8.660254037844386);
	\coordinate (simplex2I10-1-0) at (5.0, 8.660254037844386);
	\coordinate (simplex2I11-0-0) at (5.5, 9.526279441628825);
	
	\coordinate (simplex2I-3-7-7) at (5.5, -2.598076211353316);
	\coordinate (simplex2I7--3-7) at (10.5, 6.06217782649107);
	\coordinate (simplex2I7-7--3) at (0.5, 6.06217782649107);
	
	\draw[gridline](simplex2I1-3-7) -- (simplex2I1-7-3);
	\draw[gridline](simplex2I2-2-7) -- (simplex2I2-7-2);
	\draw[gridline](simplex2I3-1-7) -- (simplex2I3-7-1);
	\draw[gridline](simplex2I4-1-6) -- (simplex2I4-6-1);
	\draw[gridline](simplex2I5-1-5) -- (simplex2I5-5-1);
	\draw[gridline](simplex2I6-1-4) -- (simplex2I6-4-1);
	\draw[gridline](simplex2I7-1-3) -- (simplex2I7-3-1);
	
	\draw[gridline](simplex2I3-1-7) -- (simplex2I7-1-3);
	\draw[gridline](simplex2I2-2-7) -- (simplex2I7-2-2);
	\draw[gridline](simplex2I1-3-7) -- (simplex2I7-3-1);
	\draw[gridline](simplex2I1-4-6) -- (simplex2I6-4-1);
	\draw[gridline](simplex2I1-5-5) -- (simplex2I5-5-1);
	\draw[gridline](simplex2I1-6-4) -- (simplex2I4-6-1);
	\draw[gridline](simplex2I1-7-3) -- (simplex2I3-7-1);
	
	\draw[gridline](simplex2I3-7-1) -- (simplex2I7-3-1);
	\draw[gridline](simplex2I2-7-2) -- (simplex2I7-2-2);
	\draw[gridline](simplex2I1-7-3) -- (simplex2I7-1-3);
	\draw[gridline](simplex2I1-6-4) -- (simplex2I6-1-4);
	\draw[gridline](simplex2I1-5-5) -- (simplex2I5-1-5);
	\draw[gridline](simplex2I1-4-6) -- (simplex2I4-1-6);
	\draw[gridline](simplex2I1-3-7) -- (simplex2I3-1-7);
	
	\node[vertexterminal] (vertexI-3-7-7) at (simplex2I-3-7-7) {};
	\node[vertexterminal] (vertexI7--3-7) at (simplex2I7--3-7) {};
	\node[vertexterminal] (vertexI7-7--3) at (simplex2I7-7--3) {};
	
	\draw[gridline](vertexI-3-7-7) -- (simplex2I1-7-3);
	\draw[gridline](vertexI-3-7-7) -- (simplex2I1-6-4);
	\draw[gridline](vertexI-3-7-7) -- (simplex2I1-5-5);
	\draw[gridline](vertexI-3-7-7) -- (simplex2I1-4-6);
	\draw[gridline](vertexI-3-7-7) -- (simplex2I1-3-7);
	
	\draw[gridline](vertexI7--3-7) -- (simplex2I7-1-3);
	\draw[gridline](vertexI7--3-7) -- (simplex2I6-1-4);
	\draw[gridline](vertexI7--3-7) -- (simplex2I5-1-5);
	\draw[gridline](vertexI7--3-7) -- (simplex2I4-1-6);
	\draw[gridline](vertexI7--3-7) -- (simplex2I3-1-7);
	
	\draw[gridline](vertexI7-7--3) -- (simplex2I7-3-1);
	\draw[gridline](vertexI7-7--3) -- (simplex2I6-4-1);
	\draw[gridline](vertexI7-7--3) -- (simplex2I5-5-1);
	\draw[gridline](vertexI7-7--3) -- (simplex2I4-6-1);
	\draw[gridline](vertexI7-7--3) -- (simplex2I3-7-1);
	
	\node[vertexinternalzero] (vertexI0-3-6) at (simplex0I0-3-6) {};
	\node[vertexinternalzero] (vertexI0-4-5) at (simplex0I0-4-5) {};
	\node[vertexinternalzero] (vertexI0-5-4) at (simplex0I0-5-4) {};
	\node[vertexinternalzero] (vertexI0-6-3) at (simplex0I0-6-3) {};
	\node[vertexinternalzero] (vertexI1-2-6) at (simplex0I1-2-6) {};
	\node[vertexinternalzero] (vertexI1-3-5) at (simplex0I1-3-5) {};
	\node[vertexinternalzero] (vertexI1-4-4) at (simplex0I1-4-4) {};
	\node[vertexinternalzero] (vertexI1-5-3) at (simplex0I1-5-3) {};
	\node[vertexinternalzero] (vertexI1-6-2) at (simplex0I1-6-2) {};
	\node[vertexinternalzero] (vertexI2-1-6) at (simplex0I2-1-6) {};
	\node[vertexinternalzero] (vertexI2-2-5) at (simplex0I2-2-5) {};
	\node[vertexinternalzero] (vertexI2-3-4) at (simplex0I2-3-4) {};
	\node[vertexinternalzero] (vertexI2-4-3) at (simplex0I2-4-3) {};
	\node[vertexinternalzero] (vertexI2-5-2) at (simplex0I2-5-2) {};
	\node[vertexinternalzero] (vertexI2-6-1) at (simplex0I2-6-1) {};
	\node[vertexinternalzero] (vertexI3-0-6) at (simplex0I3-0-6) {};
	\node[vertexinternalzero] (vertexI3-1-5) at (simplex0I3-1-5) {};
	\node[vertexinternalzero] (vertexI3-2-4) at (simplex0I3-2-4) {};
	\node[vertexinternalzero] (vertexI3-3-3) at (simplex0I3-3-3) {};
	\node[vertexinternalzero] (vertexI3-4-2) at (simplex0I3-4-2) {};
	\node[vertexinternalzero] (vertexI3-5-1) at (simplex0I3-5-1) {};
	\node[vertexinternalzero] (vertexI3-6-0) at (simplex0I3-6-0) {};
	\node[vertexinternalzero] (vertexI4-0-5) at (simplex0I4-0-5) {};
	\node[vertexinternalzero] (vertexI4-1-4) at (simplex0I4-1-4) {};
	\node[vertexinternalzero] (vertexI4-2-3) at (simplex0I4-2-3) {};
	\node[vertexinternalzero] (vertexI4-3-2) at (simplex0I4-3-2) {};
	\node[vertexinternalzero] (vertexI4-4-1) at (simplex0I4-4-1) {};
	\node[vertexinternalzero] (vertexI4-5-0) at (simplex0I4-5-0) {};
	\node[vertexinternalzero] (vertexI5-0-4) at (simplex0I5-0-4) {};
	\node[vertexinternalzero] (vertexI5-1-3) at (simplex0I5-1-3) {};
	\node[vertexinternalzero] (vertexI5-2-2) at (simplex0I5-2-2) {};
	\node[vertexinternalzero] (vertexI5-3-1) at (simplex0I5-3-1) {};
	\node[vertexinternalzero] (vertexI5-4-0) at (simplex0I5-4-0) {};
	\node[vertexinternalzero] (vertexI6-0-3) at (simplex0I6-0-3) {};
	\node[vertexinternalzero] (vertexI6-1-2) at (simplex0I6-1-2) {};
	\node[vertexinternalzero] (vertexI6-2-1) at (simplex0I6-2-1) {};
	\node[vertexinternalzero] (vertexI6-3-0) at (simplex0I6-3-0) {};
	
	\node[vertexinternalone] (vertexI0-4-6) at (simplex1I0-4-6) {};
	\node[vertexinternalone] (vertexI0-5-5) at (simplex1I0-5-5) {};
	\node[vertexinternalone] (vertexI0-6-4) at (simplex1I0-6-4) {};
	\node[vertexinternalone] (vertexI1-3-6) at (simplex1I1-3-6) {};
	\node[vertexinternalone] (vertexI1-4-5) at (simplex1I1-4-5) {};
	\node[vertexinternalone] (vertexI1-5-4) at (simplex1I1-5-4) {};
	\node[vertexinternalone] (vertexI1-6-3) at (simplex1I1-6-3) {};
	\node[vertexinternalone] (vertexI2-2-6) at (simplex1I2-2-6) {};
	\node[vertexinternalone] (vertexI2-3-5) at (simplex1I2-3-5) {};
	\node[vertexinternalone] (vertexI2-4-4) at (simplex1I2-4-4) {};
	\node[vertexinternalone] (vertexI2-5-3) at (simplex1I2-5-3) {};
	\node[vertexinternalone] (vertexI2-6-2) at (simplex1I2-6-2) {};
	\node[vertexinternalone] (vertexI3-1-6) at (simplex1I3-1-6) {};
	\node[vertexinternalone] (vertexI3-2-5) at (simplex1I3-2-5) {};
	\node[vertexinternalone] (vertexI3-3-4) at (simplex1I3-3-4) {};
	\node[vertexinternalone] (vertexI3-4-3) at (simplex1I3-4-3) {};
	\node[vertexinternalone] (vertexI3-5-2) at (simplex1I3-5-2) {};
	\node[vertexinternalone] (vertexI3-6-1) at (simplex1I3-6-1) {};
	\node[vertexinternalone] (vertexI4-0-6) at (simplex1I4-0-6) {};
	\node[vertexinternalone] (vertexI4-1-5) at (simplex1I4-1-5) {};
	\node[vertexinternalone] (vertexI4-2-4) at (simplex1I4-2-4) {};
	\node[vertexinternalone] (vertexI4-3-3) at (simplex1I4-3-3) {};
	\node[vertexinternalone] (vertexI4-4-2) at (simplex1I4-4-2) {};
	\node[vertexinternalone] (vertexI4-5-1) at (simplex1I4-5-1) {};
	\node[vertexinternalone] (vertexI4-6-0) at (simplex1I4-6-0) {};
	\node[vertexinternalone] (vertexI5-0-5) at (simplex1I5-0-5) {};
	\node[vertexinternalone] (vertexI5-1-4) at (simplex1I5-1-4) {};
	\node[vertexinternalone] (vertexI5-2-3) at (simplex1I5-2-3) {};
	\node[vertexinternalone] (vertexI5-3-2) at (simplex1I5-3-2) {};
	\node[vertexinternalone] (vertexI5-4-1) at (simplex1I5-4-1) {};
	\node[vertexinternalone] (vertexI5-5-0) at (simplex1I5-5-0) {};
	\node[vertexinternalone] (vertexI6-0-4) at (simplex1I6-0-4) {};
	\node[vertexinternalone] (vertexI6-1-3) at (simplex1I6-1-3) {};
	\node[vertexinternalone] (vertexI6-2-2) at (simplex1I6-2-2) {};
	\node[vertexinternalone] (vertexI6-3-1) at (simplex1I6-3-1) {};
	\node[vertexinternalone] (vertexI6-4-0) at (simplex1I6-4-0) {};
	
	\node[vertexterminal] (vertexI9-0-0) at (simplex0I9-0-0) {};
	\node[vertexterminal] (vertexI0-9-0) at (simplex0I0-9-0) {};
	\node[vertexterminal] (vertexI0-0-9) at (simplex0I0-0-9) {};
	
	\draw[edgeinternal] (vertexI0-4-6) -- (vertexI0-3-6);
	\draw[edgeinternal] (vertexI0-4-6) -- (vertexI0-4-5);
	\draw[edgeinternal] (vertexI0-5-5) -- (vertexI0-4-5);
	\draw[edgeinternal] (vertexI0-5-5) -- (vertexI0-5-4);
	\draw[edgeinternal] (vertexI0-6-4) -- (vertexI0-5-4);
	\draw[edgeinternal] (vertexI0-6-4) -- (vertexI0-6-3);
	\draw[edgeinternal] (vertexI1-3-6) -- (vertexI0-3-6);
	\draw[edgeinternal] (vertexI1-3-6) -- (vertexI1-2-6);
	\draw[edgeinternal] (vertexI1-3-6) -- (vertexI1-3-5);
	\draw[edgeinternal] (vertexI1-4-5) -- (vertexI0-4-5);
	\draw[edgeinternal] (vertexI1-4-5) -- (vertexI1-3-5);
	\draw[edgeinternal] (vertexI1-4-5) -- (vertexI1-4-4);
	\draw[edgeinternal] (vertexI1-5-4) -- (vertexI0-5-4);
	\draw[edgeinternal] (vertexI1-5-4) -- (vertexI1-4-4);
	\draw[edgeinternal] (vertexI1-5-4) -- (vertexI1-5-3);
	\draw[edgeinternal] (vertexI1-6-3) -- (vertexI0-6-3);
	\draw[edgeinternal] (vertexI1-6-3) -- (vertexI1-5-3);
	\draw[edgeinternal] (vertexI1-6-3) -- (vertexI1-6-2);
	\draw[edgeinternal] (vertexI2-2-6) -- (vertexI1-2-6);
	\draw[edgeinternal] (vertexI2-2-6) -- (vertexI2-1-6);
	\draw[edgeinternal] (vertexI2-2-6) -- (vertexI2-2-5);
	\draw[edgeinternal] (vertexI2-3-5) -- (vertexI1-3-5);
	\draw[edgeinternal] (vertexI2-3-5) -- (vertexI2-2-5);
	\draw[edgeinternal] (vertexI2-3-5) -- (vertexI2-3-4);
	\draw[edgeinternal] (vertexI2-4-4) -- (vertexI1-4-4);
	\draw[edgeinternal] (vertexI2-4-4) -- (vertexI2-3-4);
	\draw[edgeinternal] (vertexI2-4-4) -- (vertexI2-4-3);
	\draw[edgeinternal] (vertexI2-5-3) -- (vertexI1-5-3);
	\draw[edgeinternal] (vertexI2-5-3) -- (vertexI2-4-3);
	\draw[edgeinternal] (vertexI2-5-3) -- (vertexI2-5-2);
	\draw[edgeinternal] (vertexI2-6-2) -- (vertexI1-6-2);
	\draw[edgeinternal] (vertexI2-6-2) -- (vertexI2-5-2);
	\draw[edgeinternal] (vertexI2-6-2) -- (vertexI2-6-1);
	\draw[edgeinternal] (vertexI3-1-6) -- (vertexI2-1-6);
	\draw[edgeinternal] (vertexI3-1-6) -- (vertexI3-0-6);
	\draw[edgeinternal] (vertexI3-1-6) -- (vertexI3-1-5);
	\draw[edgeinternal] (vertexI3-2-5) -- (vertexI2-2-5);
	\draw[edgeinternal] (vertexI3-2-5) -- (vertexI3-1-5);
	\draw[edgeinternal] (vertexI3-2-5) -- (vertexI3-2-4);
	\draw[edgeinternal] (vertexI3-3-4) -- (vertexI2-3-4);
	\draw[edgeinternal] (vertexI3-3-4) -- (vertexI3-2-4);
	\draw[edgeinternal] (vertexI3-3-4) -- (vertexI3-3-3);
	\draw[edgeinternal] (vertexI3-4-3) -- (vertexI2-4-3);
	\draw[edgeinternal] (vertexI3-4-3) -- (vertexI3-3-3);
	\draw[edgeinternal] (vertexI3-4-3) -- (vertexI3-4-2);
	\draw[edgeinternal] (vertexI3-5-2) -- (vertexI2-5-2);
	\draw[edgeinternal] (vertexI3-5-2) -- (vertexI3-4-2);
	\draw[edgeinternal] (vertexI3-5-2) -- (vertexI3-5-1);
	\draw[edgeinternal] (vertexI3-6-1) -- (vertexI2-6-1);
	\draw[edgeinternal] (vertexI3-6-1) -- (vertexI3-5-1);
	\draw[edgeinternal] (vertexI3-6-1) -- (vertexI3-6-0);
	\draw[edgeinternal] (vertexI4-0-6) -- (vertexI3-0-6);
	\draw[edgeinternal] (vertexI4-0-6) -- (vertexI4-0-5);
	\draw[edgeinternal] (vertexI4-1-5) -- (vertexI3-1-5);
	\draw[edgeinternal] (vertexI4-1-5) -- (vertexI4-0-5);
	\draw[edgeinternal] (vertexI4-1-5) -- (vertexI4-1-4);
	\draw[edgeinternal] (vertexI4-2-4) -- (vertexI3-2-4);
	\draw[edgeinternal] (vertexI4-2-4) -- (vertexI4-1-4);
	\draw[edgeinternal] (vertexI4-2-4) -- (vertexI4-2-3);
	\draw[edgeinternal] (vertexI4-3-3) -- (vertexI3-3-3);
	\draw[edgeinternal] (vertexI4-3-3) -- (vertexI4-2-3);
	\draw[edgeinternal] (vertexI4-3-3) -- (vertexI4-3-2);
	\draw[edgeinternal] (vertexI4-4-2) -- (vertexI3-4-2);
	\draw[edgeinternal] (vertexI4-4-2) -- (vertexI4-3-2);
	\draw[edgeinternal] (vertexI4-4-2) -- (vertexI4-4-1);
	\draw[edgeinternal] (vertexI4-5-1) -- (vertexI3-5-1);
	\draw[edgeinternal] (vertexI4-5-1) -- (vertexI4-4-1);
	\draw[edgeinternal] (vertexI4-5-1) -- (vertexI4-5-0);
	\draw[edgeinternal] (vertexI4-6-0) -- (vertexI3-6-0);
	\draw[edgeinternal] (vertexI4-6-0) -- (vertexI4-5-0);
	\draw[edgeinternal] (vertexI5-0-5) -- (vertexI4-0-5);
	\draw[edgeinternal] (vertexI5-0-5) -- (vertexI5-0-4);
	\draw[edgeinternal] (vertexI5-1-4) -- (vertexI4-1-4);
	\draw[edgeinternal] (vertexI5-1-4) -- (vertexI5-0-4);
	\draw[edgeinternal] (vertexI5-1-4) -- (vertexI5-1-3);
	\draw[edgeinternal] (vertexI5-2-3) -- (vertexI4-2-3);
	\draw[edgeinternal] (vertexI5-2-3) -- (vertexI5-1-3);
	\draw[edgeinternal] (vertexI5-2-3) -- (vertexI5-2-2);
	\draw[edgeinternal] (vertexI5-3-2) -- (vertexI4-3-2);
	\draw[edgeinternal] (vertexI5-3-2) -- (vertexI5-2-2);
	\draw[edgeinternal] (vertexI5-3-2) -- (vertexI5-3-1);
	\draw[edgeinternal] (vertexI5-4-1) -- (vertexI4-4-1);
	\draw[edgeinternal] (vertexI5-4-1) -- (vertexI5-3-1);
	\draw[edgeinternal] (vertexI5-4-1) -- (vertexI5-4-0);
	\draw[edgeinternal] (vertexI5-5-0) -- (vertexI4-5-0);
	\draw[edgeinternal] (vertexI5-5-0) -- (vertexI5-4-0);
	\draw[edgeinternal] (vertexI6-0-4) -- (vertexI5-0-4);
	\draw[edgeinternal] (vertexI6-0-4) -- (vertexI6-0-3);
	\draw[edgeinternal] (vertexI6-1-3) -- (vertexI5-1-3);
	\draw[edgeinternal] (vertexI6-1-3) -- (vertexI6-0-3);
	\draw[edgeinternal] (vertexI6-1-3) -- (vertexI6-1-2);
	\draw[edgeinternal] (vertexI6-2-2) -- (vertexI5-2-2);
	\draw[edgeinternal] (vertexI6-2-2) -- (vertexI6-1-2);
	\draw[edgeinternal] (vertexI6-2-2) -- (vertexI6-2-1);
	\draw[edgeinternal] (vertexI6-3-1) -- (vertexI5-3-1);
	\draw[edgeinternal] (vertexI6-3-1) -- (vertexI6-2-1);
	\draw[edgeinternal] (vertexI6-3-1) -- (vertexI6-3-0);
	\draw[edgeinternal] (vertexI6-4-0) -- (vertexI5-4-0);
	\draw[edgeinternal] (vertexI6-4-0) -- (vertexI6-3-0);
	
	\draw[edgeterminal] (vertexI9-0-0) -- (vertexI6-0-3);
	\draw[edgeterminal] (vertexI9-0-0) -- (vertexI6-1-2);
	\draw[edgeterminal] (vertexI9-0-0) -- (vertexI6-2-1);
	\draw[edgeterminal] (vertexI9-0-0) -- (vertexI6-3-0);
	\draw[edgeterminal] (vertexI0-9-0) -- (vertexI0-6-3);
	\draw[edgeterminal] (vertexI0-9-0) -- (vertexI1-6-2);
	\draw[edgeterminal] (vertexI0-9-0) -- (vertexI2-6-1);
	\draw[edgeterminal] (vertexI0-9-0) -- (vertexI3-6-0);
	\draw[edgeterminal] (vertexI0-0-9) -- (vertexI0-3-6);
	\draw[edgeterminal] (vertexI0-0-9) -- (vertexI1-2-6);
	\draw[edgeterminal] (vertexI0-0-9) -- (vertexI2-1-6);
	\draw[edgeterminal] (vertexI0-0-9) -- (vertexI3-0-6);
	\end{tikzpicture}
	\caption{
		\(\operatorname{SI}_{2, s, \delta}\) (solid) and \(\operatorname{SI}_{2, s, \delta}^{\operatorname{dual}}\) (dashed)\\ for \(s = 9\) and \(\delta = 3\)
	}
	\label{fig:simplified-simplex-terminal-instance-multiway-cut-problem}
	\label{fig:multiway-cut-problem-planar-duality}
\end{wrapfigure}
If \(d = 2\), then \SIds\ and in particular \SIdsdelta\ are planar.
Moreover, in the canonical embedding the required vertices lie on the outer face.
For this characteristic there is some kind of duality known for \nameref{problem:STP} and \nameref{problem:MCP}, which is for example noted in \cite{BHKM12}:
Adding at each required vertex a half-line divides the outer face into new \(|R|\) faces.
Using the notion of planar dual graphs we obtain another Steiner tree instance, where the new required vertices are given by the new \(|R|\) outer faces.
The edge costs are given by the canonical edge bijection.
It is easy to verify that each solution for \nameref{problem:STP} induces a solution for \nameref{problem:MCP} of same cost and vice versa.
Note that by our current definition the solution of \nameref{problem:STP} is an edge-set.
%An example is presented in \cref{fig:planar-dual}.

For \(\operatorname{SI}_{2, s, \delta}\) this dual Steiner tree instance is given by  \(\bm{{\operatorname{SI}}_{2, s, \delta}^{{\operatorname{dual}}}} := ((V_{2, s, \delta} \cup R_{2, s, \delta}, E_{2, s, \delta} \cup F_{2, s, \delta}), R_{2, s, \delta}, c_{{\operatorname{dual}}})\) defined as follows:
\begin{align*}
	V_{2, s, \delta} & := \left\{ v \in \triangle_{2, 2 \cdot s - 3 \cdot \delta + 1}^{\mathbb{Z}} : \max_{i \in [3]} v_{i} \leq s - \delta \right\}\\
	R_{2, s, \delta} & := \left\{ r_{i} : r_{i} = (2 \cdot s - 3 \cdot \delta + 1) \cdot e_{i} \in \triangle_{2, 2 \cdot s - 3 \cdot \delta + 1}^{\mathbb{Z}} \land i \in [3]\right\}\\
	E_{2, s, \delta} & := \left\{ \{v, w\} : v, w \in V_{2, s, \delta} \land \|v - w\|_1 = 2 \right\}\\
	F_{2, s, \delta} & := \left\{ \{r_{i}, v\} : r_{i} \in R_{2, s, \delta}, v \in V_{2, s, \delta} \land v_{i} = s - \delta \right\}
\end{align*}
\(c_{\operatorname{dual}}\) is given by the canonical edge-bijection.
An example is given in \cref{fig:simplified-simplex-terminal-instance-multiway-cut-problem}.
Note that we merged parallel edges, as we could also replace the Steiner vertices with exactly two incident edges by an edge connecting the two neighbouring vertices in \(\operatorname{SI}_{2, s, \delta}\) without changing the problem for \nameref{problem:STP} and the LP-relaxations in \BCRclass.

%\subfile{figures/figure-planar-duality}\todo{reformat pictures}
%\subfile{figures/figure-simplified-simplex-terminal-instance-multiway-cut-problem}

Since all vertices are contained in \(\triangle_{2, 2 \cdot s - 3 \cdot \delta + 1}^{\mathbb{Z}}\) and the required vertices \(R_{2, s, \delta}\) are the outermost points, by scaling the coordinates of the vertices, we obtain a canonical solution \(x : V \to \triangle_{2, 1}^{\mathbb{R}}\) for \nameref{lp:CKR}.

Recalling our simplified solution \(u : [2] \to \mathbb{R}_{\geq 0}\) for \(\operatorname{SI}_{2, s, \delta}\), where we set
\[u(1) = \frac{s - 2 \cdot \delta + 1}{2 \cdot s - 3 \cdot \delta + 1} \quad , \qquad u(2) = \frac{1}{2 \cdot s - 3 \cdot \delta + 1}\]
we see
\[\frac{\|x(v) - x(w)\|_{1}}{2} = \begin{cases}
	u(2) & \mbox{if } \{v, w\} \in E_{2, s, \delta}\\
	u(1) & \mbox{if } \{v, w\} \in F_{2, s, \delta}
\end{cases}\]
and therefore the objective value is the same.

Namely for \(s = 3 \cdot \delta - \alpha\) with \(\alpha \in \{0, 1, 2\}\) we obtain by \cref{thm:existence-solution-simplified-simplex-instance}:
\[
	\begin{aligned}
		c(u)
		& = 3 \cdot s + \frac{3}{2} \cdot \frac{(s - \delta) \cdot (s - \delta + 1)}{2 \cdot s - 3 \cdot \delta + 1}\\
		& = 3 \cdot s + \frac{1}{6} \cdot \frac{(3 \cdot s - 3 \cdot \delta) \cdot (3 \cdot s - 3 \cdot \delta + 3)}{2 \cdot s - 3 \cdot \delta + 1}\\
		& = 3 \cdot s + \frac{1}{6} \cdot \frac{(3 \cdot s - (s + \alpha)) \cdot (3 \cdot s - (s + \alpha) + 3)}{2 \cdot s - (s + \alpha) + 1}\\
		& = 3 \cdot s + \frac{1}{6} \cdot \frac{(2 \cdot s - \alpha) \cdot (2 \cdot s - \alpha + 3)}{s - \alpha + 1}\\
		& = 3 \cdot s + \frac{1}{3} \cdot \frac{2 \cdot s^{2} + (3 - 2 \cdot \alpha) \cdot s + \frac{\alpha^{2} - 3 \cdot \alpha}{2}}{s - \alpha + 1}\\
		& = 3 \cdot s + \frac{1}{3} \cdot \begin{cases}
			\frac{2 \cdot s^{2} + 3 \cdot s}{s + 1} & \mbox{if } \alpha = 0\\
			\frac{2 \cdot s^{2} + s - 1}{s} & \mbox{if } \alpha = 1\\
			\frac{2 \cdot s^{2} - s - 1}{s - 1} & \mbox{if } \alpha = 2
		\end{cases}
		\qquad = \begin{cases}
			\frac{11 \cdot s^{2} + 12 \cdot s}{3 \cdot (s + 1)} & \mbox{if } \alpha = 0\\[3pt]
			\frac{11 \cdot s^{2} + s - 1}{3 \cdot s} & \mbox{if } \alpha = 1\\[3pt]
			\frac{11 \cdot s + 1}{3} & \mbox{if } \alpha = 2
		\end{cases}
	\end{aligned}
\]
By \cref{thm:simplex-terminal-instance-optimum-solution} and the earlier remarks on the relation of \nameref{problem:STP} and \nameref{problem:MCP} we have\\ \(\operatorname{opt}_{\operatorname{MCP}}({\operatorname{SI}}_{2, s, \delta}^{{\operatorname{dual}}}) = \operatorname{opt}_{\operatorname{STP}}(\operatorname{SI}_{2, s, \delta}) = 4 \cdot s\).
Thus:
\[\operatorname{gap}_{\operatorname{CKR}, \operatorname{MCP}}({\operatorname{SI}}_{2, s, \delta}^{{\operatorname{dual}}}) = \operatorname{gap}_{\operatorname{BCR}, \operatorname{STP}}({\operatorname{SI}}_{2, s, \delta}^{{\operatorname{dual}}}) = \begin{cases}
	\frac{12 \cdot s + 12}{11 \cdot s + 12} & \mbox{if } \alpha = 0\\[3pt]
	\frac{12 \cdot s^{2}}{11 \cdot s^{2} + s - 1} & \mbox{if } \alpha = 1\\[3pt]
	\frac{12 \cdot s}{11 \cdot s + 1} & \mbox{if } \alpha = 2
\end{cases}\]
Let \(q := 2 \cdot s - 3 \cdot \delta + 1 = s - \alpha + 1\).
Then we obtain:
\[\operatorname{gap}_{\operatorname{CKR}, \operatorname{MCP}}({\operatorname{SI}}_{2, s, \delta}^{{\operatorname{dual}}}) = \begin{cases}
	\frac{12 \cdot (q - 1) + 12}{11 \cdot (q - 1) + 12} & \mbox{if } \alpha = 0\\[3pt]
	\frac{12 \cdot q^{2}}{11 \cdot q^{2} + q - 1} & \mbox{if } \alpha = 1\\[3pt]
	\frac{12 \cdot (q + 1)}{11 \cdot (q + 1) + 1} & \mbox{if } \alpha = 2
\end{cases} = \begin{cases}
	\frac{12 \cdot q}{11 \cdot q + 1} & \mbox{if } q \equiv 1 \mod 3\\[3pt]
	\frac{12 \cdot q^{2}}{11 \cdot q^{2} + q - 1} & \mbox{if } q \equiv 2 \mod 3\\[3pt]
	\frac{12 \cdot q + 12}{11 \cdot q + 12} & \mbox{if } q \equiv 0 \mod 3
\end{cases}\]
This shows that our simplex based Steiner tree instances infer worst case instances for \nameref{lp:CKR} restricted to Steiner tree instances with \(3\) required vertices.

	\pagebreak

\section{Final remarks}\label{sec:conclusion}

We have seen that our simplex based Steiner tree instances yield large gaps regarding the optimum value for \BCRclass\ towards \BCRimprovedclass, i.e.\ \BCRclass\ enhanced with Steiner vertex degree constraints.
Using these instances we are able to improve the best known lower bounds of the gaps of \BCRclass\ towards \BCRimprovedclass, \HYP\ and \nameref{problem:STP}.

\begin{wraptable}{r}{0.35\textwidth}
	\centering
	\begin{tabular}{ccc}
		%d (s = d) & \(\operatorname{gap}_{\operatorname{\mathcal{BCR}}, \operatorname{\mathcal{BCR}^{+}}}(\operatorname{SI}_{d, d})\) & \(\operatorname{gap}_{\operatorname{\mathcal{BCR}}, \operatorname{\mathcal{BCR}^{+}}}(\operatorname{SI}_{d, d}^{l \leq 2})\)\\
		& \multicolumn{2}{c}{\(\operatorname{gap}_{\operatorname{\mathcal{BCR}}, \operatorname{\mathcal{BCR}^{+}}}(\cdot)\)}\\[3pt]
		d (s = d) & \(\operatorname{SI}_{d, d}\) & \(\operatorname{SI}_{d, d}^{l \leq 2}\)\\[6pt]
		1 & 1 & 1\\
		2 & 1.06666 & 1.06666\\
		3 & 1.09459 & 1.09090\\
		4 & 1.12116 & 1.10344\\
		5 & 1.13939 & 1.12612\\
		6 & 1.15042 & 1.13513\\
		7 & 1.16094 & 1.13953\\
		8 & 1.16883 & 1.14927\\
		9 & 1.17340 & 1.15384
	\end{tabular}
	\caption{}
	\label{table:simplex-instances-integrality-gaps-comparison}
\end{wraptable}

Recall that we were not able to prove the optimum solution values of \BCRclass\ on our instances as the underlying algebraic terms in general optimum solutions are apparently highly interdependent and complicated.
Instead we determined solutions of small value using only a restricted set of edges, namely the ones on level \(1\) and \(2\).
LP-solver solutions for small \(s\) suggest that this restriction actually makes a difference for the achieved gap of \BCRclass\ towards \BCRimprovedclass\ as can be seen in \cref{table:simplex-instances-integrality-gaps-comparison} (the values are all truncated after the fifth decimal place).
Here we denote by \(\operatorname{SI}_{d, d}^{l \leq 2}\) the Steiner tree instance \(\operatorname{SI}_{d, d}\) restricted to edges only on level at most \(2\).
As already mentioned, we were only able to compute solution values up to \(d = 9\).

On a different note, the questions whether \(\operatorname{gap}_{\operatorname{\mathcal{BCR}}, \operatorname{STP}} < 2\) and \(\operatorname{gap}_{\operatorname{\mathcal{BCR}^{+}}, \operatorname{STP}} < 2\) are still open.
To the best of our knowledge, it is even not known whether \(\operatorname{gap}_{\operatorname{\mathcal{BCR}}, \operatorname{\mathcal{BCR}^{+}}} < 2\) and \(\operatorname{gap}_{\operatorname{\mathcal{BCR}}, \operatorname{\mathcal{HYP}}} < 2\).
For a positive answer one would also have to show this for our simplex instances \SIds.
Hence, a natural question is whether there exists an \(\varepsilon > 0\) such that for all \(d, s \in \mathbb{N}\) we have:
\[\operatorname{gap}_{\operatorname{\mathcal{BCR}}, \operatorname{\mathcal{BCR}^{+}}}(\operatorname{SI}_{d, s}) = \operatorname{gap}_{\operatorname{\mathcal{BCR}}, \operatorname{\mathcal{HYP}}}(\operatorname{SI}_{d, s}) = \operatorname{gap}_{\operatorname{\mathcal{BCR}}, \operatorname{STP}}(\operatorname{SI}_{d, s}) \leq 2 - \varepsilon\]

Indeed, as \SIds\ seem to be instances with large gaps of \BCRclass\ towards \BCRimprovedclass, \HYP\ and \nameref{problem:STP} this might involve some key arguments for general instances.

	\pagebreak

\appendix

\section{Set cover based Steiner tree instances\texorpdfstring{\\}{ }with large integrality gaps}\label{sec:set-cover-based-steiner-tree-instances}

In this section we generalize the instances in \cite{BGRS13} to arbitrary set cover instances and adapt the proofs in \cite{BGRS13} to \(\operatorname{\mathcal{BCR}^{+}}\), which is straightforward.

In the NP-hard \textit{set cover problem} one is given a finite set \(\mathcal{S}\) consisting of finite non-empty sets \(S \in \mathcal{S}\) and asked for a subset \(I \subseteq \mathcal{S}\) of minimum cardinality such that \(\bigcup_{S \in I} S = \bigcup_{S \in \mathcal{S}} S\), i.e.\ the chosen sets in \(I\) already cover all elements of the sets in \(\mathcal{S}\).
We call \(\mathcal{S}\) a \textit{set cover instance} and define the set \(\mathcal{U}_{\mathcal{S}} := \bigcup_{S \in \mathcal{S}} S\) which contains all elements.
\(\mathcal{U}_{\mathcal{S}}\) is often called \textit{universe}.
For analyzing the set cover problem a useful notion is the \textit{frequency} of an element, namely the number of sets which contain this element.
More formally, for each \(e \in \mathcal{U}_{\mathcal{S}}\) we define \(f_{\mathcal{S}}(e) := |\{S \in \mathcal{S} : e \in S\}|\).
For further information about the set cover problem one can consult for example \cite[Ch.\ 2]{Vaz01}.

In the following we will use the notation \(x|e\) to extend a vector \(x\) by adding \(e\) as the last entry, i.e.\ \(x|e\) has one dimension more than \(x\).
Furthermore, we use the convention that \(A^{0}\) for set \(A\) is the set consisting only of an empty vector.

\begin{definition}\label{def:set-cover-graph-and-terminal-instance-of-size-p}
	Let \(\mathcal{S}\) be a set cover instance and \(p \in \mathbb{N}\).
	We define the \textbf{set cover graph of size \(\bm{p}\)} to be the undirected graph \(\bm{{\operatorname{SCG}}_{\mathcal{S}, p}} = (V, E)\), given as follows:
	\[
	\begin{aligned}
	V_{0} & := \{r\}\\
	V_{i} & := {\mathcal{U}_{\mathcal{S}}}^{i - 1} \times \mathcal{S}  & i \in [p]\\
	V_{p + 1} & := {\mathcal{U}_{\mathcal{S}}}^{p - 1} \times \mathcal{U}_{\mathcal{S}}
	\end{aligned}
	\]
	\[
	\begin{aligned}
	E_{0} & := \{\ \{r, (x, S)\} && : \ (x, S) \in {\mathcal{U}_{\mathcal{S}}}^{0} \times \mathcal{S} \ \}\\
	E_{i} & := \{\ \{(x, S), (x|e, S')\} && : \ (x, S) \in {\mathcal{U}_{\mathcal{S}}}^{i - 1} \times \mathcal{S} \land e \in S \land S' \in \mathcal{S} \ \} & i \in [p - 1]\\
	E_{p} & := \{\ \{(x, S), (x, e)\} && : \ (x, S) \in {\mathcal{U}_{\mathcal{S}}}^{p - 1} \times \mathcal{S} \land e \in S \ \}
	\end{aligned}
	\]
	
	\[
	\begin{aligned}
	V & := \bigcup_{i = 0}^{p + 1} V_{i} & \qquad E & := \bigcup_{i = 0}^{p} E_{i}
	\end{aligned}
	\]
	
	We define the \textbf{set cover based Steiner tree instance of size \(\bm{p}\)} to be the following Steiner tree instance:
	\[\bm{{\operatorname{SCI}}_{\mathcal{S}, p}} = (\operatorname{SCG}_{\mathcal{S}, p}, V_{0} \cup V_{p + 1}, c \equiv 1)\]
\end{definition}

\begin{figure}
	\centering

	\begin{tikzpicture}[scale=0.42]
		\tikzset{square/.style={regular polygon, regular polygon sides=4}}
		
		\tikzset{vertex/.style={circle, thick, draw=black, fill=white, minimum size=0pt, inner sep=2pt, outer sep=3pt}}
		\tikzset{terminal/.style={square, line width=1, fill=black}}

		\pgfmathsetmacro{\heightlevelzero}{20}
		\pgfmathsetmacro{\heightlevelone}{16}
		\pgfmathsetmacro{\heightleveltwo}{10}
		\pgfmathsetmacro{\heightlevelthree}{4}
		
		\node[vertex, terminal] (r) at (0, \heightlevelzero) {};
		
		%
		% level p = 1
		%
		\node[vertex] (ES12) at (-12, \heightlevelone) {};
		\node[vertex] (ES13) at (+0, \heightlevelone) {};
		\node[vertex] (ES23) at (+12, \heightlevelone) {};
			
		%
		% level p = 2
		%
		\node[vertex] (E1S12) at (-16, \heightleveltwo) {};
		\node[vertex] (E1S13) at (-12, \heightleveltwo) {};
		\node[vertex] (E1S23) at (-8, \heightleveltwo) {};
		
		\node[vertex] (E2S12) at (-4, \heightleveltwo) {};
		\node[vertex] (E2S13) at (+0, \heightleveltwo) {};
		\node[vertex] (E2S23) at (+4, \heightleveltwo) {};
		
		\node[vertex] (E3S12) at (+8, \heightleveltwo) {};
		\node[vertex] (E3S13) at (+12, \heightleveltwo) {};
		\node[vertex] (E3S23) at (+16, \heightleveltwo) {};
		
		%
		% level p = 3
		%
		\node[vertex] (E11S12) at (-17, \heightlevelthree) {};
		\node[vertex] (E11S13) at (-16, \heightlevelthree) {};
		\node[vertex] (E11S23) at (-15, \heightlevelthree) {};
		
		\node[vertex] (E12S12) at (-13, \heightlevelthree) {};
		\node[vertex] (E12S13) at (-12, \heightlevelthree) {};
		\node[vertex] (E12S23) at (-11, \heightlevelthree) {};
		
		\node[vertex] (E13S12) at (-9, \heightlevelthree) {};
		\node[vertex] (E13S13) at (-8, \heightlevelthree) {};
		\node[vertex] (E13S23) at (-7, \heightlevelthree) {};
		
		\node[vertex] (E21S12) at (-5, \heightlevelthree) {};
		\node[vertex] (E21S13) at (-4, \heightlevelthree) {};
		\node[vertex] (E21S23) at (-3, \heightlevelthree) {};
		
		\node[vertex] (E22S12) at (-1, \heightlevelthree) {};
		\node[vertex] (E22S13) at (+0, \heightlevelthree) {};
		\node[vertex] (E22S23) at (+1, \heightlevelthree) {};
		
		\node[vertex] (E23S12) at (+3, \heightlevelthree) {};
		\node[vertex] (E23S13) at (+4, \heightlevelthree) {};
		\node[vertex] (E23S23) at (+5, \heightlevelthree) {};
		
		\node[vertex] (E31S12) at (+7, \heightlevelthree) {};
		\node[vertex] (E31S13) at (+8, \heightlevelthree) {};
		\node[vertex] (E31S23) at (+9, \heightlevelthree) {};
		
		\node[vertex] (E32S12) at (+11, \heightlevelthree) {};
		\node[vertex] (E32S13) at (+12, \heightlevelthree) {};
		\node[vertex] (E32S23) at (+13, \heightlevelthree) {};
		
		\node[vertex] (E33S12) at (+15, \heightlevelthree) {};
		\node[vertex] (E33S13) at (+16, \heightlevelthree) {};
		\node[vertex] (E33S23) at (+17, \heightlevelthree) {};
		
		%
		% level p + 1 = 4, terminals
		%
		\node[vertex, terminal] (E11E1) at (-17, 0) {};
		\node[vertex, terminal] (E11E2) at (-16, 0) {};
		\node[vertex, terminal] (E11E3) at (-15, 0) {};
		
		\node[vertex, terminal] (E12E1) at (-13, 0) {};
		\node[vertex, terminal] (E12E2) at (-12, 0) {};
		\node[vertex, terminal] (E12E3) at (-11, 0) {};
		
		\node[vertex, terminal] (E13E1) at (-9, 0) {};
		\node[vertex, terminal] (E13E2) at (-8, 0) {};
		\node[vertex, terminal] (E13E3) at (-7, 0) {};
		
		\node[vertex, terminal] (E21E1) at (-5, 0) {};
		\node[vertex, terminal] (E21E2) at (-4, 0) {};
		\node[vertex, terminal] (E21E3) at (-3, 0) {};
		
		\node[vertex, terminal] (E22E1) at (-1, 0) {};
		\node[vertex, terminal] (E22E2) at (+0, 0) {};
		\node[vertex, terminal] (E22E3) at (+1, 0) {};
		
		\node[vertex, terminal] (E23E1) at (+3, 0) {};
		\node[vertex, terminal] (E23E2) at (+4, 0) {};
		\node[vertex, terminal] (E23E3) at (+5, 0) {};
		
		\node[vertex, terminal] (E31E1) at (+7, 0) {};
		\node[vertex, terminal] (E31E2) at (+8, 0) {};
		\node[vertex, terminal] (E31E3) at (+9, 0) {};
		
		\node[vertex, terminal] (E32E1) at (+11, 0) {};
		\node[vertex, terminal] (E32E2) at (+12, 0) {};
		\node[vertex, terminal] (E32E3) at (+13, 0) {};
		
		\node[vertex, terminal] (E33E1) at (+15, 0) {};
		\node[vertex, terminal] (E33E2) at (+16, 0) {};
		\node[vertex, terminal] (E33E3) at (+17, 0) {};
		
		%
		% edges between level 0 and 1
		%
		\draw[] (r) -- (ES12);
		\draw[] (r) -- (ES13);
		\draw[] (r) -- (ES23);
		
		%
		% edges between level 1 and 2
		%
		
		% ES12
		\draw[] (ES12) -- (E1S12);
		\draw[] (ES12) -- (E1S13);
		\draw[] (ES12) -- (E1S23);
		
		\draw[] (ES12) -- (E2S12);
		\draw[] (ES12) -- (E2S13);
		\draw[] (ES12) -- (E2S23);
		
		% ES13
		\draw[] (ES13) -- (E1S12);
		\draw[] (ES13) -- (E1S13);
		\draw[] (ES13) -- (E1S23);
		
		\draw[] (ES13) -- (E3S12);
		\draw[] (ES13) -- (E3S13);
		\draw[] (ES13) -- (E3S23);
		
		% ES23
		\draw[] (ES23) -- (E2S12);
		\draw[] (ES23) -- (E2S13);
		\draw[] (ES23) -- (E2S23);
		
		\draw[] (ES23) -- (E3S12);
		\draw[] (ES23) -- (E3S13);
		\draw[] (ES23) -- (E3S23);
		
		%
		% edges between level 2, 3
		%
		
		% S12
		\foreach \i in {1, ..., 3} {
			\draw[] (E\i S12) -- (E\i1S12);
			\draw[] (E\i S12) -- (E\i1S13);
			\draw[] (E\i S12) -- (E\i1S23);
			
			\draw[] (E\i S12) -- (E\i2S12);
			\draw[] (E\i S12) -- (E\i2S13);
			\draw[] (E\i S12) -- (E\i2S23);
		}
	
		% S13
		\foreach \i in {1, ..., 3} {
			\draw[] (E\i S13) -- (E\i1S12);
			\draw[] (E\i S13) -- (E\i1S13);
			\draw[] (E\i S13) -- (E\i1S23);
			
			\draw[] (E\i S13) -- (E\i3S12);
			\draw[] (E\i S13) -- (E\i3S13);
			\draw[] (E\i S13) -- (E\i3S23);
		}
	
		% S23
		\foreach \i in {1, ..., 3} {
			\draw[] (E\i S23) -- (E\i2S12);
			\draw[] (E\i S23) -- (E\i2S13);
			\draw[] (E\i S23) -- (E\i2S23);
			
			\draw[] (E\i S23) -- (E\i3S12);
			\draw[] (E\i S23) -- (E\i3S13);
			\draw[] (E\i S23) -- (E\i3S23);
		}
	
		%
		% edges between level 2, 3
		%
		
		% S12
		\foreach \i in {1, ..., 3} {
			\foreach \j in {1, ..., 3} {
				\draw[] (E\i\j S12) -- (E\i\j E1);
				\draw[] (E\i\j S12) -- (E\i\j E2);
			}
		}
		
		% S13
		\foreach \i in {1, ..., 3} {
			\foreach \j in {1, ..., 3} {
				\draw[] (E\i\j S13) -- (E\i\j E1);
				\draw[] (E\i\j S13) -- (E\i\j E3);
			}
		}
		
		% S23
		\foreach \i in {1, ..., 3} {
			\foreach \j in {1, ..., 3} {
				\draw[] (E\i\j S23) -- (E\i\j E2);
				\draw[] (E\i\j S23) -- (E\i\j E3);
			}
		}
		
		% separating lines
		\draw[dashed, thick] ($0.5*(E23E3) + 0.5*(E31E1) - (0, 1)$) -- ($0.5*(E23E3) + 0.5*(E31E1) + (0, \heightleveltwo) + (0, 1)$);
		\draw[dashed, thick] ($0.5*(E13E3) + 0.5*(E21E1) - (0, 1)$) -- ($0.5*(E13E3) + 0.5*(E21E1) + (0, \heightleveltwo) + (0, 1)$);
		
		\draw[dashed, thick] ($0.5*(E11E3) + 0.5*(E12E1) - (0, 1)$) -- ($0.5*(E11E3) + 0.5*(E12E1) + (0, \heightlevelthree) + (0, 1)$);
		\draw[dashed, thick] ($0.5*(E12E3) + 0.5*(E13E1) - (0, 1)$) -- ($0.5*(E12E3) + 0.5*(E13E1) + (0, \heightlevelthree) + (0, 1)$);
		
		\draw[dashed, thick] ($0.5*(E21E3) + 0.5*(E22E1) - (0, 1)$) -- ($0.5*(E21E3) + 0.5*(E22E1) + (0, \heightlevelthree) + (0, 1)$);
		\draw[dashed, thick] ($0.5*(E22E3) + 0.5*(E23E1) - (0, 1)$) -- ($0.5*(E22E3) + 0.5*(E23E1) + (0, \heightlevelthree) + (0, 1)$);
		
		\draw[dashed, thick] ($0.5*(E31E3) + 0.5*(E32E1) - (0, 1)$) -- ($0.5*(E31E3) + 0.5*(E32E1) + (0, \heightlevelthree) + (0, 1)$);
		\draw[dashed, thick] ($0.5*(E32E3) + 0.5*(E33E1) - (0, 1)$) -- ($0.5*(E32E3) + 0.5*(E33E1) + (0, \heightlevelthree) + (0, 1)$);
	\end{tikzpicture}

	\caption{
		\(\operatorname{SCI}_{\mathcal{S}, p}\) for the set cover instance \(\mathcal{S} = \{\{1, 2\}, \{1, 3\}, \{2, 3\}\}\) and \(p = 3\)
	}

	\label{fig:set-cover-instance-p-3}

\end{figure}
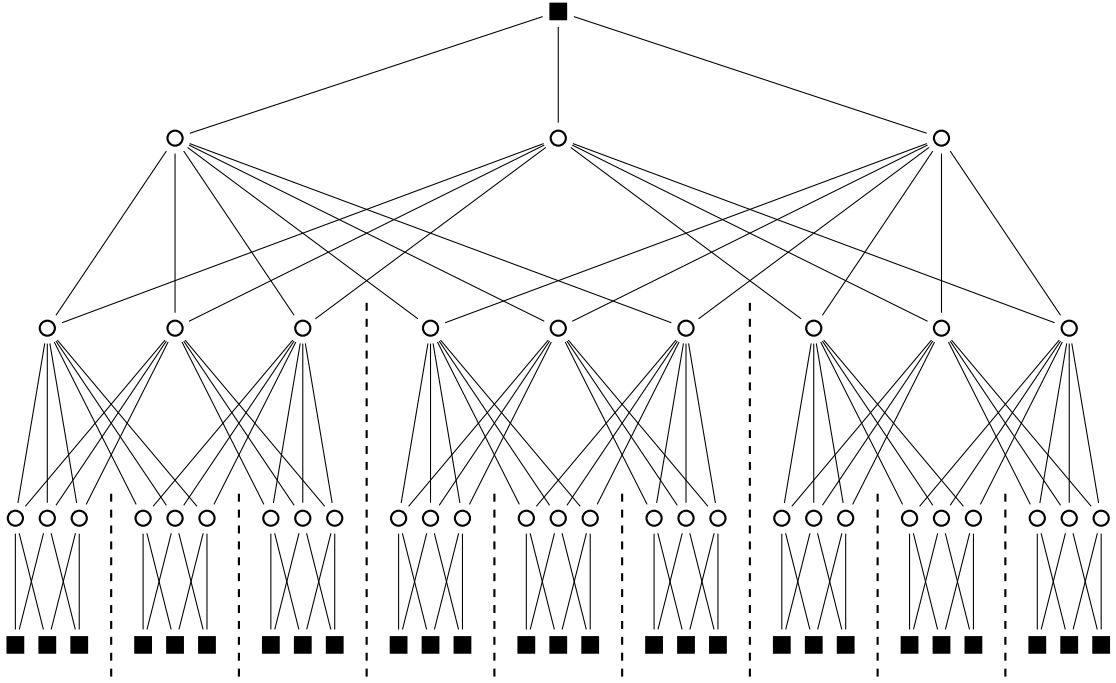

An example is given in \cref{fig:set-cover-instance-p-3}.
With each vertex \(v \in V\) of a set cover graph we can naturally associate a \textit{level}.
The vertex \(r\) is on level \(0\) and the other required vertices on level \(p + 1\).
By definition the set cover graph only contains edges between consecutive levels.
When considering the bidirected graph that arises from the set cover graph we call edges directed from level \(i\) to \(i + 1\) \textit{downward edges} and all other edges \textit{upward edges}.

\begin{lemma}\label{thm:set-cover-terminal-instance-optimum-integer-value}
	Let \(\operatorname{SCI}_{\mathcal{S}, p}\) be a set cover based Steiner tree instance and \(I \subseteq \mathcal{S}\) be an optimum solution for the set cover instance \(\mathcal{S}\).
	Then:
	\[\operatorname{opt}_{\operatorname{STP}}(\operatorname{SCI}_{\mathcal{S}, p}) = \left( 1 + \frac{1}{|\mathcal{U}_{\mathcal{S}}| - 1} \cdot |I| \right) \cdot \left( |\mathcal{U}_{\mathcal{S}}|^{p} - 1 \right) + 1\]
\end{lemma}
\begin{proof}
	Let \(T\) be a minimum Steiner tree for \(\operatorname{SCI}_{\mathcal{S}, p}\).
	
	We can assume w.l.o.g.\ that the arborescence \(T^{r\to}\) rooted in \(r\) contains only downward edges:
	It is clear that no vertex on level \(0\), \(1\) or \(p+1\) needs to be the head of an upward edge contained in \(T^{r\to}\).
	Assume \((x|e, S)\) is the head of an upward edge on level \(1 < i < p + 1\).
	Since \(T^{r\to}\) is connected there has to be a downward edge \(((x, S'), (x|e, S''))\) with \(e \in S'\).
	Remove the upward edge and add the downward edge \(((x, S'), (x|e, S))\).
	This preserves the connectivity and solution cost as \(c \equiv 1\).
	
	Hence we have \(|V_{p + 1}| = |\mathcal{U}_{\mathcal{S}}|^{p}\) edges in \(T\) between level \(p + 1\) and \(p\).
	
	Let \(1 \leq i \leq p\) and consider the vertices \((x, S) \in V_{i}\) for a fixed \(x \in {\mathcal{U}_{\mathcal{S}}}^{i - 1}\).
	Then there have to be at least \(|I|\) incoming downward edges in \(T^{r\to}\) to be able to connect all required vertices at level \(p + 1\), otherwise \(I\) would be not an optimum solution for the set cover instance \(\mathcal{S}\).
	On the other hand this is sufficient for connectivity by the definition of a set cover solution.
	
	Now we only have to count the edges in \(T^{r\to}\) as \(c \equiv 1\):
	\begin{align*}
		|E(T^{r\to})| & = |\mathcal{U}_{\mathcal{S}}|^{p} + \sum_{i = 1}^{p} |\mathcal{U}_{\mathcal{S}}|^{i - 1} \cdot |I|\\
		%& = |\mathcal{U}_{\mathcal{S}}|^{p} + |I| \cdot \sum_{i = 0}^{p - 1} |\mathcal{U}_{\mathcal{S}}|^{i}\\
		& = |\mathcal{U}_{\mathcal{S}}|^{p} + |I| \cdot \frac{|\mathcal{U}_{\mathcal{S}}|^{p} - 1}{|\mathcal{U}_{\mathcal{S}}| - 1}
		= \left( 1 + \frac{1}{|\mathcal{U}_{\mathcal{S}}| - 1} \cdot |I| \right) \cdot \left( |\mathcal{U}_{\mathcal{S}}|^{p} - 1 \right) + 1 \qedhere
	\end{align*}
\end{proof}

\begin{lemma}\label{thm:set-cover-terminal-instance-fractional-value-bound}
	Let \(\operatorname{SCI}_{\mathcal{S}, p}\) be a set cover based Steiner tree instance.
	If all elements in \(\mathcal{U}_{\mathcal{S}}\) have the same frequency \(f_{\mathcal{S}}\), then:
	\[\operatorname{opt}_{\operatorname{BCR}^{+}}(\operatorname{SCI}_{\mathcal{S}, p}) \leq \left( 1 + \frac{1}{|\mathcal{U}_{\mathcal{S}}| - 1} \cdot \frac{|\mathcal{S}|}{f_{\mathcal{S}}} \right) \cdot \left( |\mathcal{U}_{\mathcal{S}}|^{p} - 1 \right) + 1\]
\end{lemma}
\begin{proof}
	Write \(\operatorname{SCI}_{\mathcal{S}, p} = ((V, E), R, c)\).
	We will define \(u : E \to \mathbb{R}_{\geq 0}\), \(f_{r} : E_{\leftrightarrow} \to \mathbb{R}_{\geq 0}\) and \(g : R \setminus \{r\} \times E_{\leftrightarrow} \to \mathbb{R}_{\geq 0}\) such that \((u, f_{r}, g)\) is a solution of \MCFRimproved\ with the claimed upper bound as cost proving the inequality.
	We define \(u\), \(f_{r}\) and \(g\) as follows:\\
	The root is \(r\) as already indicated.
	All upward edges have value \(0\) with respect to \(u\), \(f_{r}\) and \(g\).
	Downward edges between level \(0\) and level \(1\) and between \(p\) and \(p + 1\) each have value \(f_{\mathcal{S}}^{-1}\) under \(u\) and \(f_{r}\).
	The other downward edges each have value \(f_{\mathcal{S}}^{-2}\) under \(u\) and \(f_{r}\).
	
	For \((x, e) \in {\mathcal{U}_{\mathcal{S}}}^{p - 1} \times \mathcal{U}_{\mathcal{S}}\) we define \(g_{(x, e)}\) as follows:
	\begin{align*}
		g_{(x, e)}(r, (x, S)) & = \llbracket e \in S \rrbracket \cdot f_{\mathcal{S}}^{-1} & \{r, (x, S)\} & \in E_{0}\\
		g_{(x, e)}((x, S), (x|e, S')) & = \llbracket e \in S \land e \in S' \rrbracket \cdot f_{\mathcal{S}}^{-2} & \{(x, S), (x|e, S')\} & \in E_{i}\\
		g_{(x, e)}((x, S), (x, e)) & = \llbracket e \in S \rrbracket \cdot f_{\mathcal{S}}^{-1} & \{(x, S), (x, e)\} & \in E_{p}
	\end{align*}
	It is easy to see that \(g_{(x, e)}\) is a balance-flow for \(b_{(x, e)}(v) := \llbracket v = r \rrbracket - \llbracket v = (x, e) \rrbracket\) respecting the edge-usage bounds \(f_{r}\) as each element has frequency \(f_{\mathcal{S}}\).
	Therefore, \((u, f_{r}, g)\) is a solution of \nameref{lp:MCFR}.
	
	For proving that \((u, f_{r}, g)\) is a solution of \MCFRimproved\ we need to show the flow-balance constraints at each Steiner vertex:
	For each \((x, S) \in V_{i} = {\mathcal{U}_{\mathcal{S}}}^{i - 1} \times \mathcal{S}\) with \(i \in [p]\) we have \(f_{r}(\delta_{\operatorname{SCG}_{\mathcal{S}, p}}^{-}((x, S))) = f_{\mathcal{S}}^{-1}\) and for the outgoing edges we compute:
	\[
	f_{r}(\delta_{\operatorname{SCG}_{\mathcal{S}, p}}^{+}((x, S))) = \begin{cases}
	|S| \cdot |\mathcal{S}| \cdot f_{\mathcal{S}}^{-2} & \mbox{if } i \neq p\\
	|S| \cdot f_{\mathcal{S}}^{-1} & \mbox{if } i = p
	\end{cases}
	\]
	As \(|\mathcal{S}| \geq f_{\mathcal{S}}\) and \(|S| \geq 1\) we obtain \(f_{r}(\delta_{\operatorname{SCG}_{\mathcal{S}, p}}^{+}((x, S))) \geq f_{r}(\delta_{\operatorname{SCG}_{\mathcal{S}, p}}^{-}((x, S)))\).\\
	Hence, \((u, f_{r}, g)\) is a solution of \MCFRimproved.
	For the solution cost we compute:
	\begin{align*}
		\sum_{e \in E} c(e) \cdot u(e)
		= \sum_{i = 1}^{p} \sum_{e \in E_{i}} u(e)
		& = f_{\mathcal{S}}^{-1} \cdot (|E_{0}| + |E_{p}|) + f_{\mathcal{S}}^{-2} \cdot \sum_{i = 1}^{p - 1} |E_{i}|\\
		& = f_{\mathcal{S}}^{-1} \cdot (|V_{1}| + |V_{p + 1}| \cdot f_{\mathcal{S}}) + f_{\mathcal{S}}^{-2} \cdot \sum_{i = 1}^{p - 1} |V_{i + 1}| \cdot f_{\mathcal{S}}\\
		& = f_{\mathcal{S}}^{-1} \cdot |\mathcal{S}| + |\mathcal{U}_{\mathcal{S}}|^{p} + f_{\mathcal{S}}^{-1} \cdot |\mathcal{S}| \cdot \sum_{i = 1}^{p - 1} |\mathcal{U}_{\mathcal{S}}|^{i}\\
		& = f_{\mathcal{S}}^{-1} \cdot |\mathcal{S}| + |\mathcal{U}_{\mathcal{S}}|^{p} + f_{\mathcal{S}}^{-1} \cdot |\mathcal{S}| \cdot \left( \frac{|\mathcal{U}_{\mathcal{S}}|^{p} - 1}{|\mathcal{U}_{\mathcal{S}}| - 1} - 1 \right)\\
		& = |\mathcal{U}_{\mathcal{S}}|^{p} + f_{\mathcal{S}}^{-1} \cdot |\mathcal{S}| \cdot \frac{|\mathcal{U}_{\mathcal{S}}|^{p} - 1}{|\mathcal{U}_{\mathcal{S}}| - 1} \qedhere%\\
		%& = \left( 1 + \frac{1}{|\mathcal{U}_{\mathcal{S}}| - 1} \cdot \frac{|\mathcal{S}|}{f_{\mathcal{S}}} \right) \cdot \left( |\mathcal{U}_{\mathcal{S}}|^{p} - 1 \right) + 1 \qedhere
	\end{align*}
\end{proof}

\begin{lemma}\label{thm:set-cover-based-steiner-tree-instance-gap-lower-bound}
	Let \(\operatorname{SCI}_{\mathcal{S}, p}\) be a set cover based Steiner tree instance.
	Then:
	\[\operatorname{opt}_{\operatorname{\mathcal{BCR}^{+}}}(\operatorname{SCI}_{\mathcal{S}, p}) \leq \left( 1 + \frac{1}{|\mathcal{U}_{\mathcal{S}}| - 1} \cdot \frac{|\mathcal{S}|}{\min_{e \in \mathcal{U}} f_{\mathcal{S}}(e)} \right) \cdot \left( |\mathcal{U}_{\mathcal{S}}|^{p} - 1 \right) + 1\]
	Let \(I \subseteq \mathcal{S}\) an optimum solution for the set cover instance \(\mathcal{S}\).
	Then:
	\[\operatorname{gap}_{\operatorname{\mathcal{BCR}^{+}}, \operatorname{STP}}(\operatorname{SCI}_{\mathcal{S}, p}) \geq \frac{\left( 1 + \frac{1}{|\mathcal{U}_{\mathcal{S}}| - 1} \cdot |I| \right) \cdot \left( |\mathcal{U}_{\mathcal{S}}|^{p} - 1 \right) + 1}{\left( 1 + \frac{1}{|\mathcal{U}_{\mathcal{S}}| - 1} \cdot \frac{|\mathcal{S}|}{\min_{e \in \mathcal{U}} f_{\mathcal{S}}(e)} \right) \cdot \left( |\mathcal{U}_{\mathcal{S}}|^{p} - 1 \right) + 1}\]
\end{lemma}
\begin{proof}
	For a given set cover instance \(\mathcal{S}\) we obtain a new set cover instance \(\mathcal{S}'\) where all elements have the same frequency by choosing for each element \(\min_{e \in \mathcal{U}_{\mathcal{S}}} f_{\mathcal{S}}(e)\) sets in \(\mathcal{S}\) which contain that element, removing the element from all other sets which contain it and discarding resulting empty sets.
	The set cover graph obtained from this new set cover instance is then subgraph of the original one.
	The two bounds then follow with \cref{thm:set-cover-terminal-instance-optimum-integer-value}, \cref{thm:set-cover-terminal-instance-fractional-value-bound} and the fact that \(|\mathcal{S}'| \leq |\mathcal{S}|\) and \(\mathcal{U}_{\mathcal{S}'} = \mathcal{U}_{\mathcal{S}}\).
\end{proof}

%\paragraph{Remark.} For the case \(p = 1\) we obtain the following bound:
%\[\operatorname{gap}_{\operatorname{\mathcal{BCR}^{+}}, \operatorname{STP}}(\operatorname{SCI}_{\mathcal{S}, 1}) \geq \frac{|\mathcal{U}_{\mathcal{S}}| + |I|}{|\mathcal{U}_{\mathcal{S}}| + \frac{|\mathcal{S}|}{\min_{e \in \mathcal{U}} f_{\mathcal{S}}(e)}}\]

\begin{theorem}\label{thm:set-cover-bcr-improved-integrality-gap-lower-bound}
	Let \(\mathcal{S}\) be a set cover instance and \(I \subseteq \mathcal{S}\) be an optimum solution for the set cover instance \(\mathcal{S}\).
	Then
	\[\operatorname{gap}_{\operatorname{\mathcal{BCR}^{+}}, \operatorname{STP}} \geq \frac{|\mathcal{U}_{\mathcal{S}}| - 1 + |I|}{|\mathcal{U}_{\mathcal{S}}| - 1 + \frac{|\mathcal{S}|}{\min_{e \in \mathcal{U}} f_{\mathcal{S}}(e)}}\]
\end{theorem}
\begin{proof}
	This follows by \cref{thm:set-cover-based-steiner-tree-instance-gap-lower-bound} with \(\operatorname{gap}_{\operatorname{\mathcal{BCR}^{+}}, \operatorname{STP}} \geq \lim_{p \to \infty} \operatorname{gap}_{\operatorname{\mathcal{BCR}^{+}}, \operatorname{STP}}(\operatorname{SCI}_{\mathcal{S}, p})\).
%	\begin{align*}
%		\lim_{p \to \infty} \operatorname{gap}_{\operatorname{\mathcal{BCR}^{+}}, \operatorname{STP}}(\operatorname{SCI}_{\mathcal{S}, p})
%		& \geq \lim_{p \to \infty} \frac{\left( 1 + \frac{1}{|\mathcal{U}_{\mathcal{S}}| - 1} \cdot |I| \right) \cdot \left( |\mathcal{U}_{\mathcal{S}}|^{p} - 1 \right) + 1}{\left( 1 + \frac{1}{|\mathcal{U}_{\mathcal{S}}| - 1} \cdot \frac{|\mathcal{S}|}{\min_{e \in \mathcal{U}} f_{\mathcal{S}}(e)} \right) \cdot \left( |\mathcal{U}_{\mathcal{S}}|^{p} - 1 \right) + 1}\\
%		& = \frac{1 + \frac{1}{|\mathcal{U}_{\mathcal{S}}| - 1} \cdot |I|}{1 + \frac{1}{|\mathcal{U}_{\mathcal{S}}| - 1} \cdot \frac{|\mathcal{S}|}{\min_{e \in \mathcal{U}} f_{\mathcal{S}}(e)}} \qedhere
%	\end{align*}
\end{proof}

In \cite{BGRS13} the current best lower bound for \(\operatorname{gap}_{\operatorname{\mathcal{BCR}}, \operatorname{STP}}\) is proved via \cref{thm:set-cover-bcr-improved-integrality-gap-lower-bound}.
The set cover instance used is \(\mathcal{S}_{n} = \{S_{x} : x \in \{0, 1\}^{n} \land x \neq 0\}\) for \(n = 3\) where \(S_{x} = \{y \in \{0, 1\}^{n} : x^{t} \cdot y \equiv 1 \mod 2\}\).
This instance for \(n \to \infty\) yields a lower bound on the integrality gap of a classical set cover LP \cite[Ch.\ 13]{Vaz01}.
We obtain:
\begin{corollary}
	\(\operatorname{gap}_{\operatorname{\mathcal{BCR}^{+}}, \operatorname{STP}} \geq \frac{36}{31} \approx 1.161\)
\end{corollary}
\begin{proof}
	We have \(|\mathcal{S}_{3}| = |\mathcal{U}_{\mathcal{S}_{3}}| = 7\) and it can be easily shown that for each \(e \in \mathcal{U}_{\mathcal{S}_{3}}\) \(f_{\mathcal{S}_{3}}(e) = 4\) and an optimum solution for the set cover instance has cardinality \(3\).
	Hence by \cref{thm:set-cover-bcr-improved-integrality-gap-lower-bound}:
	\[\operatorname{gap}_{\operatorname{\mathcal{BCR}^{+}}, \operatorname{STP}} \geq \frac{|\mathcal{U}_{\mathcal{S}_{3}}| - 1 + |I|}{|\mathcal{U}_{\mathcal{S}_{3}}| - 1 + \frac{|\mathcal{S}_{3}|}{\min_{e \in \mathcal{U}_{\mathcal{S}_{3}}} f_{\mathcal{S}_{3}}(e)}} = \frac{7 - 1 + 3}{7 - 1 + \frac{7}{4}} = \frac{36}{31} \approx 1.161 \qedhere\]
\end{proof}

The case \(p = 1\) is Skutella's instance and for that we obtain with \cref{thm:set-cover-based-steiner-tree-instance-gap-lower-bound}:
\[\operatorname{gap}_{\operatorname{\mathcal{BCR}^{+}}, \operatorname{STP}}(\operatorname{SCI}_{\mathcal{S}_{3}, 1}) \geq \frac{|\mathcal{U}_{\mathcal{S}_{3}}| + |I|}{|\mathcal{U}_{\mathcal{S}_{3}}| + \frac{|\mathcal{S}_{3}|}{\min_{e \in \mathcal{U}} f_{\mathcal{S}_{3}}(e)}} = \frac{7 + 3}{7 + \frac{7}{4}} = \frac{8}{7} \approx 1.142\]

In \cite{FKOS16} it is shown that \(\operatorname{gap}_{\operatorname{\mathcal{BCR}}, \operatorname{\mathcal{HYP}}} \geq \frac{8}{7}\).
This is proved via the set cover based Steiner tree instance \(\operatorname{SCI}_{\mathcal{S}_{2}, p}\) with the set cover instance \(\mathcal{S}_{n}\) from above with \(n = 2\) with an additional vertex \(r'\) that is connected to  \(\operatorname{SCG}_{\mathcal{S}_{2}, p}\) by the new edge \(\{r, r'\}\) and replacing the \(r\) by \(r'\) in the set of required vertices.
Equivalently one can use the set cover instance \(\mathcal{S} = \{\{1, 2\}, \{1, 3\}, \{2, 3\}\}\) (see \cref{fig:set-cover-instance-p-3}).
It is easy to see that one can obtain the same ratio by considering the LP-relaxations in \BCRimprovedclass\ and therefore:

\begin{corollary}
	\(\operatorname{gap}_{\operatorname{\mathcal{BCR}^{+}}, \operatorname{\mathcal{HYP}}} \geq \frac{8}{7} \approx 1.142\)
\end{corollary}

%\subfile{figures/figure-set-cover-instance-example-02}

\paragraph{Remark.}
In \cite[Sec.\ 4.9]{Pritchard2009} the set cover based Steiner tree instance \(\operatorname{SCI}_{\mathcal{S}, 1}\) for the set cover instance \(\mathcal{S} = \{[n] \setminus \{i\} : i \in [n]\}\) is considered and it is shown that the solution from the proof of \cref{thm:set-cover-terminal-instance-fractional-value-bound} is an extreme point of \nameref{lp:BCR}.
Therefore the number of edges with positive solution value is in \(\Omega(n^{2}) = \Omega(|V|^{2})\), although the integral optimum solution only needs \(|n + 2| \in \mathcal{O}(|V|)\) edges.
%An example for \(n = 4\) is presented in \cref{fig:dense-extreme-point-instance-bcr}.

	\pagebreak
	
	\begingroup
	\small
	\bibliographystyle{alpha}
	\bibliography{vicari-mathematics-references}
	\endgroup
	
%	\pagebreak
%	
%	\tableofcontents
\end{document}